\documentclass[
 aps,
 prx,
 superscriptaddress,
 amsmath,amssymb,
 twocolumn,
 reprint,
 longbibliography
]{revtex4-2}

\usepackage{graphicx}% Include figure files
\usepackage{dcolumn}% Align table columns on decimal point
\usepackage{physics}
\usepackage{newtxtext}
\usepackage{xcolor}
\usepackage{zxjatype}
\usepackage{bm}
\usepackage{braket}
\usepackage[legacycolonsymbols]{mathtools}

\usepackage[
    colorlinks = true,
    linkcolor = blue,
    citecolor=blue,
    urlcolor = blue
]{hyperref}

\graphicspath{{main_figures/}} % This is where you have saved the images

\begin{document}

\title{Quantitative imaging of nonlinear spin-wave propagation using diamond quantum sensors}

\author{Kensuke Ogawa}
\email{kensuke.ogawa.phys@gmail.com}
\affiliation{Department of Physics, \href{https://ror.org/057zh3y96}{The University of Tokyo}, Bunkyo-ku, Tokyo, 113-0033, Japan}
\author{Moeta Tsukamoto}
\affiliation{Department of Physics, \href{https://ror.org/057zh3y96}{The University of Tokyo}, Bunkyo-ku, Tokyo, 113-0033, Japan}
\author{Yusuke Mori} 
\affiliation{Department of Physics, \href{https://ror.org/035t8zc32}{Osaka University}, Toyonaka, Osaka 560-0043, Japan}
\author{Daigo Takafuji} 
\affiliation{Department of Physics, \href{https://ror.org/035t8zc32}{Osaka University}, Toyonaka, Osaka 560-0043, Japan}
\author{Junichi Shiogai}
\affiliation{Department of Physics, \href{https://ror.org/035t8zc32}{Osaka University}, Toyonaka, Osaka 560-0043, Japan}
\affiliation{Division of Spintronics Research Network, Institute for Open and Transdisciplinary Research Initiatives, \href{https://ror.org/035t8zc32}{Osaka University}, Suita,
Osaka 565-0871, Japan}
\author{Kohei Ueda}
\affiliation{Department of Physics, \href{https://ror.org/035t8zc32}{Osaka University}, Toyonaka, Osaka 560-0043, Japan}
\affiliation{Division of Spintronics Research Network, Institute for Open and Transdisciplinary Research Initiatives, \href{https://ror.org/035t8zc32}{Osaka University}, Suita,
Osaka 565-0871, Japan}
\author{Jobu Matsuno}
\affiliation{Department of Physics, \href{https://ror.org/035t8zc32}{Osaka University}, Toyonaka, Osaka 560-0043, Japan}
\affiliation{Division of Spintronics Research Network, Institute for Open and Transdisciplinary Research Initiatives, \href{https://ror.org/035t8zc32}{Osaka University}, Suita,
Osaka 565-0871, Japan}
\author{Jun-ichiro Ohe}
\affiliation{Department of Physics, \href{https://ror.org/02hcx7n63}{Toho University}, 2-2-1 Miyama, Funabashi 274-8510, Japan}
\author{Kento Sasaki}
\affiliation{Department of Physics, \href{https://ror.org/057zh3y96}{The University of Tokyo}, Bunkyo-ku, Tokyo, 113-0033, Japan}
\author{Kensuke Kobayashi}
\affiliation{Department of Physics, \href{https://ror.org/057zh3y96}{The University of Tokyo}, Bunkyo-ku, Tokyo, 113-0033, Japan}
\affiliation{Institute for Physics of Intelligence, \href{https://ror.org/057zh3y96}{The University of Tokyo}, Bunkyo-ku, Tokyo, 113-0033, Japan}
\affiliation{Trans-Scale Quantum Science Institute, \href{https://ror.org/057zh3y96}{The University of Tokyo}, Bunkyo-ku, Tokyo, 113-0033, Japan}

\begin{abstract}
Spin waves propagating in magnetic materials exhibit nonlinear behavior at large amplitudes due to the competition between excitation and relaxation, providing an attractive platform for exploring nonlinear wave dynamics. In particular, spin waves with a non-zero wavenumber that carry momentum undergo nonlinear relaxation and experience wavenumber modulation in the nonlinear regime. This nonlinearity has been observed experimentally—for example, in \href{https://link.aps.org/doi/10.1103/PhysRevApplied.17.034010}{S. R. Lake \textit{et al}., Phys. Rev. Appl. 17, 034010 (2022)}— but a quantitative comparison with theory has not yet been carried out. Here, We image nonlinear spin-wave propagation in two yttrium iron garnet thin films with distinct spin-wave decay rates using a wide-field quantum diamond microscope. We obtain quantitative distributions of spin-wave amplitude and phase as a function of the excitation microwave strength. As a result, we observe a threshold in the spin-wave amplitude beyond which nonlinear effects become evident and confirm that this threshold is consistent with theoretical predictions based on four-magnon scattering for both samples. Moreover, as the amplitude of the spin waves increases, we observe modulation of the wavenumber across the field of view. We attribute this modulation primarily to a reduction in the saturation magnetization caused by incoherent spin waves generated by multi-magnon scattering. Our quantitative measurements provide a pathway for visualizing nonlinear spin-wave dynamics and are crucial for deepening our understanding of the underlying mechanisms.
\end{abstract}

\maketitle
\date{\today}

\section{Introduction}
\label{sec:intro}
Spin waves, collective excitations of ordered magnets, possess an intrinsic nonlinearity \cite{Suhl1957209, Krivosik2010}. This nonlinearity arises from the competition between spin-wave excitation and relaxation and becomes prominent when the spin-wave amplitude exceeds a certain threshold. The nonlinear dynamics of the uniform mode (i.e., ferromagnetic resonance, FMR) have been comprehensively studied since the 1950s. In these studies, various nonlinear processes have been explored, including parametric excitation \cite{Bloembergen1952relaxation, Damon1953relaxation, Bloembergen1954relaxation}, Bose-Einstein condensation of magnons \cite{Demokritov2006Bose}, and the first and second Suhl instabilities \cite{Suhl1957209}. Recently, significant interest has focused on the dynamics of spin waves with non-zero wavenumbers due to their rich physics and potential applications in information devices \cite{Pirro2021a}. The nonlinear behavior of these spin waves has been investigated both theoretically and experimentally \cite{Schultheiss2012, Tobias2020, Mohseni2021, Boardman1988, Mark2004, Lake2022}, in which the initially excited coherent spin waves are scattered by multi-magnon processes, thereby magnons with various wavenumbers and frequencies are generated. In addition, when the spin-wave amplitude becomes large, an effective reduction in the static magnetization leads to a frequency shift of the spin waves \cite{Krivosik2010, Lake2022, Merbouche2022}.
\par  
These nonlinear dynamics of the spin waves are primarily investigated using a vector network analyzer (VNA) \cite{Mark2004}, magneto-optical effects \cite{Nembach2011}, and, especially in recent years, Brillouin light scattering (BLS) \cite{Schultheiss2012, Tobias2020, Mohseni2021,Lake2022}. BLS enables time-, space-, and frequency-resolved detection of spin-wave propagation, making it a valuable technique for visualizing these dynamics \cite{Sebastian2015micro}. However, the signal obtained via BLS is the intensity of the scattered light arising from Raman scattering processes. While this intensity is proportional to the magnon intensity, there is no direct theoretical relationship that links it to the spin-wave amplitude. Although Ref.~\cite{Lake2022} systematically investigates the nonlinearity of surface spin waves propagating on a magnetic waveguide, discusses the underlying mechanism, and estimates the precession angle at which nonlinearity emerges by calibrating the spin-wave amplitude from the variation in wavenumber, quantitative comparisons with theoretical predictions have not been carried out, and thus the validity of the obtained amplitude has not yet been verified.
\par  
Recently, nitrogen-vacancy (NV) centers in diamonds \cite{doherty2013nitrogen} have gained significant attention as quantitative sensors for microwave amplitude. With a resonant frequency of approximately
$3 \, \mathrm{GHz}$, they are particularly sensitive to microwave fields in this frequency range, making them well suited for sensing spin waves, whose frequencies typically lie in the several GHz range. In previous studies, it has been demonstrated that both the amplitude and phase of coherently propagating spin waves in magnetic films can be quantitatively imaged \cite{van2015nanometre,Andrich2017,kikuchi2017long,bertelli2020magnetic,zhou2021magnon}.
\par
Although some previous studies employing NV centers has observed nonlinear spin-wave dynamics, such as the generation of incoherent spin waves at the resonant frequencies of the NV spins when exciting the FMR mode \cite{Wolfe2014offresonant,LeeWong2020nanoscale} and the frequency conversion of coherent spin waves by spin-wave mixing through four-magnon scattering \cite{Carmiggelt2023}, no study to date has quantitatively examined the nonlinear dynamics of propagating spin waves with non-zero wavenumbers themselves.
\par

\begin{figure}[t]
    \centering
    \includegraphics[width=\linewidth]{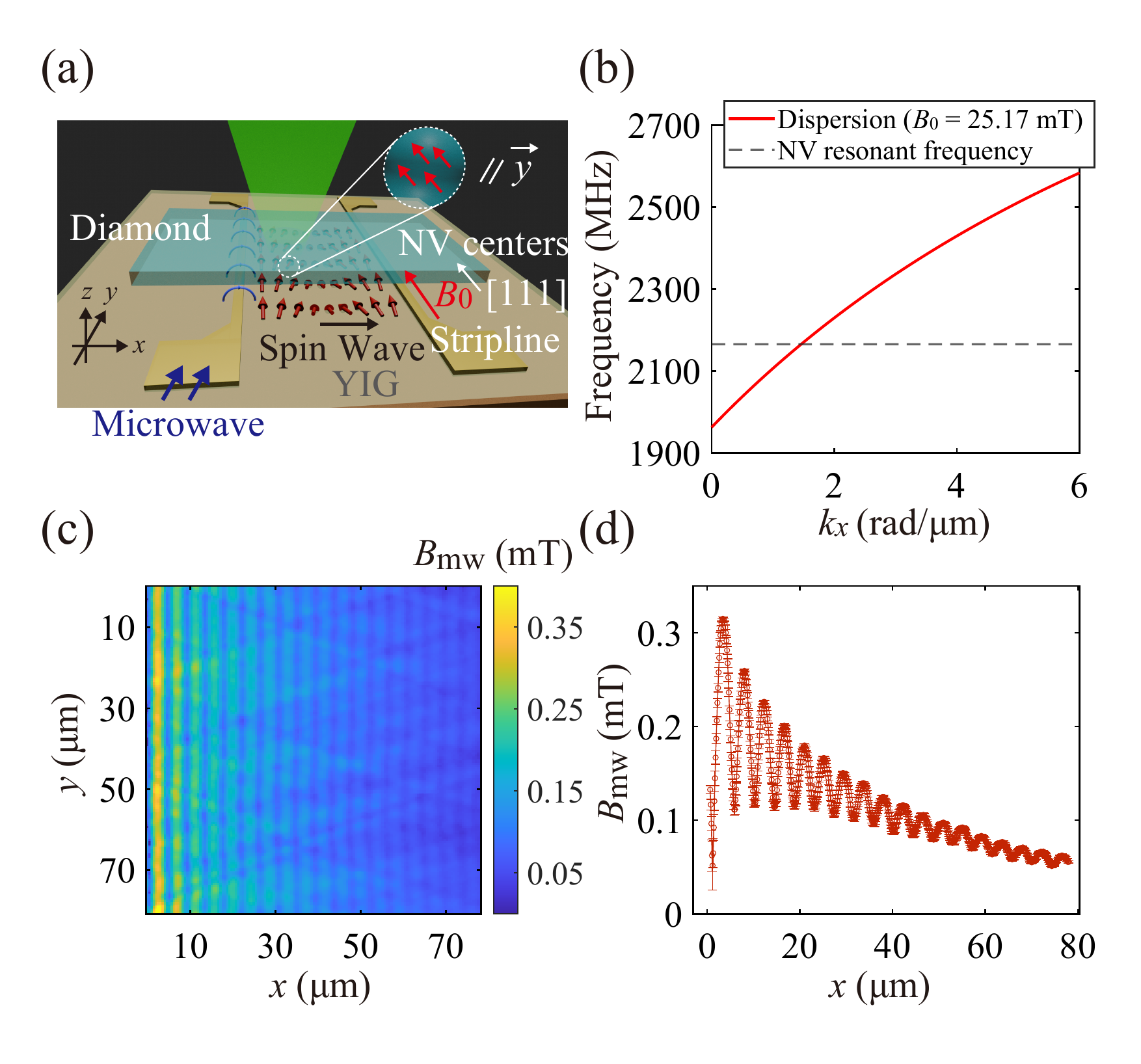}
    \caption{
    (a) Schematic of the experiment.
    (b) Dispersion relation of the surface spin waves for YIG-A at an external magnetic field $B_{0} = 25.17 \, \mathrm{mT}$. The horizontal dashed line corresponds to the resonance frequency of the NV spin.
    (c) Two-dimensional image of the microwave amplitude at an external magnetic field $B_{0} = 25.17 \, \mathrm{mT}$ and input microwave power $P_{\mathrm{mw}} = 11 \, \mathrm{dBm}$ for YIG-A.
    (d) One-dimensional microwave amplitude distribution $B_{\mathrm{mw}}(x)$ for the data in (c).
    }
    \label{fig:exp_setup}
\end{figure}

In this paper, we present results on imaging the amplitude and phase of surface spin waves in two yttrium iron garnet (YIG) thin films with distinct characteristics using a wide-field quantum diamond microscope (QDM), while systematically varying the excitation microwave amplitude. First, we show that the spin waves in the first YIG thin film, grown by sputtering, exhibit a significant decay beyond a characteristic amplitude threshold. We confirm that this threshold is consistent with theoretical predictions based on four-magnon scattering. In addition, as the amplitude of the spin waves increases, we observe modulation of the wavenumber within the field of view, and our quantitative analysis reveals that this modulation largely originates from incoherent spin waves generated by multi-magnon scattering. Finally, to validate our findings, we show that the second YIG thin film, which is grown by liquid phase epitaxy (LPE) and characterized by a smaller relaxation rate than the first YIG film, exhibits a distinct spin-wave amplitude threshold, also in agreement with theoretical expectations. This achievement demonstrates the usefulness of quantitative QDM investigations for probing nonlinear spin-wave dynamics and contributes to a systematic understanding of nonlinear spin waves.
\par
This paper is organized as follows. We describe the experimental setup and principle in Sec.~\ref{sec:exp_setup}. 
In Sec.~\ref{sec:toyonaka}, we show spin-wave imaging in the linear regime for the first YIG film.  We then present the nonlinear behavior that emerges as the spin-wave amplitude is increased in Sec.~\ref{sec:from_linear_to_nonlinear}, and quantitatively extract the spin-wave amplitude in Sec.~\ref{sec:nonlinear_behavior}. Then, we introduce the theoretical formulation to discuss the threshold between the linear and nonlinear regimes and apply it to our experimental findings in Sec.~\ref{sec:theoretical_interpretation}. We address the modulation of the wavenumber in Sec.~\ref{sec:magnetization_reduction}. Sec.~\ref{sec:comparison_with_a_different_YIG} is devoted to comparing with the second YIG film.
Finally, we conclude the paper with a summary and outlook in Sec.~\ref{sec:conclusion}.

\section{Methods}
\label{sec:exp_setup}
The experimental setup is depicted in Fig.~\ref{fig:exp_setup}(a), where we employ a QDM as the measurement system. In this setup, a green laser (wavelength: $515 \, \mathrm{nm}$) irradiates an ensemble of NV centers, and their red fluorescence is collected using an sCMOS camera \cite{ogawa2024wideband}. The magnification of the microscope is $100\times$, and the field of view is approximately $100 \, \mathrm{\mu m}$ square.
\par
The diamond chip used in this study is fabricated by cutting a (111) oriented Ib-type diamond single crystal (Element Six, MD111) into a substrate with a thickness of $38 \, \mathrm{\mu m}$ in the (110) orientation. 
An NV ensemble layer is formed by carbon ion implantation ($30 \, \mathrm{keV}$, $1 \times 10^{12} \, \mathrm{cm^{-2}}$) and subsequent anneals \cite{tetienne2018spin,ogawa2024wideband}. This process creates an NV layer at a depth of around $30 \, \mathrm{nm}$ from the surface. We utilize NV centers oriented in the in-plane direction [along the $y$-axis; see Fig.~\ref{fig:exp_setup}(a)].
\par
We use two YIG thin films with different growth methods as our magnetic materials. The first film, ``YIG-A'', is epitaxially grown on (111)-oriented gadolinium gallium garnet (GGG) substrate using RF sputtering in the group of the co-authors of this paper \cite{fukushima2022spin}, and has a thickness of $d = 54 \, \mathrm{nm}$. The second film, ``YIG-B'', is epitaxially grown on a (111)-oriented GGG substrate by LPE and is commercially obtained from Matesy GmbH, with a thickness of $d = 109 \, \mathrm{nm}$. 
For YIG-A, the measurement setup is identical to that described in Ref.~\cite{ogawa2024wideband} except for the configuration of the microwave circuits.
\par
Striplines with a width of approximately $20 \, \mathrm{\mu m}$ and a length of $2 \, \mathrm{mm}$, with $500 \, \mathrm{\mu m} \times 500 \, \mathrm{\mu m} $ pads for wiring at both ends, are deposited on the film surface for both YIG samples. Microwaves are input to the stripline on the left side to excite spin waves [Fig. 1(a)]. The striplines on YIG-A consist of Ti/Cu/Au ($10 \, \mathrm{nm}/190 \, \mathrm{nm}/9 \, \mathrm{nm}$), and those on YIG-B consist of Ti/Cu/Au ($10 \, \mathrm{nm}/97 \, \mathrm{nm}/9 \, \mathrm{nm}$). Microwaves are first amplified from the signal generator through an amplifier and are then fed into the coplanar waveguide; from there, they are delivered to the stripline by wire bonding gold wires to the pad. The other side is also connected to the coplanar waveguide and terminated with $50 \, \Omega$.
We define the input microwave power $P_{\mathrm{mw}}$ as the power supplied to the coplanar waveguide. Consequently, due to differences in losses, the actual microwave power reaching the stripline may differ between YIG-A and YIG-B.
\par
The diamond chips are attached to the YIG films using varnish, with the surface containing the NV layer facing downward. The distance $z_{d}$ between the NV layer and the surface of the magnetic film is estimated by evaluating the distribution of the stray magnetic field from the stripline on the surface of the YIG film under current application \cite{Tsukamoto2021, ogawa2024wideband}. For YIG-A and YIG-B, the distance $z_{d}$ is estimated to be $(878 \pm 20) \, \mathrm{nm}$ and $z_{d} = (860 \pm 18) \, \mathrm{nm}$, respectively (see Supplemental Materials for details \cite{SM}).
\par
A static magnetic field is applied in the film plane along the stripline ($y$-axis) using a Helmholtz coil. In this configuration, surface spin waves propagating perpendicular to the magnetic field are excited \cite{Serga2010}.
\par
Using Rabi oscillation measurements, we detect spin waves whose frequency matches the resonance between the $m_{s} = 0$ and $m_{s} = -1$ states of the NV spins  \cite{Jelezko2004, Andrich2017, bertelli2020magnetic}. In each region of interest, we fit the Rabi oscillation signals and obtain the Rabi frequency, which is then converted to the amplitude of the circularly polarized microwave field $B_{\mathrm{mw}}$ \cite{ogawa2024wideband}.
\par

\section{Results and Discussions}
\subsection{Spin-wave imaging in the linear regime}
\label{sec:toyonaka}
First, we present the results for YIG-A. The external magnetic field is set to $B_{0} = 25.17 \, \mathrm{mT}$. Figure~\ref{fig:exp_setup}(b) shows the dispersion relation of the surface spin waves under this magnetic field as well as the resonance frequency of the NV spins. The dispersion relation is given by the following equation \cite{kalinikos1986theory, Yu2019, Bertelli2021}:
\begin{equation}
\begin{split}
    f_{\mathrm{sw}}(k_{x}) = \gamma_{e} \mu_{0}  &\sqrt{\qty(H_{0} + M_{\mathrm{eff}}(1-g(k_{x}d))) \qty(H_{0} + M_{\mathrm{eff}}g(k_{x}d))}, \\
    g(k_{x}d) &= 1 - \frac{1-e^{-k_{x}d}}{k_{x}d},
    \label{eq:dispersion_analysis}
\end{split}
\end{equation}
where $\gamma_{e}$ is the gyromagnetic ratio of an electron spin, $\mu_{0}$ is the permeability of vacuum, $M_{\mathrm{eff}} \, (= 169.6 \, \mathrm{mT})$ is the effective magnetization \cite{ogawa2024wideband}, and $d = 54 \, \mathrm{nm}$ is the thickness of the YIG film. Under this magnetic field, the resonance frequency of the NV spins $f_{\mathrm{NV}}$ is $2165.8 \, \mathrm{MHz}$, and the wavenumber of the surface spin waves at this frequency is around $1.4 \, \mathrm{rad/\mu m}$.
\par
Figure~\ref{fig:exp_setup}(c) shows the two-dimensional distribution of the microwave field amplitude at an input microwave power of $P_{\mathrm{mw}} = 11 \, \mathrm{dBm}$. In the analysis, signals from each $520 \, \mathrm{nm}$ in the $x$- and $y$-directions are integrated and treated as a single pixel. The microwave amplitude oscillates periodically in the $x$-direction due to the interference between the microwave field generated by the spin waves propagating in the $x$-direction and the reference microwave with a spatially uniform phase across the entire field of view \cite{bertelli2020magnetic, zhou2021magnon, ogawa2024wideband}. In addition to the propagation in the $x$-direction, an increase in the microwave amplitude is observed at a specific diagonal angle. This phenomenon, known as “caustics,” arises from the anisotropy of the spin-wave dispersion relation \cite{Schneider2010, bertelli2020magnetic, ogawa2024wideband}. 
\par
Figure~\ref{fig:exp_setup}(d) displays the one-dimensional microwave amplitude distribution along the $x$-direction $B_{\mathrm{mw}}(x)$. Here, signals are integrated over a width of $13 \, \mathrm{\mu m}$ from $y = 31.87 \, \mathrm{\mu m}$ to $y = 44.87 \, \mathrm{\mu m}$ in the $y$-direction. The leftmost $x$-coordinate of the experimental data is $x = 0.65 \, \mathrm{\mu m}$. The microwave amplitude at the position of the NV centers results from the interference between the microwave field generated by the spin waves and that from the stripline, and can be expressed as \cite{bertelli2020magnetic, zhou2021magnon,ogawa2024wideband}

\begin{align}
\label{eq:Bmw_inter}
    &B_{\mathrm{mw}}(x) \\
    &= \sqrt{B^2_{\mathrm{sw}}(x) + 2  B_{\mathrm{sw}}(x) B_{\mathrm{strip}}(x) \cos (k_{x}(x)x + \theta_{0}) + B^2_{\mathrm{strip}}(x)}, \notag 
\end{align}
where $B_{\mathrm{strip}}(x)$ is the microwave amplitude distribution from the stripline, $B_{\mathrm{sw}}(x)$ is that from the spin waves, $k_{x}(x)$ is the spatial distribution of the $x$-component of the wavenumber and $\theta_{0}$ is the phase of the spin waves at the left edge of the field of view. When $B_{\mathrm{sw}}$ ($B_{\mathrm{strip}}$) is sufficiently larger than $B_{\mathrm{strip}}$ ($B_{\mathrm{sw}}$), $B_{\mathrm{sw} (\mathrm{strip})} \gg B_{\mathrm{strip} (\mathrm{sw})}$, the above equation can be approximated as
\begin{equation}
    B_{\mathrm{mw}}(x) \simeq B_{\mathrm{sw}(\mathrm{strip})}(x) + B_{\mathrm{strip}(\mathrm{sw})}(x) \cos (k_{x}(x)x + \theta_{0}).
    \label{eq:Btot_approx}
\end{equation}
For YIG-A, under low input microwave power, the condition $B_{\mathrm{sw}}(x) > B_{\mathrm{mw}}(x)$ holds in the entire field of view (see Supplemental Materials for details \cite{SM}). Consequently, the mean line (the line passing through the center of the oscillations) of $B_{\mathrm{mw}}(x)$ shown in Fig.~\ref{fig:exp_setup}(d) corresponds to $B_{\mathrm{sw}}(x)$, while the oscillatory component corresponds to $B_{\mathrm{strip}}(x)$. From the decay of this mean line, we can estimate that the decay length of the spin waves is approximately $50 \, \mathrm{\mu m}$.

\subsection{From linear to nonlinear regime}
\label{sec:from_linear_to_nonlinear}
We examine the case where the input power is varied in 10 steps from $5 \, \mathrm{dBm}$ to $25 \, \mathrm{dBm}$. Figure~\ref{fig:matsuno_psweep}(a) displays $B_{\mathrm{mw}}(x)$ at five selected power values (see Supplemental Materials for the full dataset \cite{SM}). The triangular markers with error bars represent the experimental data, while the solid black lines indicate the fitting results discussed later. The integrated region along the $y$-direction is the same as in Fig.~\ref{fig:exp_setup}(d). From $P_{\mathrm{mw}} = 11 \, \mathrm{dBm}$ to about $P_{\mathrm{mw}} = 17 \, \mathrm{dBm}$, the shape of the amplitude distribution remains unchanged; however, for input power above this range, significant modulation of the distribution is observed.

\begin{figure}[t]
    \centering
    \includegraphics[width=\linewidth]{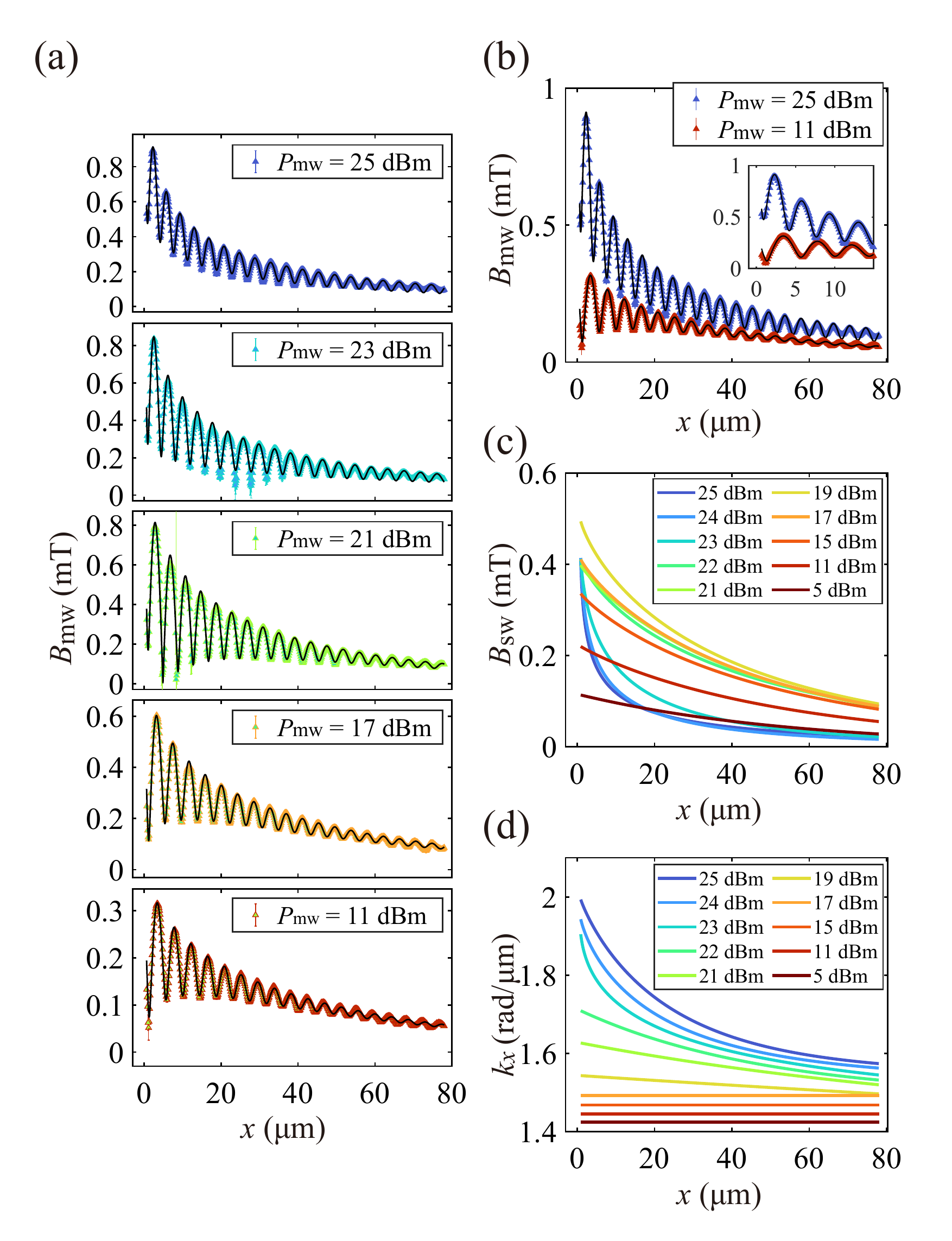}
    \caption{
    (a) One-dimensional microwave amplitude distribution $B_{\mathrm{mw}}(x)$ obtained at five different input microwave powers from $P_{\mathrm{mw}} = 11 \, \mathrm{dBm}$ to $25 \, \mathrm{dBm}$ for YIG-A. The triangular markers with error bars represent experimental data, and the solid black lines correspond to the fitting results.
    (b) Comparison of $B_{\mathrm{mw}}(x)$ for data between large input microwave power ($P_{\mathrm{mw}} = 25 \, \mathrm{dBm}$) and small input microwave power ($P_{\mathrm{mw}} = 11 \, \mathrm{dBm}$) for YIG-A.
    (c) Microwave amplitude and (d) Wavenumber distribution from the spin waves for each input microwave power calculated from the fitting results for YIG-A.
    }
    \label{fig:matsuno_psweep}
\end{figure}

Figure~\ref{fig:matsuno_psweep}(b) compares $B_{\mathrm{mw}}(x)$ measured at $P_{\mathrm{mw}} = 11 \, \mathrm{dBm}$ and $P_{\mathrm{mw}} = 25 \, \mathrm{dBm}$. The first notable feature is that the shapes of the mean lines of $B_{\mathrm{mw}}(x)$ differ between the two power levels. This variation arises because $B_{\mathrm{strip}}(x)$ increases linearly with the input microwave power, whereas $B_{\mathrm{sw}}(x)$ exhibits a nonlinear response. According to Eq.~(\ref{eq:Btot_approx}), at low input powers, the mean line primarily reflects $B_{\mathrm{sw}}(x)$. In contrast, at high input powers, the mean line is dominated by the contribution from the stripline (see Supplemental Materials for details \cite{SM}). Furthermore, in the region of smaller $x$-coordinate [see the inset of Fig.~\ref{fig:matsuno_psweep}(b)], the spacing between the peaks of the microwave amplitude oscillations is noticeably shorter at $P_{\mathrm{mw}} = 25 \, \mathrm{dBm}$ compared to $11 \, \mathrm{dBm}$. This increase in the wavenumber is likely due to a modulation of the dispersion relation, caused by a reduction in the static magnetization along the $y$-direction resulting from the increase in the spin-wave amplitude \cite{Lake2022}.
\par
Next, we quantitatively analyze the observed nonlinear dynamics by fitting the measured microwave amplitude distribution. Based on Eq.~(\ref{eq:Bmw_inter}), we extract the microwave amplitude distribution from the spin waves $B_{\mathrm{sw}}(x)$, the wavenumber distribution of the spin waves $k_{x}(x)$, and the microwave amplitude distribution from the stripline $B_{\mathrm{strip}}(x)$. For the fitting, we assume the following expressions for $B_{\mathrm{sw}}(x)$, $k_{x}(x)$, and $B_{\mathrm{strip}}(x)$:
\begin{equation}
\begin{split}
    B_{\mathrm{sw}}(x) &= B^{0}_{\mathrm{sw}} e^{-\qty(\frac{x-x_{0}}{l_{\mathrm{sw}}})^{p_{\mathrm{sw}}}}, \\ 
    k_{x}(x) &= k_{0} + k_{1} e^{-\qty(\frac{x-x_{0}}{l_{\mathrm{k}}})^{p_{\mathrm{k}}}}, \\
    B_{\mathrm{strip}}(x) &= B^{0}_{\mathrm{strip}} e^{-\qty(\frac{x-x_{0}}{l_{\mathrm{strip}}})^{p_{\mathrm{strip}}}},
\end{split}
\label{eq:Bmw_fit_comp}
\end{equation}
where $x_{0}$ denotes the coordinate of the left edge of the field of view ($x_{0} = 0.65 \, \mathrm{\mu m}$ in this measurement). For $B_{\mathrm{sw}}(x)$ and $B_{\mathrm{strip}}(x)$, we adopt stretched exponential functions. The function for the spin waves (stripline) is characterized by the amplitude at $x = x_{0}$, $B^{0}_{\mathrm{sw}} = B_{\mathrm{sw}}(x_0)$ ($B^{0}_{\mathrm{strip}} = B_{\mathrm{strip}}(x_0)$), the decay length $l_{\mathrm{sw}}$ ($l_{\mathrm{strip}}$), and the stretch factor $p_{\mathrm{sw}}$ ($p_{\mathrm{strip}}$). In the linear regime, $B_{\mathrm{sw}}(x)$ generally decays exponentially; however, in the nonlinear regime, rapid decay of the amplitude near the stripline has been reported \cite{Lake2022}. Therefore, we introduce a stretch factor into the exponential function to capture this behavior (see Supplemental Materials for details \cite{SM}). For the wavenumber distribution, we adopt a function consisting of a stretched exponential function corresponding to the spatial modulation of the wavenumber and a constant term $k_{0}$. The stretched  exponential function is characterized by the wavenumber modulation at $x = x_{0}$, $k_{1}$, the decay length $l_{k}$, and the stretch factor $p_{\mathrm{k}}$. 
\par
When performing the fitting, we note that Eq.~(\ref{eq:Bmw_inter}) is symmetric with respect to $B_{\mathrm{sw}}(x)$ and $B_{\mathrm{strip}}(x)$. Starting from random initial values makes it impossible to distinguish between these two contributions. To resolve this ambiguity, we use the microwave amplitude distribution obtained under the same input microwave power but with the direction of the external magnetic field reversed as a reference. When the magnetic field is reversed, the amplitude of the spin waves propagating in the $+x$-direction decreases due to a reduction of the spin-wave excitation efficiency of the stripline \cite{bertelli2020magnetic, Simon2021}. Moreover, the microwave field from the spin waves propagating in the $+x$-direction in this configuration does not induce transitions between the NV spins' $m_{s} = 0$ and $m_{s} = -1$ states because of its polarization \cite{bertelli2020magnetic, Rustagi2020sensing}. Thus, we assume that the reference microwave amplitude distribution arises primarily from the stripline. Accordingly, we first fit the reference distribution and then use the obtained fitting parameters as the initial values for $B_{\mathrm{strip}}(x)$ in Eq.~(\ref{eq:Bmw_fit_comp}) (see Supplemental Materials for details \cite{SM}).
\par
The fitted curves are shown in Fig.~\ref{fig:matsuno_psweep}(a) as solid black lines. For each input microwave power, the fitted curve reproduces the experimental data well. In Fig.~\ref{fig:matsuno_psweep}(c), we present the estimated microwave amplitude distribution from the spin waves $B_{\mathrm{sw}}(x)$ obtained from these fitting results. Here, we define the left end of the plot at $x = 0.91 \, \mathrm{\mu m}$, corresponding to the region where the amplitude variation from the stripline becomes relatively small. For input powers ranging from $P_{\mathrm{mw}} = 5 \, \mathrm{dBm}$ to approximately $19 \, \mathrm{dBm}$, $B_{\mathrm{sw}}(x)$ increases with increasing input power and decays gradually over a similar decay length within the field of view. However, at microwave powers above this range, it no longer increases and instead decays rapidly near the stripline. In Fig.~\ref{fig:matsuno_psweep}(d), we show the wavenumber distribution $k_{x}(x)$ estimated by fitting for each input microwave power. As the input power increases, the wavenumber also increases. While no significant spatial variation is observed within the field of view at lower input powers, noticeable spatial variations emerge from around $P_{\mathrm{mw}} = 19 \, \mathrm{dBm}$. At the maximum power of $25 \, \mathrm{dBm}$, the wavenumber varies by more than 20\% across the field of view.
\par

\subsection{Nonlinear behavior of spin-wave amplitude}
\label{sec:nonlinear_behavior}
\begin{figure}[htbp]
    \centering
    \includegraphics[width=\linewidth]{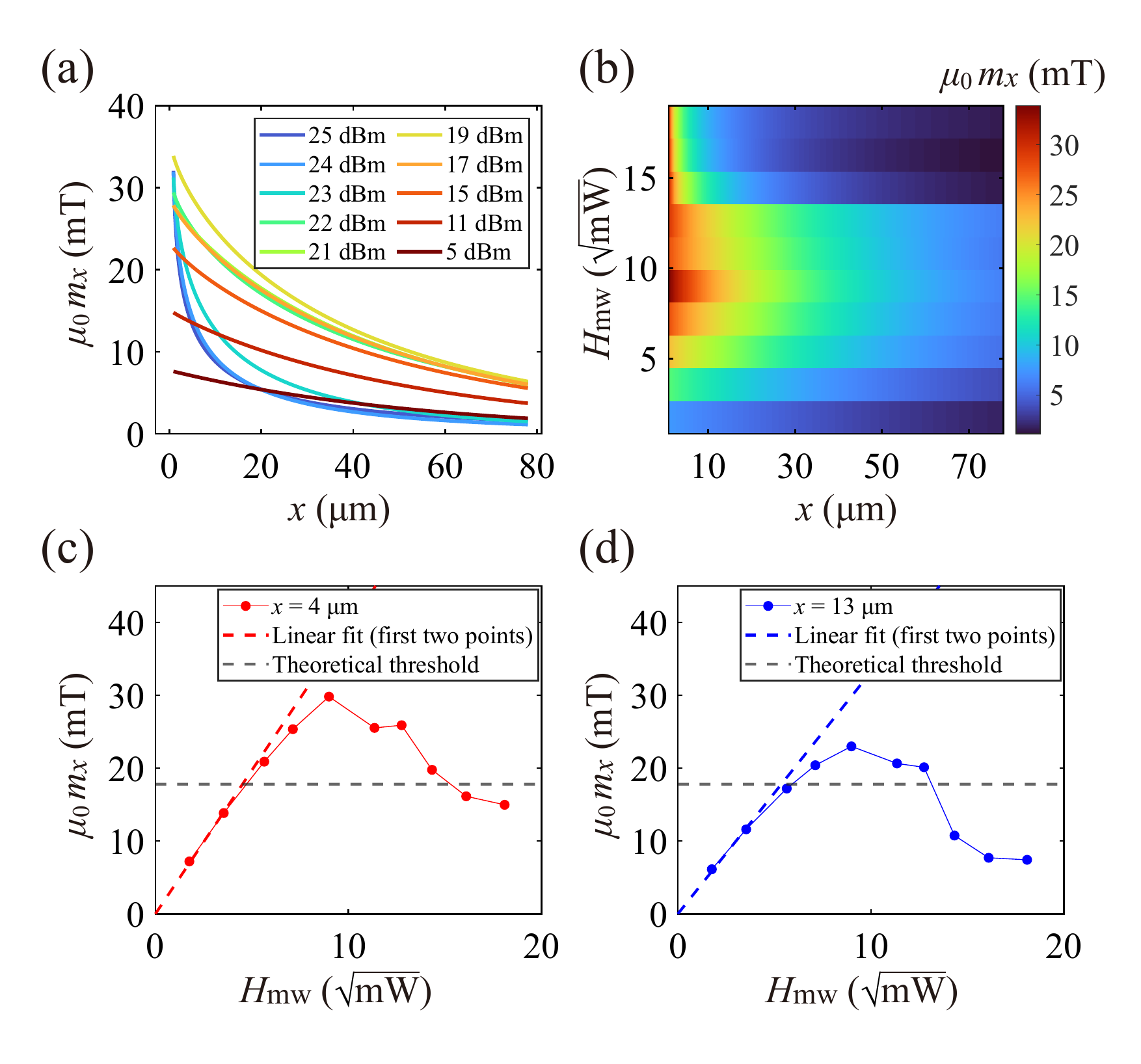}
    \caption{(a) Calculated spin-wave amplitude distribution $m_{x}(x)$ for each input microwave power for YIG-A.
    (b) Heatmap of the spin-wave amplitude distribution as a function of the input microwave amplitude.
    (c)(d) Dependence of the spin-wave amplitude on input microwave amplitude (c) at $x = 4 \, \mathrm{\mu m}$ and (d) at $x = 13 \, \mathrm{\mu m}$.}
    \label{fig:matsuno_m}
\end{figure}

Now, we conduct a quantitative analysis of the spin-wave amplitude in the nonlinear regime. We convert the obtained microwave amplitude distribution from spin waves $B_{\mathrm{sw}}(x)$ into the corresponding spin-wave amplitude distribution. $B_{\mathrm{sw}}(x)$ is related to the spin-wave amplitude distribution in the $x$-direction $m_{x}(x)$ by the following equation \cite{bertelli2020magnetic,zhou2021magnon,ogawa2024wideband}:
\begin{equation}
\label{eq:Bsw_nv}
    B_{\mathrm{sw}}(x) = \frac{\mu_{0}}{2} e^{-k_{x}(x)z_{d}} (1-e^{-k_{x}(x)d}) (1  + \eta_{k_{x}(x)}) m_{x}(x), \\
\end{equation}
where $\eta_{k_{x}(x)}$ is the ellipticity of the spin-wave precession, which can be obtained analytically from the experimental conditions (See Appendix B of Ref.~\cite{ogawa2024wideband} for details).
Figure~\ref{fig:matsuno_m}(a) shows the results of converting $B_{\mathrm{sw}}(x)$, shown in Fig.~\ref{fig:matsuno_psweep}(a), into $m_{x}(x)$ using Eq.~(\ref{eq:Bsw_nv}). As with the microwave amplitude distribution, the decay of the spin-wave amplitude near the stripline becomes more pronounced above $P_{\mathrm{mw}} = 19 \, \mathrm{dBm}$. The nonlinearity of the spin-wave amplitude is also reproduced by numerical simulations (see Supplemental Materials for details \cite{SM}), which validates our results. 
\par
Figure~\ref{fig:matsuno_m}(b) displays a heatmap of $m_{x}(x)$ from Fig.~\ref{fig:matsuno_m}(a) as a function of the input microwave amplitude $H_{\mathrm{mw}} \, (\coloneqq \sqrt{P_{\mathrm{mw}}})$. Up to approximately $H_{\mathrm{mw}} \sim 10 \, \sqrt{\mathrm{mW}}$, the spin-wave amplitude around the left edge of the field of view (small $x$ region) increases with the increase of $H_{\mathrm{mw}}$, and gradually decays along the propagation direction. Beyond this range, however, the amplitude no longer increases and instead decays rapidly from the left edge, signaling peculiar nonlinearity.

Figure~\ref{fig:matsuno_m}(c) shows the spin-wave amplitude at $x = 4 \, \mathrm{\mu m}$ as a function of the input microwave amplitude. The red dashed line represents a linear fit through the origin based on data from the two smallest input powers ($P_{\mathrm{mw}} = 5, 11 \, \mathrm{dBm}$). The spin-wave amplitude begins to deviate from linear behavior around $\mu_{0} m_{x} = 20 \, \mathrm{mT}$, increases up to approximately $30 \, \mathrm{mT}$, and then begins to decay. This deviation from linearity at a certain microwave power is consistent with previous studies using VNA and BLS \cite{Mark2004, Schultheiss2012, Tobias2020, Lake2022}. Figure~\ref{fig:matsuno_m}(d) shows the dependence of the spin-wave amplitude at $x = 13 \, \mathrm{\mu m}$. The amplitude deviates from linear behavior at the same $H_{\mathrm{mw}}$ as $x = 4 \, \mathrm{\mu m}$, saturates around $20 \, \mathrm{mT}$, and then begins to decay.

\subsection{Theoretical interpretation: four-magnon scattering}
\label{sec:theoretical_interpretation}
The observed nonlinear decay of the spin-wave amplitude can be attributed to multi-magnon scattering. In this process, excited magnons characterized by frequency $f_{\mathrm{sw}}$ and wavevector $\bm{k}$ are scattered into magnons with different frequencies and wavevectors while conserving the total energy and wavevector, and this scattering process becomes pronounced once the spin-wave amplitude exceeds a certain threshold. In this subsection, following Refs.~\cite{HolsteinMagnon, Rodrigo1999Extrinsic, rezende2020fundamentals, vov1987nonlinear, Boardman1988, Krivosik2010, Nembach2011}, we first formulate the time evolution of magnon operators under the Hamiltonian that includes nonlinear terms and calculate the threshold magnon amplitude at which nonlinearity becomes significant. We then perform a quantitative comparison with the experimental results.
\par
The magnon Hamiltonian $\hat{H}$, including nonlinear terms, can be expressed as \cite{Nembach2011, 
Krivosik2010}

\begin{align}
    \label{eq:H_all_bogo}
    \hat{H} = \hbar \sum_{\bm{k}} &\omega_{\bm{k}} \hat{c}_{\bm{k}}^{\dag} \hat{c}_{\bm{k}} \\ 
    &+ \sum_{\bm{k_{1}}+\bm{k_{2}}=\bm{k_{3}}+\bm{k_{4}}} H_{\bm{k_{1}},\bm{k_{2}},\bm{k_{3}},\bm{k_{4}}} \hat{c}_{\bm{k_{1}}}^{\dag} \hat{c}_{\bm{k_{2}}}^{\dag} \hat{c}_{\bm{k_{3}}} \hat{c}_{\bm{k_{4}}}, \notag
\end{align}
where the first term represents the linear term describing the dispersion relation of the spin waves, while the second term corresponds to the nonlinear four-magnon scattering process characterized by the coupling strength $H_{\bm{k_{1}},\bm{k_{2}},\bm{k_{3}},\bm{k_{4}}}$ (see Supplemental Materials for the explicit expressions for each term and their derivations \cite{SM}). Although nonlinear processes such as three-magnon scattering also appear among the nonlinear terms, they do not contribute due to conservation laws governing frequency and wavevector. Therefore, we consider only the four-magnon scattering term described above (see Supplemental Materials for details \cite{SM}). \par
In the Heisenberg picture, the time evolution of the magnon operator $\hat{c}_{\bm{k}}$ follows the Heisenberg equation:
\begin{equation}
\label{eq:Heq}
    i \hbar \frac{\mathrm{d}}{\mathrm{d}t} \hat{c}_{\bm{k}} = \qty[\hat{c}_{\bm{k}}, \hat{H}].
\end{equation}
Here, we consider the situation where only magnons with wavevector $\bm{k}$ are excited by a microwave field. These excited magnons subsequently undergo four-magnon scattering such that $\bm{k_{1}} = \bm{k} + \bm{q}$, $\bm{k_{2}} = \bm{k} - \bm{q}$, and $\bm{k_{3}} = \bm{k_{4}} = \bm{k}$ in Eq.~(\ref{eq:H_all_bogo}) \cite{vov1987nonlinear, Boardman1988}. In the rotating frame defined as $\hat{c}^{'}_{\bm{k}-\bm{q}} \, (= \hat{c}_{\bm{k}-\bm{q}} e^{i\omega_{\bm{k}-\bm{q}} t})$, the time evolution of the magnon annihilation operator obeys the following differential equation:
\begin{equation}
\label{eq:4mag_trans}
    \frac{\mathrm{d}}{\mathrm{d}t} \hat{c}^{'}_{\bm{k}-\bm{q}} = - \Gamma_{\bm{k}-\bm{q}} \hat{c}^{'}_{\bm{k}-\bm{q}} - i H_{\bm{k}-\bm{q},\bm{k}+\bm{q},\bm{k},\bm{k}} \hat{c}_{\bm{k}+\bm{q}}^{' \dag} \hat{c}^{'}_{\bm{k}} \hat{c}^{'}_{\bm{k}},
\end{equation}
where we have introduced the decay term $\Gamma_{\bm{k}-\bm{q}}$. Following the same discussion, the time evolution of $\hat{c}^{' \dag}_{\bm{k}+\bm{q}}$ obeys the following differential equation:
\begin{equation}
\label{eq:4mag_trans_dag}
    \frac{\mathrm{d}}{\mathrm{d}t} \hat{c}^{' \dag}_{\bm{k}+\bm{q}} = - \Gamma_{\bm{k}+\bm{q}} \hat{c}^{' \dag}_{\bm{k}+\bm{q}} + i H^{*}_{\bm{k}+\bm{q},\bm{k}-\bm{q},\bm{k},\bm{k}} \hat{c}_{\bm{k}-\bm{q}}^{'} \hat{c}^{' \dag}_{\bm{k}} \hat{c}^{' \dag}_{\bm{k}}.
\end{equation}
\par
For the nonlinear term in Eqs.~(\ref{eq:4mag_trans}) and (\ref{eq:4mag_trans_dag}), we assume that $\hat{c}^{'}_{\bm{k}}$ attains the steady-state value $\ev{\hat{c}^{'}_{\bm{k}}}$ under resonant microwave excitation \cite{Zagury1971}. By combining Eqs.~(\ref{eq:4mag_trans}) and (\ref{eq:4mag_trans_dag}), we derive the following differential equation for $\hat{c}^{'}_{\bm{k}-\bm{q}}$:
\begin{align}
\label{eq:4mag_trans_double}
    \qty(\frac{\mathrm{d}}{\mathrm{d}t} + \Gamma_{\bm{k}+\bm{q}}) & \qty(\frac{\mathrm{d}}{\mathrm{d}t} + \Gamma_{\bm{k}-\bm{q}}) \hat{c}^{'}_{\bm{k}-\bm{q}} \\
    &= H^{*}_{\bm{k}+\bm{q},\bm{k}-\bm{q},\bm{k},\bm{k}} H_{\bm{k}-\bm{q},\bm{k}+\bm{q},\bm{k},\bm{k}} \qty|\ev{\hat{c}^{'}_{\bm{k}}}|^{4} \hat{c}^{'}_{\bm{k}-\bm{q}}. \notag
\end{align}

Now, we define the magnon generation rate $\gamma_{\bm{k}-\bm{q}}$ by assuming that $\hat{c}^{'}_{\bm{k}-\bm{q}} \propto e^{\gamma_{\bm{k}-\bm{q}}t}$. Taking the limit $\bm{q} \rightarrow 0$ in Eq.~(\ref{eq:4mag_trans_double}), we obtain
\begin{equation}
    \gamma_{\bm{k}} = -\Gamma_{\bm{k}} + \qty|H_{\bm{k},\bm{k},\bm{k},\bm{k}}| \qty|\ev{\hat{c}^{'}_{\bm{k}}}|^{2}.
\end{equation}
This equation indicates that the generation rate of magnons produced by scattering changes from negative to positive as the magnon amplitude increases from zero. The resulting sign change corresponds to a threshold at which the excited coherent spin waves begin to decay via four-magnon scattering. At this threshold, the magnon amplitude is given by
\begin{equation}
    \qty|\ev{\hat{c}^{'}_{\bm{k}}}| =  \sqrt{\frac{\Gamma_{\bm{k}}}{\qty|H_{\bm{k},\bm{k},\bm{k},\bm{k}}|}}.
\end{equation}
By introducing a dimensionless parameter $\lambda_{\bm{k}}$ to characterize the four-magnon scattering, $H_{\bm{k},\bm{k},\bm{k},\bm{k}}$ is expressed as
\begin{equation}
\label{eq:theory_four_magnon_coeff}
    H_{\bm{k},\bm{k},\bm{k},\bm{k}} = \lambda_{\bm{k}} \frac{\hbar \gamma^{\prime 2}_{e}}{V},
\end{equation}
where $V$ is the volume of the magnetic film and $\gamma^{\prime}_{e} = 2\pi \gamma_{e}$. 
By utilizing the expression above, the spin-wave amplitude in the $x$-direction at the threshold $m^{\mathrm{th}}_{x,\bm{k}}$ can be calculated as
\begin{equation}
\label{eq:Mth}
    m^{\mathrm{th}}_{x,\bm{k}} = (u_{\bm{k}} + v_{\bm{k}}) \sqrt{\frac{2\Gamma_{\bm{k}}}{\qty|\lambda_{\bm{k}}| \gamma^{\prime}_{e} M_{\mathrm{s}}}} M_{\mathrm{s}},
\end{equation}
where $u_{\bm{k}}$ and $v_{\bm{k}}$ are dimensionless parameters of the Bogoliubov transformation \cite{rezende2020fundamentals} (see Supplemental Materials for details \cite{SM}), and $M_{\mathrm{s}}$ is the saturation magnetization of the magnetic film.
\par
We calculate the spin-wave amplitude threshold corresponding to the experimental results using this expression. Table~\ref{tb:exp_params} summarizes the relevant experimental parameters. For YIG-A, we use the parameters listed in the left column of Table~\ref{tb:exp_params}. The saturation magnetization is taken as $M_{\mathrm{s}} = 141 \, \mathrm{mT}$, as measured by SQUID (superconducting quantum interference device) magnetometry at 300 K. The anisotropy field $H_{\mathrm{ani}}$ is estimated from the difference between the effective magnetization $M_{\mathrm{eff}} = 169.6 \, \mathrm{mT}$ and the saturation magnetization as $H_{\mathrm{ani}} = M_{s} - M_{\mathrm{eff}}$. The decay rate $\Gamma_{\bm{k}}$ is obtained from the spin-wave lifetime, which is estimated by dividing the decay length $l_{d} = 54.21 \, \mathrm{\mu m}$ (determined from fitting the data at $P_{\mathrm{mw}} = 11 \, \mathrm{dBm}$) by the group velocity of the spin waves. The four-magnon scattering coefficient $\lambda_{\bm{k}}$ and the Bogoliubov transformation parameters $u_{\bm{k}}$ and $v_{\bm{k}}$ are calculated from analytical expressions using the experimental parameters (see Supplemental Materials for the explicit expressions \cite{SM}). Based on these parameters, the spin-wave amplitude threshold for the instability due to four-magnon scattering is calculated to be $\mu_{0} m^{\mathrm{th}}_{x,\bm{k}} = 17.8 \, \mathrm{mT}$. 
\par
It is important to examine how the above discussion applies to our experiment. The black dotted lines in Figs.~\ref{fig:matsuno_m}(c) and (d) indicate the threshold amplitude obtained in this way. In particular, regarding Fig.~\ref{fig:matsuno_m}(c), which displays the spin-wave amplitude at $x = 4 \, \mathrm{\mu m}$ near the stripline, the threshold for the emergence of nonlinearity and the experimental results agree satisfactorily. This result demonstrates that the nonlinear dynamics can be quantitatively visualized using NV centers. The ability to experimentally detect and quantitatively characterize the transition from the linear to the nonlinear regime, consistent with theoretical predictions, constitutes the central achievement of the present work.

\begin{table}[t]
    \begin{center}
    \begin{tabular}{|c|c|c|} \hline
       & YIG-A & YIG-B  \\ \hline
       Film thickness $d$ & $54 \, \mathrm{nm}$ & $109 \, \mathrm{nm}$ \\ \hline
       External magnetic field $B_{0}$ & $25.17 \, \mathrm{mT}$ &  $22.45 \, \mathrm{mT}$ \\ \hline
       Saturation magnetization $M_{\mathrm{s}}$ &  $141 \, \mathrm{mT}$ & $162 \, \mathrm{mT}$ \\ \hline
       Anisotropy field $H_{\mathrm{ani}}$ &  $-28.6 \, \mathrm{mT}$ & $-15.6 \, \mathrm{mT}$ \\ \hline
       Spin-wave frequency $f_{\mathrm{sw}}$ & $2166 \, \mathrm{MHz}$ & $2243.1 \, \mathrm{MHz}$ \\ \hline
       Spin-wave wavenumber $k_{x}$ & $1.45 \, \mathrm{rad/\mu m}$ & $1.27 \, \mathrm{rad/\mu m}$ \\ \hline
       Decay rate $\Gamma_{\bm{k}}$ & $12.6 \, \mathrm{MHz}$ & $4.78 \, \mathrm{MHz}$ \\ \hline
       Four-magnon scattering coefficient $\lambda_{\bm{k}}$ & -0.155 & -0.173  \\ \hline
       Bogoliucdbov transformation parameter $u_{\bm{k}}$ & 1.11 & 1.10  \\ \hline
       Bogoliubov transformation parameter $v_{\bm{k}}$ & 0.472 & 0.449 \\ \hline
    \end{tabular}
    \end{center}
    \caption{Experimental parameters of YIG-A and YIG-B.}
    \label{tb:exp_params}
\end{table}

\subsection{Magnetization reduction in the nonlinear regime}
\label{sec:magnetization_reduction}
\begin{figure}[htbp]
    \centering
    \includegraphics[width=\linewidth]{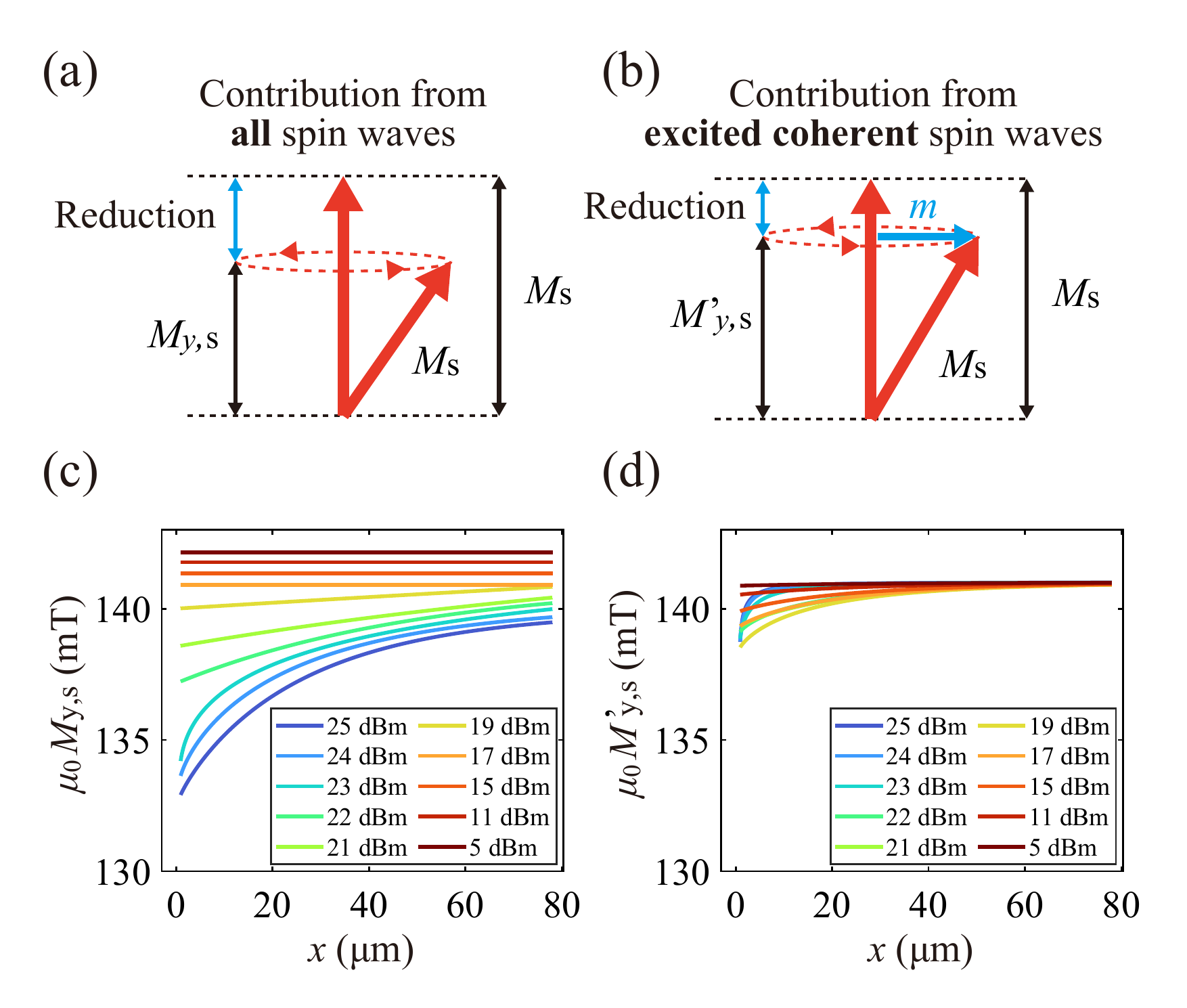}
    \caption{(a)(b) Schematic of the reduction of the static magnetization along the $y$-axis, (a) considering contributions from all spin waves. (b) considering only contributions from excited coherent spin waves.
    (c) Static magnetization distribution obtained by converting the wavenumber distribution for YIG-A. 
    (d) Estimation of the static magnetization distribution considering only contributions from the excited coherent spin waves for YIG-A.}
    \label{fig:matsuno_k}
\end{figure}

Having examined the spin-wave amplitude, we next analyze the spatial distribution of the wavenumber. The observed increase in the wavenumber reflects a decrease in the static magnetization along the $y$-direction (i.e., the direction of the saturation magnetization) $M_{y,s}$ and the corresponding effective magnetization $M_{\mathrm{eff}}$ due to an increase in the spin-wave amplitude [Fig.~\ref{fig:matsuno_k}(a)] \cite{Lake2022, Merbouche2022}.
To calculate the effective magnetization, we utilize the dispersion relation of surface spin waves [Eq.~(\ref{eq:dispersion_analysis})].
At each position, the effective magnetization $M_{\mathrm{eff}}(x)$ corresponding to the spin-wave frequency $f_{\mathrm{sw}} = 2165.8 \, \mathrm{MHz}$ is calculated from the wavenumber $k_{x}(x)$. 
As the dispersion relation holds for a small spin-wave amplitude,
there may be deviations from Eq.~(\ref{eq:dispersion_analysis}) for large amplitudes, particularly in nonlinear regions. However, the coherent spin-wave amplitude with the excitation microwave frequency decays rapidly near the stripline in the nonlinear region, and the amplitude is small compared to the saturation magnetization across most of the field of view.
Therefore, we utilize the dispersion relation of the surface spin waves in Eq.~(\ref{eq:dispersion_analysis}) to derive the static magnetization distribution approximately. 
\par
Figure \ref{fig:matsuno_k}(c) plots the resulting spatial distribution of the static magnetization $M_{y,\mathrm{s}}(x)$. 
We convert the effective magnetization distribution $M_{\mathrm{eff}}(x)$ into the static magnetization distribution $M_{\mathrm{y,s}}(x)$ by removing the contribution from the surface anisotropy $\qty(M_{\mathrm{y,s}}(x) = M_{\mathrm{eff}}(x) + H_{\mathrm{ani}})$. Figure~\ref{fig:matsuno_k}(c) shows that, in regions with small input microwave power, though there is a uniform reduction as the power increases, the static magnetization remains almost constant within the field of view. However, in addition to the uniform reduction, significant spatial variation within the field of view appears above $P_{\mathrm{mw}} = 19 \, \mathrm{dBm}$, and at the maximum input power ($P_{\mathrm{mw}} = 25 \, \mathrm{dBm}$), a variation of about $6 \, \mathrm{mT}$ within the field of view is observed.

It is instructive to examine how the surface spin waves contribute to the reduction in static magnetization [Fig.~\ref{fig:matsuno_k}(b)]. When considering only the contribution from the coherent surface spin waves with the excitation frequency, the magnetization in the $y$-direction at a specific position and time is expressed as
\begin{equation}
\begin{split}
    M_{y}(t) &= \sqrt{M^{2}_{s} - m^{2}_{x}(t) -  m^{2}_{z}(t)}, \\
\end{split}
\end{equation}
where $m_{x}(t)$ and $m_{z}(t)$ are the precession of the magnetization components along the $x$- and $z$-directions of the surface spin waves, respectively:
\begin{equation}
\begin{split}
    m_{x}(t) &= m_{x} \cos(2\pi f_{\mathrm{sw}} t), \\
    m_{z}(t) &= \eta_{k} m_{x} \sin(2\pi f_{\mathrm{sw}} t). \\
\end{split}
\end{equation}
Consequently, the averaged static magnetization $M^{'}_{y,\mathrm{s}}$, which accounts only for the contribution from the coherent surface spin waves, can be obtained by averaging the phase of the spin waves $\theta \, (= 2\pi f_{\mathrm{sw}} t)$ as
\begin{equation}
\label{eq:Mseff}
    M^{'}_{y,\mathrm{s}} = \frac{1}{2\pi} \int_{0}^{2\pi} \sqrt{M^{2}_{s} - \qty(m_{x} \cos \theta)^2 - \qty(\eta_{k} m_{x} \sin \theta)^2} \, d \theta.
\end{equation}

\par
Figure~\ref{fig:matsuno_k}(d) shows the effective static magnetization distribution $M^{'}_{y,\mathrm{s}}(x)$, which is calculated from the spin-wave amplitude distribution in Fig.~\ref{fig:matsuno_m}(a) using Eq.~(\ref{eq:Mseff}). The left edge of the plot is defined at $x = 0.91 \, \mathrm{\mu m}$. At small input  microwave powers in the linear regime, $M^{'}_{y,\mathrm{s}}$ exhibits a slight decrease only near the left edge of the field of view ($x\lesssim 10$ $\mathrm{\mu m}$), which is in clear contrast to the observation that the static magnetization derived from the wavenumber [$M_{\mathrm{y,s}}(x)$ shown in Fig.~\ref{fig:matsuno_k}(c)] shows a prominent decrease across the entire field of view. These results imply that the decay of the surface spin waves is not solely due to relaxation into the lattice but also involves scattering by other magnons, such as thermal magnons. In other words, the excited surface spin waves are transformed into other modes, leading to a persistent reduction in the static magnetization across the field of view \cite{rezende2020fundamentals}. 
\par
Around maximum input microwave powers in the nonlinear regime, the reduction in static magnetization attributable to surface spin waves becomes almost negligible at a distance of about $5 \, \mathrm{\mu m}$ from the stripline. In contrast, the static magnetization estimated from the wavenumber [Fig.~\ref{fig:matsuno_k}(c)] continues to change gradually across the field of view. This observation suggests that most of the modulation in magnetization is caused by additional magnons generated through four-magnon scattering from the excited surface spin waves.
\par
As a note, we investigate the possible role of Joule heating resulting from the application of a large-amplitude microwave, which might cause the sample to heat up. Such heating could potentially create a temperature gradient along the $x$-direction, thereby causing a spatial variation in the saturation magnetization and wavenumber. However, we can safely exclude this possibility. We perform similar measurements at a different magnetic field, where the spin-wave excitation is more inefficient and the spin waves are less likely to reach the nonlinear regime, and confirm that the wavenumber along the $x$-direction remains almost constant even at high microwave power (see Supplemental Materials for details \cite{SM}).

\subsection{Comparison with a different YIG film}
\label{sec:comparison_with_a_different_YIG}
So far, we have quantitatively imaged the amplitude and wavenumber distributions of spin waves in YIG-A over a wide range of input microwave powers. In particular, we have confirmed that the threshold of nonlinearity coincides with the threshold determined by four-magnon scattering. As shown in Eq.~(\ref{eq:Mth}), this threshold depends on the relaxation rate of the spin waves. It is therefore important to investigate how our interpretation can be generalized. One possible approach to demonstrate the validity of our results is to check whether the amplitude indirectly inferred in Ref.~\cite{Lake2022} is consistent with theory. However, a straightforward and quantitative comparison is difficult for two reasons. First, the present work investigates a two-dimensional extended YIG thin film, whereas Ref.~\cite{Lake2022} studies a waveguide. As a result, key parameters of the magnon Hamiltonian, including dipolar interactions, differ substantially, making it difficult to apply the theory used in our paper as is. Second, in Ref.~\cite{Lake2022}, the spin-wave precession angle is inferred from a numerically computed dispersion by imposing a uniform tilt $\theta$ and then allowing the spins to precess. In this procedure, however, the ellipticity of spin-wave precession is not taken into account, and it is unclear how faithfully the imposed $\theta$ maps onto the actual dynamic magnetization amplitude in the experimental system. For these reasons, the indirectly determined threshold reported in Ref.~\cite{Lake2022} cannot be assessed in a rigorous manner. Instead, to substantiate the validity of both the theory and the measurements, we present results of YIG-B, which has a different relaxation rate.
\par

The external magnetic field $B_{0}$ is set to $22.45 \, \mathrm{mT}$ when measuring YIG-B. The resonance frequency of the NV spins $f_{\mathrm{NV}}$ is $2243.1 \, \mathrm{MHz}$, and the wavenumber of the surface spin waves at this frequency is around $1.3 \, \mathrm{rad/\mu m}$ (see Supplemental Materials for details \cite{SM}). 
\par

\begin{figure}[t]
    \centering
    \includegraphics[width=\linewidth]{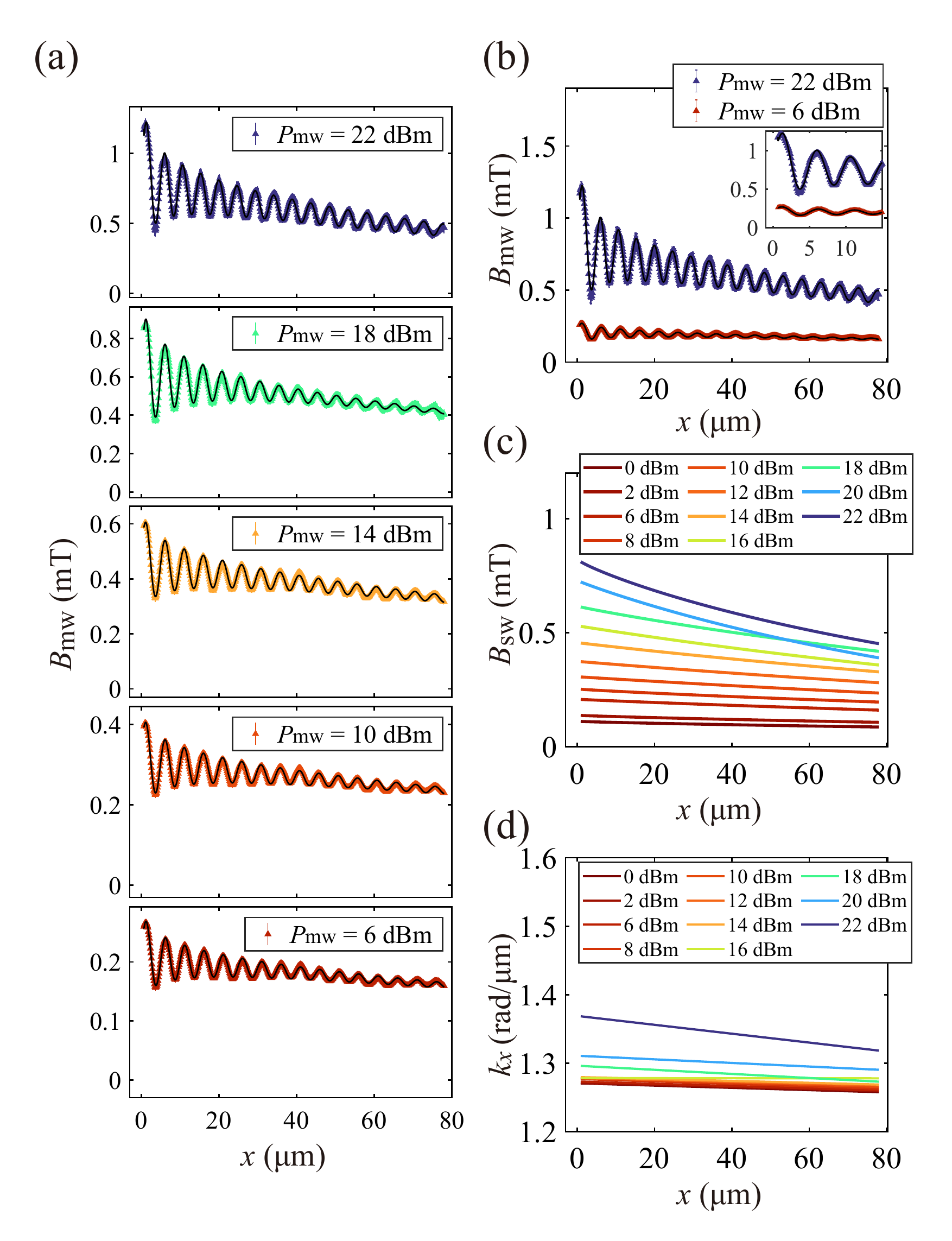}
    \caption{
    (a) Input microwave power dependence of the one-dimensional microwave amplitude distribution $B_{\mathrm{mw}}(x)$ obtained for YIG-B. The triangular markers with error bars represent experimental data, and the solid black lines correspond to the fitting results.
    (b) Comparison of $B_{\mathrm{mw}}(x)$ for data with large input microwave power ($P_{\mathrm{mw}} = 22 \, \mathrm{dBm}$) and data with small input microwave power ($P_{\mathrm{mw}} = 6 \, \mathrm{dBm}$) for YIG-B. 
    (c) Microwave amplitude and (d) Wavenumber distribution from the spin waves for each input microwave power calculated from the fitting results for YIG-B.
    }
    \label{fig:matesy_psweep}
\end{figure}

We sweep the input microwave power $P_{\mathrm{mw}}$ in 23 steps from $0 \, \mathrm{dBm}$ to $22 \, \mathrm{dBm}$. Figure~\ref{fig:matesy_psweep}(a) displays the microwave amplitude distribution $B_{\mathrm{mw}}(x)$ for five representative input powers (see Supplemental Materials for additional data \cite{SM}). The triangular markers with error bars indicate the experimental data. The integration region along the $y$-direction is the same as that used for YIG-A [see Figs.~\ref{fig:exp_setup}(d) and \ref{fig:matsuno_psweep}(a)]. 
Focusing on the data at $P_{\mathrm{mw}} = 6 \, \mathrm{dBm}$, we observe that, compared with YIG-A, the decay of $B_{\mathrm{mw}}(x)$ from the left to the right end is small, reflecting a lower relaxation rate of the spin waves. Overall, the mean line of $B_{\mathrm{mw}}(x)$ attenuates by approximately 20\% across the entire field of view, which corresponds to a decay length of about  $300 \, \mathrm{\mu m}$.

In Fig.~\ref{fig:matesy_psweep}(b), $B_{\mathrm{mw}}(x)$ at $P_{\mathrm{mw}} = 6 \, \mathrm{dBm}$ and $P_{\mathrm{mw}} = 22 \, \mathrm{dBm}$ are overlaid for direct comparison [see Fig.~\ref{fig:matsuno_psweep}(b) for YIG-A]. Although there is no pronounced decay at $P_{\mathrm{mw}} = 22 \, \mathrm{dBm}$, the amplitude of the mean line is reduced by approximately 50\% from the left to the right end of the field of view, indicating the emergence of nonlinear decay at this input power. Furthermore, when focusing on the smaller $x$-coordinate shown in the inset, a slight difference in wavenumber between the two input microwave powers is observed. However, the modulation is less significant than that seen in YIG-A.
\par
We follow the same procedure as YIG-A to fit $B_{\mathrm{mw}}(x)$, with two differences. First, for YIG-B, across all regions and at every input microwave power, the reference microwave distribution is consistently smaller than the measured microwave amplitude distribution (see Supplemental Materials for details \cite{SM}). Consequently, we employ Eq.~(\ref{eq:Btot_approx}) for the fitting. Second, because the observed wavenumber modulation is small, we adopt the following linear approximation for the wavenumber distribution to describe the minor spatial modulation:
\begin{equation} 
    k_{x}(x) = k_{0} - k_{1} \frac{(x-x_{0})}{l_{\mathrm{k}}}.
\end{equation}
For each input microwave power, the fitted curve reproduces the experimental data precisely, as shown by the solid black lines in Fig.~\ref{fig:matesy_psweep}(a).
\par
In Fig.~\ref{fig:matesy_psweep}(c), we plot the microwave amplitude distribution from spin waves $B_{\mathrm{sw}}(x)$, estimated from the fitting results at eleven different input microwave powers. The left end of the plot is set at $x = 0.91 \, \mathrm{\mu m}$. Notably, the amplitude at the left end continues to increase up to the maximum input power of $P_{\mathrm{mw}} = 22 \, \mathrm{dBm}$. In contrast, the amplitude at the right end saturates at around $B_{\mathrm{sw}} = 0.5 \, \mathrm{mT}$, indicating the onset of nonlinearity. Figure~\ref{fig:matesy_psweep}(d) shows the estimated wavenumber distribution $k_{x}(x)$ for each input microwave power. The modulation of $k_{x}(x)$ across the field of view is minimal; even at the maximum input microwave power of $P_{\mathrm{mw}} = 22 \, \mathrm{dBm}$, the variation from the left to the right end is less than $0.1 \, \mathrm{rad/\mu m}$. These observations are markedly different from those for YIG-A, as shown in Figs.~\ref{fig:matsuno_psweep}(c) and (d).
\par

\begin{figure}[t]
    \centering
    \includegraphics[width=\linewidth]{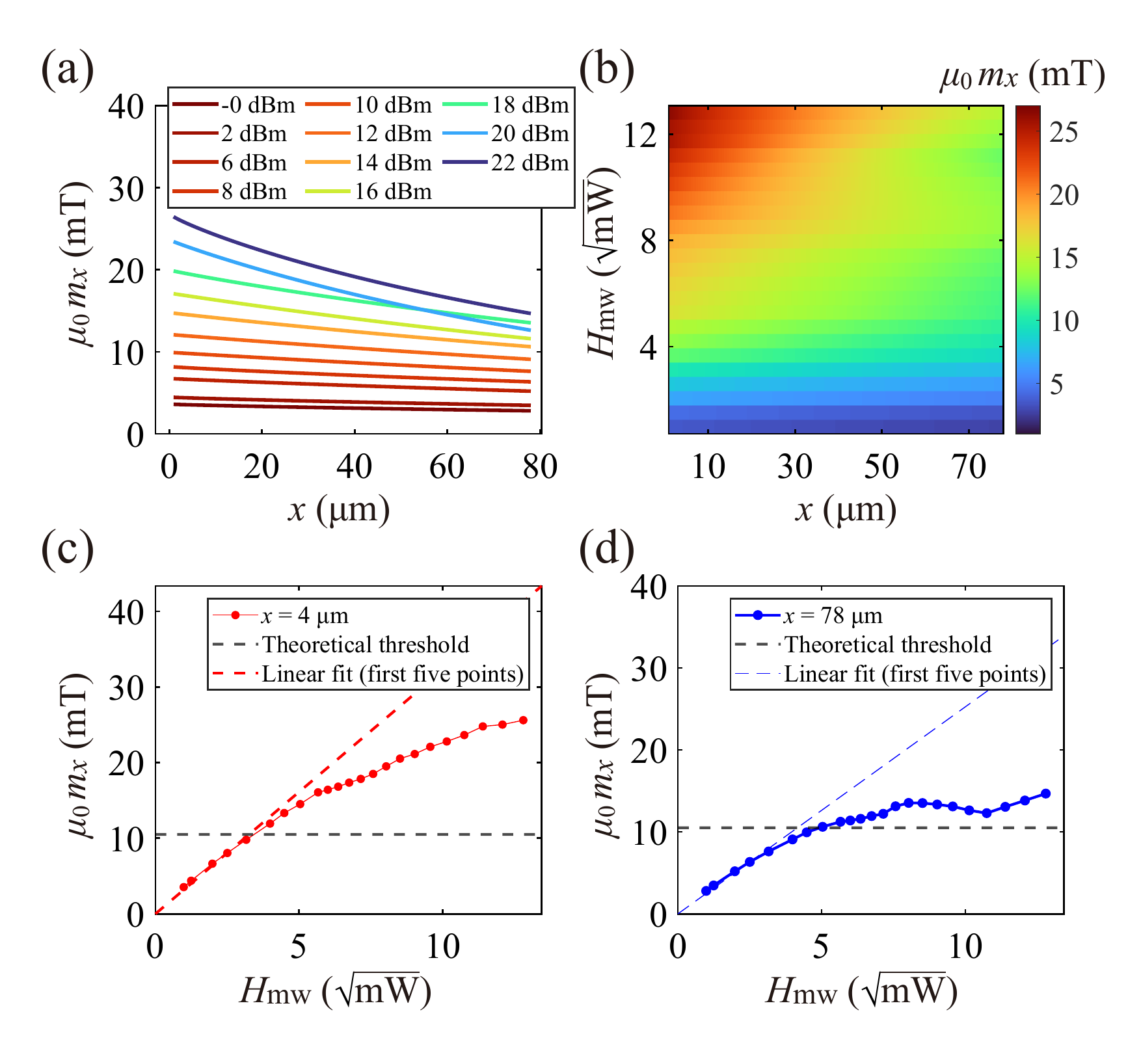}
    \caption{(a) Calculated spin-wave amplitude distribution $m_{x}(x)$ for each input microwave power for YIG-B.
    (b) The heatmap of the spin-wave amplitude distribution for each input microwave amplitude for YIG-B.
    (c) (d) Dependence of the spin-wave amplitude on input microwave amplitude (c) at $x = 4 \, \mathrm{\mu m}$ and (d) at $x = 78 \, \mathrm{\mu m}$ for YIG-B.}
    \label{fig:matesy_m}
\end{figure}

As we did for YIG-A, we use Eq.~(\ref{eq:Bsw_nv}) to convert $B_{\mathrm{sw}}(x)$, shown in Fig.~\ref{fig:matesy_psweep}(c), into the corresponding spin-wave amplitude distributions $m_{x}(x)$. Figure~\ref{fig:matesy_m}(a) shows the converted spin-wave amplitude, and Fig.~\ref{fig:matesy_m}(b) displays a heatmap of $m_{x}(x)$ for all data, together with the individual profiles from Fig.~\ref{fig:matesy_m}(a). To investigate the nonlinearity of the spin waves, we examine the dependence of the spin-wave amplitude on the input microwave amplitude at two positions: $x = 4 \, \mathrm{\mu m}$ on the left side of the field of view and $x = 78 \, \mathrm{\mu m}$ on the right side. Figure~\ref{fig:matesy_m}(c) shows the dependence of the spin-wave amplitude at $x = 4 \, \mathrm{\mu m}$ on the input microwave amplitude. The red dashed line represents a linear fit through the origin, obtained using data from the five smallest input microwave amplitudes (up to $P_{\mathrm{mw}} = 10 \, \mathrm{dBm}$). The data indicate that nonlinearity begins to occur when the spin-wave amplitude exceeds approximately $10 \, \mathrm{mT}$. Although the amplitude does not immediately saturate, it deviates from the linear trend and the gradient continues to decrease. Similarly, Fig.~\ref{fig:matesy_m}(d) shows the dependence of the spin-wave amplitude at $x = 78 \, \mathrm{\mu m}$. As for the data at $x = 4 \, \mathrm{\mu m}$, nonlinearity appears when the amplitude reaches around $10 \, \mathrm{mT}$. However, in contrast to the left side, the amplitude at $x = 78 \, \mathrm{\mu m}$ does not increase significantly in the nonlinear region and appears to saturate around $13 \, \mathrm{mT}$. These behaviors are qualitatively similar to those observed for YIG-A, but the threshold at which nonlinearity emerges is smaller than in YIG-A. 
\par
We compare the obtained results with the theoretical threshold given by Eq.~(\ref{eq:Mth}). The right column of Table~\ref{tb:exp_params} lists the parameters used for the calculation. The saturation magnetization is taken as $M_{\mathrm{s}} = 162 \, \mathrm{mT}$, as measured by SQUID magnetometry. The anisotropy field is estimated from the difference between the effective magnetization $M_{\mathrm{eff}} = 177.6 \, \mathrm{mT}$ and the saturation magnetization. The decay rate is calculated from the decay length $l_{d} = 318 \, \mathrm{\mu m}$, obtained from the fitting result at $P_{\mathrm{mw}} = 6 \, \mathrm{dBm}$. With these parameters, the threshold spin-wave amplitude for the onset of four-magnon decay is estimated to be $\mu_{0} m^{\mathrm{th}}_{x} = 10.5 \, \mathrm{mT}$. As expected, this threshold is lower than that of YIG-A, reflecting the smaller decay rate in the present case. As shown in Figs.~\ref{fig:matesy_m}(c) and (d), the experimental data indicate that nonlinearity begins to occur when the spin-wave amplitude exceeds approximately $10 \, \mathrm{mT}$ (marked by the horizontal lines in the graphs), again in good agreement with this theoretical prediction.
\par
Unlike in YIG-A, the absence of significant wavenumber modulation in YIG-B can be attributed to the smaller spin-wave amplitude, even at maximum power. As shown in Fig.~\ref{fig:matsuno_m}(c), at $x = 4 \, \mathrm{\mu m}$, the extrapolated value from the linear fit of the spin-wave amplitude is approximately $70 \, \mathrm{mT}$ at the maximum excitation microwave amplitude, whereas it is around $40 \, \mathrm{mT}$ for YIG-B. Consequently, even at maximum power, the lower amplitude of the excited spin waves in YIG-B results in only minor variations in the static magnetization across the field of view.

\section{Conclusion and Perspectives}
\label{sec:conclusion}
To conclude, we have employed QDM to investigate the nonlinear dynamics of surface spin waves quantitatively. For YIG-A, we swept the input microwave power to map the microwave amplitude distribution, revealing both nonlinear behavior in the amplitude profile and shifts in the wavenumber resulting from a reduction in the saturation magnetization. Quantitative analysis of these data showed that nonlinearity in the amplitude emerges once the spin-wave amplitude exceeds a specific threshold, consistent with theoretical predictions based on four-magnon scattering. Furthermore, as the spin-wave amplitude increases, we observed modulation of the wavenumber within the field of view, and our quantitative analysis revealed that this modulation largely originates from the spin waves generated by multi-magnon scattering. Similar measurements conducted on YIG-B, which exhibits a different relaxation rate, confirmed that the threshold for the onset of nonlinearity also agrees with the theoretical predictions.
\par
These results deepen our understanding of nonlinear spin-wave dynamics and underscore the utility of NV centers as sensors for the quantitative investigation of nonlinear spin-wave phenomena, which have previously been accessible only at a qualitative level.
\par
As a future extension of this work, one could consider the quantitative visualization of the incoherent magnons generated by four-magnon scattering. NV centers can quantitatively measure the magnetic noise at their resonance frequency via longitudinal relaxation measurements. Therefore, by exciting surface spin waves with frequencies detuned from the NV spin resonance and monitoring the relaxation rate of the NV spins while varying the input microwave power, it becomes possible to extract quantitative information about the incoherent magnons produced by nonlinear scattering \cite{Simon2021}. Furthermore, under the same conditions, quantitative detection of the coherent spin waves detuned from the NV resonance frequency is possible by employing sensing based on the AC Zeeman effect \cite{ogawa2024wideband}.
\par
In addition, nonlinear phenomena regarding spin waves are not limited to surface spin waves. For example, parametric excitation, where spin waves are generated by microwave fields at twice the frequency of the spin waves oriented along the saturation magnetization direction \cite{Bloembergen1952relaxation,Damon1953relaxation,Bloembergen1954relaxation}, is an important example. Furthermore, in the case of large-amplitude spin waves, the propagation of spin-wave bullets, which behave as solitons \cite{Sulymenko2018}, represents another notable nonlinear effect. NV centers are expected to serve as powerful probes for investigating these phenomena. 
\par
The nonlinear dynamics of spin waves are also anticipated to have applications in spintronics devices \cite{Chumak2014magnon,Ustinov2008microwave}. Recent developments of magnonics has been especially active in spin waves in the shorter-wavelength, higher-frequency regime \cite{Flebus2024}, within which nonlinearity of the spin waves in particular holds substantial potential for efficient spin-wave excitation \cite{Wang2023} and applications for logic devices \cite{Wang2020} and neuromorphic computing \cite{Papp2021}. Methods capable of quantitatively evaluating these operations can provide valuable guidelines for device design.

\section*{Acknowledgements}
This work was partially supported by JST, CREST Grant No.~JPMJCR23I2, Japan; 
Grants-in-Aid for Scientific Research (Nos.~JP22K03524, JP22KJ1058, JP22KJ1059, JP23K25800, JP23K21077, JP24K21194, JP24H01666 and JP25H01248);
the Mitsubishi Foundation (Grant No.~202310021);
Seiko Instruments; Advanced Technology Foundation Research Grants; 
the Cooperative Research Project of RIEC, Tohoku University; 
“Advanced Research Infrastructure for Materials and Nanotechnology in Japan (ARIM)” (No.~JPMXP1222UT1131) of the Ministry of Education, Culture, Sports, Science and Technology of Japan (MEXT).
K.O. and M.T. acknowledge supports from FoPM, WINGS Program, The University of Tokyo. \\

\bibliography{citation}

\end{document}

\title{Supplementary Materials: Quantitative imaging of nonlinear spin-wave propagation using diamond quantum sensors}

\author{Kensuke Ogawa} 
\affiliation{Department of Physics, \href{https://ror.org/057zh3y96}{The University of Tokyo}, Bunkyo-ku, Tokyo, 113-0033, Japan}
\author{Moeta Tsukamoto}
\affiliation{Department of Physics, \href{https://ror.org/057zh3y96}{The University of Tokyo}, Bunkyo-ku, Tokyo, 113-0033, Japan}
\author{Yusuke Mori} 
\affiliation{Department of Physics, \href{https://ror.org/035t8zc32}{Osaka University}, Toyonaka, Osaka 560-0043, Japan}
\author{Daigo Takafuji} 
\affiliation{Department of Physics, \href{https://ror.org/035t8zc32}{Osaka University}, Toyonaka, Osaka 560-0043, Japan}
\author{Junichi Shiogai}
\affiliation{Department of Physics, \href{https://ror.org/035t8zc32}{Osaka University}, Toyonaka, Osaka 560-0043, Japan}
\affiliation{Division of Spintronics Research Network, Institute for Open and Transdisciplinary Research Initiatives, \href{https://ror.org/035t8zc32}{Osaka University}, Suita,
Osaka 565-0871, Japan}
\author{Kohei Ueda}
\affiliation{Department of Physics, \href{https://ror.org/035t8zc32}{Osaka University}, Toyonaka, Osaka 560-0043, Japan}
\affiliation{Division of Spintronics Research Network, Institute for Open and Transdisciplinary Research Initiatives, \href{https://ror.org/035t8zc32}{Osaka University}, Suita,
Osaka 565-0871, Japan}
\author{Jobu Matsuno}
\affiliation{Department of Physics, \href{https://ror.org/035t8zc32}{Osaka University}, Toyonaka, Osaka 560-0043, Japan}
\affiliation{Division of Spintronics Research Network, Institute for Open and Transdisciplinary Research Initiatives, \href{https://ror.org/035t8zc32}{Osaka University}, Suita,
Osaka 565-0871, Japan}
\author{Jun-ichiro Ohe}
\affiliation{Department of Physics, \href{https://ror.org/02hcx7n63}{Toho University}, 2-2-1 Miyama, Funabashi 274-8510, Japan}
\author{Kento Sasaki}
\affiliation{Department of Physics, \href{https://ror.org/057zh3y96}{The University of Tokyo}, Bunkyo-ku, Tokyo, 113-0033, Japan}
\author{Kensuke Kobayashi}
\affiliation{Department of Physics, \href{https://ror.org/057zh3y96}{The University of Tokyo}, Bunkyo-ku, Tokyo, 113-0033, Japan}
\affiliation{Institute for Physics of Intelligence, \href{https://ror.org/057zh3y96}{The University of Tokyo}, Bunkyo-ku, Tokyo, 113-0033, Japan}
\affiliation{Trans-Scale Quantum Science Institute, \href{https://ror.org/057zh3y96}{The University of Tokyo}, Bunkyo-ku, Tokyo, 113-0033, Japan}

\maketitle
\date{\today}

\renewcommand{\thesection}{S\arabic{section}}
\renewcommand{\thefigure}{S\arabic{figure}}
\renewcommand{\theequation}{S\arabic{equation}}

\section{Quantum theory of spin waves}
The nonlinear dynamics of spin waves are typically described by considering higher-order magnon scattering processes. In this section, we formulate the spin-wave dynamics within the magnon picture and calculate the time evolution of the magnons, explicitly including these higher-order terms. We then derive the threshold spin-wave amplitude at which nonlinear decay begins to occur \cite{rezende2020fundamentals}.

\subsection{Formulation of spin waves in magnon picture}
We formulate the spin waves in the magnon picture using the Holstein-Primakoff transformation \cite{HolsteinMagnon}.
\par
Let $\hat{\bm{S}_{i}}$ denote the spin operator at the $i$-th site in a ferromagnet. For simplicity, we assume that the spin operator is dimensionless by dividing by $\hbar$. The spin operators satisfy the following commutation relations:

\begin{equation}
    \qty[\hat{S}^{\alpha}_{i}, \hat{S}^{\beta}_{j}] = i \epsilon_{\alpha \beta \gamma} \delta_{ij} \hat{S}^{\gamma}_{j}.
\end{equation}

The eigenvalues of the $z$-component of the spin operator, defined as $S^{z}_{i}$, take on the values with 
\begin{equation}
    S^{z}_{i} = -S, \, -S+1, \, ... \, , \, S-1, \, S,
\end{equation}
with corresponding eigenvectors $\ket{S^{i}_{z}}$ satisfying
\begin{equation}
    \hat{S_{i}^{z}} \ket{S^{z}_{i}} = S^{z}_{i} \ket{S^{z}_{i}}.
\end{equation}
The ladder operators, defined as $\hat{S_{i}}^{\pm} = \hat{S_{i}}^{x} \pm i \hat{S_{i}}^{y}$, raise or lower the magnetic quantum number according to
\begin{equation}
    \hat{S_{i}}^{\pm} \ket{S^{z}_{i}} = \sqrt{S(S+1) - S^{z}_{i}(S^{z}_{i} \pm 1)} \ket{S^{z}_{i} \pm 1}.
\end{equation}
\par
We consider a magnetic material that is magnetized along the $z$-direction. In the ordered state at absolute zero, every spin is aligned such that $S^{z}_{i} = S$. To describe excitations from this ordered state, we introduce the number state $\ket{n_{i}}$, where $n_{i}$ denotes the number by which the magnetic quantum number is reduced. The corresponding eigenvalue equation is
\begin{equation}
    \hat{S}_{i}^{z} \ket{n_{i}} = (S - n_{i}) \ket{n_{i}}.
\end{equation}

Furthermore, we introduce the creation and annihilation operators, $\hat{a_{i}}^{\dag}$ and $\hat{a_{i}}$, for the number states. These operators satisfy the following relations:

\begin{equation}
\begin{split}
    [\hat{a_{i}}, \hat{a_{j}}^{\dag}] = \delta_{i,j}, &\, [\hat{a_{i}}, \hat{a_{j}}] = 0, \, [\hat{a_{i}}^{\dag}, \hat{a_{j}}^{\dag}] = 0, \\
    \hat{a_{i}}^{\dag} \hat{a_{i}} \ket{n_{i}} &= n_{i} \ket{n_{i}}, \\
    \hat{a_{i}}^{\dag} \ket{n_{i}} &= \sqrt{n_{i}+1} \ket{n_{i}+1}, \\
    \hat{a_{i}} \ket{n_{i}} &= \sqrt{n_{i}} \ket{n_{i}-1}. \\
\end{split}
\end{equation}

At this point, the following relationships hold for the spin ladder operators:
\begin{equation}
\begin{split}
    \hat{S}^{-}_{i} \ket{n_{i}} &= \sqrt{2S} \sqrt{n_{i}+1} \sqrt{1-\frac{n_{i}}{2S}} \ket{n_{i}+1}, \\
    \hat{S}^{+}_{i} \ket{n_{i}} &= \sqrt{2S} \sqrt{1-\frac{n_{i}-1}{2S}} \sqrt{n_{i}} \ket{n_{i}-1}. \\
\end{split}
\end{equation}
Therefore, we can establish the following correspondence between the spin ladder operators and the bosonic creation and annihilation operators:
\begin{equation}
\begin{split}
    \hat{S_{i}^{-}} &= \sqrt{2S} \hat{a_{i}}^{\dag} \sqrt{1-\frac{\hat{a_{i}}^{\dag} \hat{a_{i}}}{2S}}, \\
    \hat{S_{i}^{+}} &= \sqrt{2S} \sqrt{1-\frac{\hat{a_{i}}^{\dag} \hat{a_{i}}}{2S}} \hat{a_{i}}. \\
\end{split}
\end{equation}
The above ladder operators can be expanded in terms of the creation and annihilation operators as follows:
\begin{equation}
\begin{split}
    \hat{S}^{-}_{i} &= \sqrt{2S} \qty(\hat{a_{i}}^{\dag} - \frac{1}{4S} \hat{a_{i}}^{\dag} \hat{a_{i}}^{\dag} \hat{a_{i}} + ...), \\
    \hat{S}^{+}_{i} &= \sqrt{2S} \qty(\hat{a_{i}} - \frac{1}{4S} \hat{a_{i}}^{\dag} \hat{a_{i}} \hat{a_{i}} + ...). \\
\end{split}
\end{equation}
When the amplitude of the spin waves is small, such that $\frac{\ev{n_{i}}}{S} \ll 1$, it is sufficient to consider only the first term in the above expansions. By including higher-order terms, we can discuss the nonlinear dynamics of the system.
\par
The spin operator in the $z$-direction, $\hat{S^{z}_{i}}$, can be expressed in terms of the creation and annihilation operators as
\begin{equation}
    \hat{S_{i}^{z}} = S - \hat{a_{i}}^{\dag} \hat{a_{i}}. \\
\end{equation}

\subsection{Magnon Hamiltonian}
\begin{figure}[b]
    \centering
    \includegraphics[width=\linewidth]{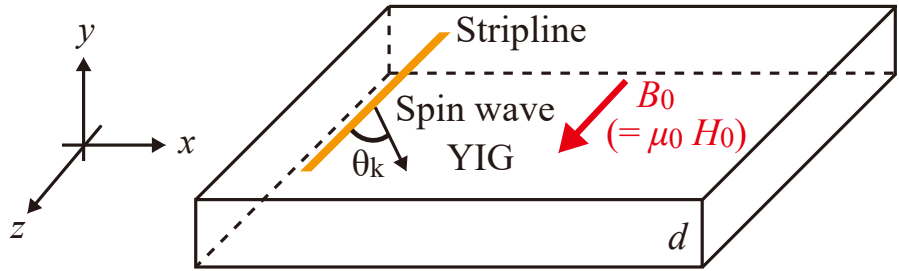}
    \caption{
    Schematic of the spin-wave propagation.
    }
    \label{fig:si_disp_setup}
\end{figure}

We consider in-plane ($zx$-plane) spin waves in a two-dimensional magnetic thin film, with an external static magnetic field applied in the $+z$-direction. The schematic is shown in Fig.~\ref{fig:si_disp_setup}. This choice of axes follows the conventional framework of magnon theory. Consequently, we note the following differences between Fig.~1(a) in the main text and Fig.~\ref{fig:si_disp_setup}: In both cases, the spin waves propagate in the $+x$-direction; however, the external magnetic field is applied along the $+y$-direction in Fig.~1(a) and in the $+z$-direction in Fig.~\ref{fig:si_disp_setup}. In addition, the magnetic thin film lies in the $xy$-plane in Fig.~1(a) but in the $zx$-plane in Fig.~\ref{fig:si_disp_setup}. These differences are taken into account when comparing the experimental and theoretical results.

The energy $E$ of the magnetic thin film, including the Zeeman energy from the static magnetic field, the exchange interaction, the dipolar interaction, and the surface anisotropy field, can be expressed as \cite{Rodrigo1999Extrinsic, Nembach2011}
\begin{equation}
   E = \mu_{0} \int \, d\bm{r} \qty(-M^{z}(\bm{r}) H_{0} - \frac{1}{2} \bm{M}(\bm{r})\cdot \bm{H}_{\mathrm{dip}}(\bm{r}) - \frac{1}{2} \bm{M}(\bm{r})\cdot \bm{H}_{\mathrm{ex}}(\bm{r}) - \frac{K_{s}}{M^{2}_{\mathrm{s}}d} \qty(M^{y}(\bm{r}))^{2}),
\end{equation}

where $\bm{M}(\bm{r})=\left(M^{x}(\bm{r}), M^{y}(\bm{r}), M^{z}(\bm{r})\right)$ is the magnetization, $H_{0} \, (= B_{0}/\mu_{0})$ is the external static magnetic field, $\bm{H}_{\mathrm{dip}}$ is the dipolar magnetic field, $\bm{H}_{\mathrm{ex}}$ is the exchange magnetic field, $K_{s}$ is the surface anisotropy constant, and $d$ is the thickness of the magnetic film. Here, $M_{\mathrm{s}}$ denotes the saturation magnetization and $\mu_{0}$ is the permeability of vacuum.
The dipolar magnetic field and the exchange magnetic field are represented as

\begin{equation}
    \bm{H}_{\mathrm{dip}}(\bm{r}) = \frac{1}{4\pi} \nabla \int \frac{\nabla' \cdot \bm{M}(\bm{r}')}{|\bm{r}-\bm{r}'|} d\bm{r}',
\end{equation}
\begin{equation}
\label{eq:ch6_Hex_classical}
    \bm{H}_{\mathrm{ex}}(\bm{r}) = \alpha_{\mathrm{ex}} \qty(\frac{\partial^2 \bm{M}(\bm{r})}{\partial x^2} + \frac{\partial^2 \bm{M}(\bm{r})}{\partial y^2} + \frac{\partial^2 \bm{M}(\bm{r})}{\partial z^2}),
\end{equation}
where $\alpha_{\mathrm{ex}}$ is the exchange stiffness coefficient.
\par

Starting from the classical energy of the magnetic film described above, we calculate the Hamiltonian by replacing the magnetization with magnon operators. The magnetization $M^{\alpha}(\bm{r})$ $(\alpha = x, \, y, \, z)$ of the magnetic material is related to the spin operator $S^{\alpha}_{i}$ through
\begin{equation}
    M^{\alpha}(\bm{r}) = \frac{g\mu_{B}N}{V} \hat{S}^{\alpha}_{i},
\end{equation}
where $N$ is the total number of spins, $V$ is the volume of the magnetic film, $g$ is the $g$-factor of an electron spin, and $\mu_{B}$ is the Bohr magneton.
\par

We define the following magnon operators by Fourier transforming the creation and annihilation operators:
\begin{equation}
\begin{split}
    \hat{a_{\bm{k}}} &= \frac{1}{\sqrt{N}} \sum_{i} e^{-i\bm{k} \cdot \bm{r_{i}}} \hat{a_{i}},  \\
    \hat{a_{\bm{k}}}^{\dag} &= \frac{1}{\sqrt{N}} \sum_{i} e^{i\bm{k} \cdot \bm{r_{i}}} \hat{a_{i}}^{\dag}.  \\
\end{split}
\end{equation}

By defining the Fourier transform of the magnetization as \footnote{For convenience, we use the three-dimensional Fourier transform here, even though we consider spin waves that propagate only in two dimensions on a thin film (i.e., $k_{y} = 0$).}

\begin{equation}
    M^{\alpha}_{\bm{k}} = \frac{1}{\sqrt{V}} \int \, d\bm{r} \, M^{\alpha}(\bm{r}) e^{-i\bm{k} \cdot \bm{r}}, \\
\end{equation}

The Fourier components of the magnetization can be expressed in terms of the magnon operators as
\begin{equation}
\label{eq:Mk}
\begin{split}
    M^{x}_{\bm{k}} &= \hbar\gamma^{\prime}_{e} \sqrt{\frac{NS}{2V}} \qty(\hat{a}_{\bm{k}} + \hat{a}_{\bm{-k}}^{\dag}) \\
    &- \frac{\hbar \gamma^{\prime}_{e}}{4}\sqrt{\frac{1}{2VNS}} \sum_{\bm{k_{1}},\bm{k_{2}},\bm{k_{3}}} \qty(\hat{a}_{\bm{k_{1}}}^{\dag} \hat{a}_{\bm{k_{2}}} \hat{a}_{\bm{k_{3}}} \delta(\bm{k} + \bm{k_{1}} - \bm{k_{2}} - \bm{k_{3}}) 
    + \hat{a}_{\bm{k_{1}}}^{\dag} \hat{a}_{\bm{k_{2}}}^{\dag} \hat{a}_{\bm{k_{3}}} \delta(\bm{k} + \bm{k_{1}} + \bm{k_{2}} - \bm{k_{3}})) + \, ...,  \\
    M^{y}_{\bm{k}} &= -i\hbar\gamma^{\prime}_{e} \sqrt{\frac{NS}{2V}} \qty(\hat{a}_{\bm{k}} - \hat{a}_{\bm{-k}}^{\dag}) \\
    &+ \frac{i \hbar \gamma^{\prime}_{e}}{4}\sqrt{\frac{1}{2VNS}} \sum_{\bm{k_{1}},\bm{k_{2}},\bm{k_{3}}} \qty(\hat{a}_{\bm{k_{1}}}^{\dag} \hat{a}_{\bm{k_{2}}} \hat{a}_{\bm{k_{3}}} \delta(\bm{k} + \bm{k_{1}} - \bm{k_{2}} - \bm{k_{3}}) 
    - \hat{a}_{\bm{k_{1}}}^{\dag} \hat{a}_{\bm{k_{2}}}^{\dag} \hat{a}_{\bm{k_{3}}} \delta(\bm{k} + \bm{k_{1}} + \bm{k_{2}} - \bm{k_{3}})) + \, ...,  \\
    M^{z}_{\bm{k}} &= \frac{\hbar \gamma^{\prime}_{e}}{\sqrt{V}} \qty(SN \delta(\bm{k}) - \sum_{\bm{k_{1}},\bm{k_{2}}} \hat{a}_{\bm{k_{1}}}^{\dag} \hat{a}_{\bm{k_{2}}} \delta(\bm{k} + \bm{k_{1}} - \bm{k_{2}})),
\end{split}
\end{equation}
where $\gamma^{\prime}_{e} \, (= 2\pi \gamma_{e})$ is the gyromagnetic ratio of the electron spin.
\par
Based on the expressions above, we compute each term of the Hamiltonian while omitting constant terms that are independent of the magnon operators.
\par
First, the Zeeman energy associated with the external magnetic field $\hat{H}_{\mathrm{Zeeman}}$ can be calculated as

\begin{equation}
\label{eq:Hze_all}
    \hat{H}_{\mathrm{Zeeman}} = \hbar \mu_{0} H_{0} \gamma^{\prime}_{e} \sum_{\bm{k}} \hat{a}_{\bm{k}}^{\dag} \hat{a}_{\bm{k}}.
\end{equation}
\par
Second, we consider the exchange energy $\hat{H}_{\mathrm{ex}}$. After Fourier transforming, the exchange energy can be written as

\begin{equation}
\label{eq:Hexk_all}
    \hat{H}_{\mathrm{ex}} = -\frac{\mu_{0}}{2} \sum_{\bm{k}} \qty(M^{x}_{\bm{k}} H^{x}_{\mathrm{ex},-\bm{k}} + M^{y}_{\bm{k}} H^{y}_{\mathrm{ex},-\bm{k}} + M^{z}_{\bm{k}} H^{z}_{\mathrm{ex},-\bm{k}}).
\end{equation}
Here, according to Eq.~(\ref{eq:ch6_Hex_classical}), the Fourier transform of the exchange field can be expressed as
\begin{equation}
\label{eq:Hexk}
    H^{\alpha}_{\mathrm{ex},\bm{k}} = -\alpha_{\mathrm{ex}} k^{2} M^{\alpha}_{\bm{k}}. 
\end{equation}
Using Eqs.~(\ref{eq:Mk}), (\ref{eq:Hexk_all}), and (\ref{eq:Hexk}), we obtain the following expression for the exchange energy up to fourth order in the magnon operators:
\begin{equation}
\label{eq:Hex_all}
\begin{split}
    \hat{H}_{\mathrm{ex}} &= \hbar \mu_{0} \gamma^{\prime}_{e} \sum_{\bm{k}} \alpha_{\mathrm{ex}} M_{\mathrm{s}} k^{2} \hat{a}_{\bm{k}}^{\dag} \hat{a}_{\bm{k}} \\
    &+ \mu_{0} \frac{(\hbar \gamma^{\prime}_{e})^{2}}{4V} \sum_{\bm{k_{1}},\bm{k_{2}, \bm{k_{3}},\bm{k_{4}}}} \alpha_{\mathrm{ex}} \qty(k^{2}_{1} + k^{2}_{3} - 4 \bm{k_{1}} \cdot \bm{k_{3}})  \hat{a}_{\bm{k1}}^{\dag}  \hat{a}_{\bm{k2}}^{\dag} \hat{a}_{\bm{k3}} \hat{a}_{\bm{k4}} \delta(\bm{k_{1}} + \bm{k_{2}} - \bm{k_{3}} - \bm{k_{4}}) + ....
\end{split}
\end{equation}

\par
Third, we consider the dipolar energy $\hat{H}_{\mathrm{dip}}$. Following the same procedure used for the exchange energy, the dipolar energy can be expressed as
\begin{equation}
\label{eq:Hdipk_all}
    \hat{H}_{\mathrm{dip}} = -\frac{\mu_{0}}{2} \sum_{\bm{k}} \qty(M^{x}_{\bm{k}} H^{x}_{\mathrm{dip},-\bm{k}} + M^{y}_{\bm{k}} H^{y}_{\mathrm{dip},-\bm{k}} + M^{z}_{\bm{k}} H^{z}_{\mathrm{dip},-\bm{k}}).
\end{equation}

Here, the Fourier transform of the dipolar magnetic field, averaged over the film thickness in a two-dimensional thin film, can be written as \cite{Yu2019, Bertelli2021}

\begin{equation}
\label{eq:Hdipk}
\begin{split}
    H_{\mathrm{dip},\bm{k}} &= \mqty(-\frac{k_{x}^2}{k^2} g(kd) & 0 & -\frac{k_{x}k_{z}}{k^2} g(kd)  \\
    0 & g(kd)-1 & 0 \\
    -\frac{k_{x}k_{z}}{k^2} g(kd) & 0 & -\frac{k_{z}^{2}}{k^2} g(kd)
    )
    \mqty(M^{x}_{\bm{k}} \\  M^{y}_{\bm{k}} \\ M^{z}_{\bm{k}}) \\
    &= \mqty(-g(kd)\sin^{2}\theta_{\bm{k}} & 0 & -g(kd)\sin\theta_{\bm{k}}\cos\theta_{\bm{k}} \\
    0 & g(kd)-1 & 0 \\
    -g(kd)\sin\theta_{\bm{k}}\cos\theta_{\bm{k}} & 0 & -g(kd)\cos^{2}\theta_{\bm{k}}
    )
    \mqty(M^{x}_{\bm{k}} \\  M^{y}_{\bm{k}} \\ M^{z}_{\bm{k}}),
\end{split}
\end{equation}
where 
\begin{equation}
    g(kd) = 1 - \frac{1-e^{-kd}}{kd},
\end{equation}
and $\theta_{\bm{k}}$ corresponds to the angle between the magnetization and the propagation direction of the spin waves (see Fig.~\ref{fig:si_disp_setup}). \\
Using Eqs.~(\ref{eq:Mk}), (\ref{eq:Hdipk_all}), and (\ref{eq:Hdipk}), the dipolar energy can be expressed up to fourth order in the magnon operators as \cite{Nembach2011}
\begin{equation}
\label{eq:Hdip_all}
\begin{split}
    \hat{H}_{\mathrm{dip}} &= \frac{\hbar \gamma^{\prime}_{e} \mu_{0} M_{\mathrm{s}}}{2} \sum_{\bm{k}} \left\{(g(kd)\sin^{2}\theta_{\bm{k}} + 1 - g(kd)) \hat{a}_{\bm{k}}^{\dag} \hat{a}_{\bm{k}} \right. \\
    &\hspace{5cm} + \frac{1}{2} \qty(g(kd)\sin^{2}\theta_{\bm{k}} - (1 - g(kd))) \qty(\hat{a}_{\bm{k}} \hat{a}_{\bm{-k}} + \hat{a}_{\bm{k}}^{\dag} \hat{a}_{\bm{-k}}^{\dag}) \Big. \Big\} \\
    &- \mu_{0} \frac{(\hbar \gamma^{\prime}_{e})^2}{V} \sqrt{\frac{NS}{2}} \sum_{\bm{k_{1}},\bm{k_{2}},\bm{k_{3}}} g(k_{1}d) \sin \theta_{\bm{k_{1}}} \cos \theta_{\bm{k_{1}}} \qty(\hat{a}_{\bm{k_{1}}}^{\dag} \hat{a}_{\bm{k_{2}}}^{\dag} \hat{a}_{\bm{k_{3}}} + \hat{a}_{\bm{k_{3}}} \hat{a}_{\bm{k_{1}}} \hat{a}_{\bm{k_{2}}}) \delta(\bm{k_{1}} + \bm{k_{2}} - \bm{k_{3}})  \\
    &+ \mu_{0} \frac{(\hbar \gamma^{\prime}_{e})^{2}}{4V} \sum_{\bm{k_{1}},\bm{k_{2}},\bm{k_{3}}, \bm{k_{4}}} \qty(2g(|\bm{k_{1}}-\bm{k_{3}}|d)\cos^{2}\theta_{\bm{k_{1}}-\bm{k_{3}}} + g(k_{1}d) \cos^{2} \theta_{\bm{k_{1}}} - 1) \\
    &\hspace{9cm} \times \hat{a}_{\bm{k_{1}}}^{\dag} \hat{a}_{\bm{k_{2}}}^{\dag} \hat{a}_{\bm{k_{3}}} \hat{a}_{\bm{k_{4}}} \delta(\bm{k_{1}} + \bm{k_{2}} - \bm{k_{3}} - \bm{k_{4}}) \\
    &+ \mu_{0} \frac{(\hbar \gamma^{\prime}_{e})^{2}}{8V} \sum_{\bm{k_{1}},\bm{k_{2}},\bm{k_{3}}, \bm{k_{4}}} \qty(1 - g(k_{1}d) - g(k_{1}d) \sin^{2} \theta_{\bm{k_{1}}}) \\
    &\hspace{6cm} \times \qty(\hat{a}_{\bm{k_{1}}}^{\dag} \hat{a}_{\bm{k_{2}}}^{\dag} \hat{a}_{\bm{k_{3}}}^{\dag} \hat{a}_{\bm{k_{4}}} + \hat{a}_{\bm{k_{4}}}^{\dag} \hat{a}_{\bm{k_{1}}} \hat{a}_{\bm{k_{2}}} \hat{a}_{\bm{k_{3}}}) \delta(\bm{k_{1}} + \bm{k_{2}} + \bm{k_{3}} - \bm{k_{4}}) \\
    &+ ...
\end{split}
\end{equation}
\par
Finally, we calculate the surface anisotropy energy $\hat{H}_{\mathrm{ani}}$. The surface anisotropy energy can be expressed as

\begin{equation}
\label{eq:Hani_fourier}
    \hat{H}_{\mathrm{ani}} = -\frac{\mu_{0}H_{\mathrm{ani}}}{2 M_{\mathrm{s}}} \sum_{\bm{k}} M^{y}_{\bm{k}} M^{y}_{\bm{-k}},
\end{equation}
where $H_{\mathrm{ani}} = \frac{2 K_{\mathrm{s}}}{M_{\mathrm{s}}d}$ is the anisotropy field. By substituting Eq.~(\ref{eq:Mk}) into Eq.~(\ref{eq:Hani_fourier}), the surface anisotropy energy can be evaluated up to fourth order as
\begin{equation}
\label{eq:Hani_all}
\begin{split}
    \hat{H}_{\mathrm{ani}} &= \frac{\hbar \gamma^{\prime}_{e} \mu_{0} H_{\mathrm{ani}}}{2} \sum_{\bm{k}} \qty{-\hat{a}_{\bm{k}}^{\dag} \hat{a}_{\bm{k}} + \frac{1}{2}\qty(\hat{a}_{\bm{k}} \hat{a}_{\bm{-k}} + \hat{a}_{\bm{k}}^{\dag} \hat{a}_{\bm{-k}}^{\dag})} \\
    &+ \mu_{0} \frac{(\hbar \gamma^{\prime}_{e})^{2}H_{\mathrm{ani}}}{4VM_{\mathrm{s}}} \sum_{\bm{k_{1}},\bm{k_{2}},\bm{k_{3}}, \bm{k_{4}}} \hat{a}_{\bm{k_{1}}}^{\dag} \hat{a}_{\bm{k_{2}}}^{\dag} \hat{a}_{\bm{k_{3}}} \hat{a}_{\bm{k_{4}}} \delta(\bm{k_{1}} + \bm{k_{2}} - \bm{k_{3}} - \bm{k_{4}}) \\
    &- \mu_{0} \frac{(\hbar \gamma^{\prime}_{e})^{2} H_{\mathrm{ani}}}{8VM_{\mathrm{s}}} \sum_{\bm{k_{1}},\bm{k_{2}},\bm{k_{3}}, \bm{k_{4}}} \qty(\hat{a}_{\bm{k_{1}}}^{\dag} \hat{a}_{\bm{k_{2}}}^{\dag} \hat{a}_{\bm{k_{3}}}^{\dag} \hat{a}_{\bm{k_{4}}} + \hat{a}_{\bm{k_{4}}}^{\dag} \hat{a}_{\bm{k_{1}}} \hat{a}_{\bm{k_{2}}} \hat{a}_{\bm{k_{3}}}) \delta(\bm{k_{1}} + \bm{k_{2}} + \bm{k_{3}} - \bm{k_{4}}) \\
    &+ ....
\end{split}
\end{equation}

\par
By combining Eqs.~(\ref{eq:Hze_all}), (\ref{eq:Hex_all}), (\ref{eq:Hdip_all}), and (\ref{eq:Hani_all}), the total spin-wave Hamiltonian $\hat{H}$ can be written in terms of magnon operators up to fourth order as
\begin{equation}
\label{eq:H_all}
\begin{split}
    \hat{H} &= \hbar \sum_{\bm{k}} \qty(A_{\bm{k}} \hat{a}_{\bm{k}}^{\dag} \hat{a}_{\bm{k}} + \frac{1}{2} B_{\bm{k}} \hat{a}_{\bm{k}} \hat{a}_{\bm{-k}} + \frac{1}{2} B_{\bm{k}} \hat{a}_{\bm{k}}^{\dag} \hat{a}_{\bm{-k}}^{\dag} ) + \hbar \sum_{\bm{k_{1}},\bm{k_{2}},\bm{k_{3}}} \qty(C_{\bm{k_{1}}} \hat{a}_{\bm{k_{1}}}^{\dag} \hat{a}_{\bm{k_{2}}}^{\dag} \hat{a}_{\bm{k_{3}}} + \mathrm{c.c}) \delta(\bm{k_{1}} + \bm{k_{2}} - \bm{k_{3}}) \\
    &+ \hbar \sum_{\bm{k_{1}},\bm{k_{2}},\bm{k_{3}}, \bm{k_{4}}} D_{\bm{k_{1}},\bm{k_{3}}} \hat{a}_{\bm{k_{1}}}^{\dag} \hat{a}_{\bm{k_{2}}}^{\dag} \hat{a}_{\bm{k_{3}}} \hat{a}_{\bm{k_{4}}} \delta(\bm{k_{1}} + \bm{k_{2}} - \bm{k_{3}} - \bm{k_{4}}) \\
    &+ \hbar \sum_{\bm{k_{1}},\bm{k_{2}},\bm{k_{3}}, \bm{k_{4}}} \qty(E_{\bm{k_{1}}} \hat{a}_{\bm{k_{1}}}^{\dag} \hat{a}_{\bm{k_{2}}}^{\dag} \hat{a}_{\bm{k_{3}}}^{\dag} \hat{a}_{\bm{k_{4}}} + \mathrm{c.c}) \delta(\bm{k_{1}} + \bm{k_{2}} + \bm{k_{3}} - \bm{k_{4}}) \\
    &+ ...,
\end{split}
\end{equation}
where
\begin{equation}
\label{eq:H_element}
\begin{split}
    A_{\bm{k}} &= \mu_{0} \gamma^{\prime}_{e} \qty(H_{0} + \alpha_{\mathrm{ex}} k^{2} + \frac{M_{\mathrm{s}}}{2}g(kd) \sin^{2}(\theta_{\bm{k}}) + \frac{M_{\mathrm{s}}}{2}(1-g(kd)) - \frac{H_{\mathrm{ani}}}{2}), \\
    B_{\bm{k}} &= \mu_{0} \gamma^{\prime}_{e} \qty(\frac{M_{\mathrm{s}}}{2} g(kd) \sin^{2} \theta_{\bm{k}} - \frac{M_{\mathrm{s}}}{2} \qty(1 - g(kd)) + \frac{H_{\mathrm{ani}}}{2}), \\
    C_{\bm{k}} &= - \frac{\mu_{0} \hbar \gamma^{\prime 2}_{e}}{V} \sqrt{\frac{NS}{2}} g(kd) \sin \theta_{\bm{k}} \cos \theta_{\bm{k}}, \\
    D_{\bm{k_{1}},\bm{k_{2}}} &= \mu_{0} \frac{\hbar \gamma^{\prime 2}_{e}}{4V} \qty(2g(|\bm{k_{1}}-\bm{k_{2}}|d)\cos^{2}\theta_{\bm{k_{1}}-\bm{k_{2}}} + g(k_{1}d) \cos^{2} \theta_{\bm{k_{1}}} - 1 + \alpha_{\mathrm{ex}} \qty(k^{2}_{1} + k^{2}_{2} - 4 \bm{k_{1}} \cdot \bm{k_{2}}) + \frac{H_{\mathrm{ani}}}{M_{\mathrm{s}}}), \\
    E_{\bm{k}} &= \mu_{0} \frac{\hbar \gamma^{\prime 2}_{e}}{8V} \qty(1 - g(kd) - g(kd) \sin^{2} \theta_{\bm{k}} - \frac{H_{\mathrm{ani}}}{M_{\mathrm{s}}}). \\
\end{split}
\end{equation}

\subsection{Derivation of the spin-wave dispersion relation}
While Eq.~(\ref{eq:H_all}) yields the Hamiltonian describing the magnon dynamics, due to the presence of the dipolar interaction terms $B_{\bm{k}}$, the second-order Hamiltonian, which governs the linear dynamics of the magnons, is not diagonal.

\par
Here, we diagonalize the Hamiltonian by applying an appropriate transformation to the magnon operators, derive the dispersion relation of the spin waves, and subsequently calculate the higher-order terms in the diagonalized basis.

\par
We consider the following Bogoliubov transformation \cite{rezende2020fundamentals}:
\begin{equation}
\begin{split}
\label{eq:bogoliubov}
    \hat{a}_{\bm{k}} &= u_{\bm{k}} \hat{c}_{\bm{k}} + v_{\bm{k}} \hat{c}_{\bm{-k}}^{\dag}, \\
    \hat{a}_{\bm{-k}}^{\dag} &= v_{\bm{k}} \hat{c}_{\bm{k}} + u_{\bm{k}} \hat{c}_{\bm{-k}}^{\dag}, \\
\end{split}
\end{equation}
where $u_{\bm{k}}$ and $v_{\bm{k}}$ satisfy $u^{2}_{\bm{k}}-v^{2}_{\bm{k}} = 1$.
This transformation is unitary, and $\hat{c}_{\bm{k}}$ and $\hat{c}^{\dag}_{\bm{k}}$ obey the following commutation relations:
\begin{equation}
    [\hat{c}_{\bm{k_{1}}}, \hat{c}_{\bm{k_{2}}}^{\dag}] = \delta(\bm{k_{1}}-\bm{k_{2}}), \, [\hat{c}_{\bm{k_{1}}}, \hat{c}_{\bm{k_{2}}}] = 0, \, [\hat{c}^{\dag}_{\bm{k_{1}}}, \hat{c}^{\dag}_{\bm{k_{2}}}] = 0.
\end{equation}
Through this transformation, the Hamiltonian up to the second order terms can be transformed into
\begin{equation}
\label{eq:H_bogo}
    \hat{H} = \sum_{\bm{k}} \qty{A_{\bm{k}}\qty(u^{2}_{\bm{k}} + v^{2}_{\bm{k}}) \hat{c}_{\bm{k}}^{\dag} \hat{c}_{\bm{k}} + \qty(A_{\bm{k}} u_{\bm{k}} v_{\bm{k}} + \frac{B_{\bm{k}}}{2} \qty(u^{2}_{\bm{k}} + v^{2}_{\bm{k}})) \qty(\hat{c}_{\bm{k}}^{\dag} \hat{c}_{\bm{k}}^{\dag} + \hat{c}_{\bm{k}} \hat{c}_{\bm{k}})}.
\end{equation}
Here, by setting $u_{\bm{k}}$ and $v_{\bm{k}}$ as
\begin{equation}
\begin{split}
    u_{\bm{k}} &= \sqrt{\frac{A_{\bm{k}}+\omega_{\bm{k}}}{2\omega_{\bm{k}}}}, \\ 
    v_{\bm{k}} &= -\frac{B_{k}}{|B_{\bm{k}}|} \sqrt{\frac{A_{\bm{k}}-\omega_{\bm{k}}}{2\omega_{\bm{k}}}}, \\ 
    \omega_{\bm{k}} &= \sqrt{A^{2}_{k} - B^{2}_{k}}, \\
\end{split}
\end{equation}
the non-diagonal terms of the Hamiltonian [Eq.~(\ref{eq:H_bogo})] are canceled out, yielding the following diagonalized Hamiltonian:
\begin{equation}
    \hat{H} = \sum_{\bm{k}} \omega_{\bm{k}} \hat{c}_{\bm{k}}^{\dag} \hat{c}_{\bm{k}},
\end{equation}
where
\begin{equation}
\label{eq:ch6_dispersion}
    \omega_{\bm{k}} = \mu_{0} \gamma^{\prime}_{e} \sqrt{\qty(H_{0} + \alpha_{\mathrm{ex}} k^{2} + M_{\mathrm{s}}(1-g(kd)) - H_{\mathrm{ani}}) \qty(H_{0} + \alpha_{\mathrm{ex}} k^{2} + M_{\mathrm{s}}g(kd) \sin^{2}\theta_{\bm{k}}) }
\end{equation}
corresponds to the dispersion relation of the spin waves. This expression is consistent with the classical result obtained by solving the Landau-Lifshitz-Gilbert (LLG) equation for spin waves propagating in a thin film \cite{Yu2019,Bertelli2021}.
\par
Higher-order terms can be obtained by substituting the Bogoliubov transformation into Eq.~(\ref{eq:H_all}). Consequently, the total Hamiltonian up to fourth order can be written as
\begin{equation}
\label{eq:H_all_bogo}
\begin{split}
    \hat{H} &= \hbar \sum_{\bm{k}} \omega_{\bm{k}} \hat{c}_{\bm{k}}^{\dag} \hat{c}_{\bm{k}} + \sum_{\bm{k_{1}},\bm{k_{2}},\bm{k_{3}}} \qty(F_{\bm{k_{1}},\bm{k_{2}},\bm{k_{3}}} \hat{c}_{\bm{k_{1}}} \hat{c}_{\bm{k_{2}}} \hat{c}_{\bm{k_{3}}} + \mathrm{c.c}) \delta(\bm{k_{1}} + \bm{k_{2}} + \bm{k_{3}}) \\
    &+ \sum_{\bm{k_{1}},\bm{k_{2}},\bm{k_{3}}} \qty(G_{\bm{k_{1}},\bm{k_{2}},\bm{k_{3}}} \hat{c}_{\bm{k_{1}}}^{\dag} \hat{c}_{\bm{k_{2}}}^{\dag} \hat{c}_{\bm{k_{3}}} + \mathrm{c.c}) \delta(\bm{k_{1}} + \bm{k_{2}} - \bm{k_{3}}) \\
    &+ \sum_{\bm{k_{1}},\bm{k_{2}},\bm{k_{3}}, \bm{k_{4}}} H_{\bm{k_{1}},\bm{k_{2}},\bm{k_{3}},\bm{k_{4}}} \hat{c}_{\bm{k_{1}}}^{\dag} \hat{c}_{\bm{k_{2}}}^{\dag} \hat{c}_{\bm{k_{3}}} \hat{c}_{\bm{k_{4}}} \delta(\bm{k_{1}} + \bm{k_{2}} - \bm{k_{3}} -  \bm{k_{4}}) \\
    &+ \sum_{\bm{k_{1}},\bm{k_{2}},\bm{k_{3}}, \bm{k_{4}}} \qty(I_{\bm{k_{1}},\bm{k_{2}},\bm{k_{3}},\bm{k_{4}}} \hat{c}_{\bm{k_{1}}}^{\dag} \hat{c}_{\bm{k_{2}}}^{\dag} \hat{c}_{\bm{k_{3}}}^{\dag} \hat{c}_{\bm{k_{4}}} + \mathrm{c.c}) \delta(\bm{k_{1}} + \bm{k_{2}} + \bm{k_{3}} -  \bm{k_{4}}) \\
    &+ \sum_{\bm{k_{1}},\bm{k_{2}},\bm{k_{3}}, \bm{k_{4}}} \qty(J_{\bm{k_{1}},\bm{k_{2}},\bm{k_{3}},\bm{k_{4}}} \hat{c}_{\bm{k_{1}}} \hat{c}_{\bm{k_{2}}} \hat{c}_{\bm{k_{3}}} \hat{c}_{\bm{k_{4}}} + \mathrm{c.c}) \delta(\bm{k_{1}} + \bm{k_{2}} + \bm{k_{3}} + \bm{k_{4}}) \\
    &+ ...
\end{split}
\end{equation}
\par
In our experiments, the dominant contribution to the nonlinear dynamics arises from four-magnon scattering, which involves the annihilation of two magnons and the creation of two others. Therefore, we present only the expression for the term $H_{\bm{k_{1}},\bm{k_{2}},\bm{k_{3}},\bm{k_{4}}}$, which can be calculated as \cite{Nembach2011, Krivosik2010}
\begin{equation}
\begin{split}
    H_{\bm{k_{1}},\bm{k_{2}},\bm{k_{3}},\bm{k_{4}}} &= D_{\bm{k_{1}},\bm{k_{3}}} u_{\bm{k_{1}}} u_{\bm{k_{2}}} u_{\bm{k_{3}}} u_{\bm{k_{4}}} + D_{\bm{k_{1}},\bm{k_{3}}} u_{\bm{k_{1}}} v_{\bm{k_{2}}} u_{\bm{k_{3}}} v_{\bm{k_{4}}} + D_{\bm{k_{1}},\bm{-k_{2}}} u_{\bm{k_{1}}} v_{\bm{k_{2}}} v_{\bm{k_{3}}} v_{\bm{k_{4}}} \\
    &+ D_{\bm{-k_{4}},\bm{k_{3}}} v_{\bm{k_{1}}} u_{\bm{k_{2}}} u_{\bm{k_{3}}} v_{\bm{k_{4}}}
    + D_{\bm{-k_{3}},-\bm{k_{1}}} v_{\bm{k_{1}}} v_{\bm{k_{2}}} v_{\bm{k_{3}}} v_{\bm{k_{4}}}
    + D_{\bm{-k_{3}},-\bm{k_{1}}} v_{\bm{k_{1}}} u_{\bm{k_{2}}} v_{\bm{k_{3}}} u_{\bm{k_{4}}} \\
    &+ 2 \qty(2 E_{\bm{k_{1}}} + E_{\bm{-k_{4}}}) u_{\bm{k_{1}}} u_{\bm{k_{2}}} u_{\bm{k_{3}}} v_{\bm{k_{4}}} + 2\qty(E_{\bm{k_{1}}} + 2 E_{\bm{-k_{4}}}) u_{\bm{k_{1}}} v_{\bm{k_{2}}} v_{\bm{k_{3}}} v_{\bm{k_{4}}}.
\end{split}
\end{equation}

\subsection{Magnon coherent states}
The eigenstates of the diagonalized Hamiltonian $\ket{n_{\bm{k}}}$ correspond to the magnon number states. However, these number states satisfy
\begin{equation}
\begin{split}
    \braket{n_{\bm{k}}|\hat{c}_{\bm{k}}|n_{\bm{k}}} &= 0, \\
    \braket{n_{\bm{k}}|\hat{c}_{\bm{k}}^{\dag}|n_{\bm{k}}} &= 0. \\ 
\end{split}
\end{equation}
From Eq.~(\ref{eq:Mk}), it follows that the average values of the spin-wave components in the $x$- and $y$-directions vanish at all times. Consequently, the number state by itself does not represent the quantum state corresponding to the classical magnetization dynamics.
\par
The coherent state, on the other hand, is the quantum state corresponding to the classical magnetization dynamics \cite{Zagury1971, rezende2020fundamentals}. It is defined via the displacement operator:
\begin{equation}
    \hat{D}(\alpha_{\bm{k}}) = e^{\alpha_{\bm{k}} \hat{c}_{\bm{k}}^{\dag} - \alpha^{*}_{\bm{k}}  \hat{c}_{\bm{k}}}.
\end{equation}
This displacement operator shifts the magnon creation (annihilation) operator by $\alpha \, (\alpha^{*})$:
\begin{equation}
\begin{split}
    \hat{D}^{\dag}(\alpha_{\bm{k}}) \hat{c}_{\bm{k}} \hat{D}(\alpha_{\bm{k}}) &= \hat{c}_{\bm{k}} + \alpha, \\
    \hat{D}^{\dag}(\alpha_{\bm{k}}) \hat{c}^{\dag}_{\bm{k}} \hat{D}(\alpha_{\bm{k}}) &= \hat{c}^{\dag}_{\bm{k}} + \alpha^{*}. \\
\end{split}
\end{equation}
The coherent state $\ket{\alpha_{\bm{k}}}$ is obtained by applying the displacement operator to the magnon vacuum state:
\begin{equation}
 \ket{\alpha_{\bm{k}}} = \hat{D}(\alpha_{\bm{k}}) \ket{0}.  \\
\end{equation}
This state is an eigenstate of the annihilation operator with eigenvalue $\alpha_{\bm{k}}$:
\begin{equation}
  \hat{c}_{\bm{k}} \ket{\alpha_{\bm{k}}} = \alpha_{\bm{k}} \ket{\alpha_{\bm{k}}}.
\end{equation}
It can be expressed in terms of magnon number states as
\begin{equation}
    \ket{\alpha_{\bm{k}}} = e^{-\frac{|\alpha_{\bm{k}}|^{2}}{2}} \sum_{n_{\bm{k}}} \frac{\alpha^{n_{\bm{k}}}_{\bm{k}}}{\sqrt{n_{\bm{k}}!}} \ket{n_{\bm{k}}},
\end{equation}
and the probability distribution for the magnon number $P(n_{\bm{k}})$ is given by
\begin{equation}
    P(n_{\bm{k}}) = |\braket{n_{\bm{k}}|\alpha_{\bm{k}}}|^{2} = \frac{|\alpha_{\bm{k}}|^{2n_{\bm{k}}}}{n_{\bm{k}}!} e^{-|\alpha_{\bm{k}}|^{2}},
\end{equation}
which follows a Poisson distribution with both the mean and variance equal to $|\alpha_{\bm{k}}|^{2}$
\par
For the coherent state $\ket{\alpha_{\bm{k}}} \, (\alpha_{\bm{k}} = |\alpha_{\bm{k}}|e^{i\phi_{\bm{k}}})$, using Eqs.~(\ref{eq:Mk}) and (\ref{eq:bogoliubov}), the expectation values of the classical magnetization components $M_{x,y}(\bm{r},t)$ in the $x$- and $y$-directions are obtained as \cite{rezende2020fundamentals}

\begin{equation}
\label{eq:Mclass}
\begin{split}
    \ev{M_{x}(\bm{r},t)} &= \frac{M_{\mathrm{s}}}{\sqrt{\frac{NS}{2}}} |\alpha_{\bm{k}}| (u_{\bm{k}} + v_{\bm{k}}) \cos \qty(\bm{k} \cdot \bm{r} - \omega_{\bm{k}}t + \phi_{\bm{k}}), \\
    \ev{M_{y}(\bm{r},t)} &= \frac{M_{\mathrm{s}}}{\sqrt{\frac{NS}{2}}} |\alpha_{\bm{k}}| (u_{\bm{k}} - v_{\bm{k}}) \sin \qty(\bm{k} \cdot \bm{r} - \omega_{\bm{k}}t + \phi_{\bm{k}}),
\end{split}
\end{equation}
which correspond to the macroscopic magnetization dynamics of the spin waves. \\

\subsection{Magnon dynamics in the linear region}
In this subsection, we derive the dynamics of magnons under the application of excitation microwaves in the linear regime, where terms of order higher than second in the Hamiltonian can be are neglected \cite{rezende2020fundamentals}.
\par
First, in the Heisenberg picture, we consider the time evolution of the magnon operator $\hat{c}_{\bm{k}}$. Its dynamics are governed by the Heisenberg equation:

\begin{equation}
\label{eq:Heq}
    i \hbar \frac{\mathrm{d}}{\mathrm{d}t} \hat{c}_{\bm{k}} = \qty[\hat{c}_{\bm{k}}, \hat{H}(t)],
\end{equation}
where $\hat{H}(t)$ can be decomposed as
\begin{equation}
    \hat{H}(t) = \hat{H}_{0} + \hat{H}_{\mathrm{mw}}(t),
\end{equation}
where $\hat{H}_{0}$ denotes the magnon Hamiltonian [Eq.~(\ref{eq:H_all_bogo})], and $\hat{H}_{\mathrm{mw}}(t)$ represents the Zeeman energy due to the excitation microwave fields.
\par
We assume that the microwave magnetic field $h_{\mathrm{mw}}(\bm{r},t)$ has only an $x$-component and oscillates with an angular frequency $\omega$. In addition, we assume that it is an even function with respect to the spatial coordinates.
\\
In this case, the Zeeman energy due to the microwave field $\hat{H}_{\mathrm{mw}}(t)$ is given by
\begin{equation}
\label{eq:Hmw_all}
    \hat{H}_{\mathrm{mw}}(t) = -\mu_{0} \int \, d \bm{r} M^{x}(\bm{r}) h^{x}_{\mathrm{mw}}(\bm{r}) \sin \omega t,
\end{equation}
and by performing Fourier transforms of the magnetization and microwave fields, we obtain 
\begin{equation}
\label{eq:Hmw}
    \hat{H}_{\mathrm{mw}}(t) = \sum_{\bm{k}} \qty(g_{\bm{k}} \hat{c}_{\bm{k}} e^{i\omega t} + \mathrm{c.c}),
\end{equation}
where
\begin{equation}
    g_{\bm{k}} = i \mu_{0} \hbar \gamma^{\prime}_{e} \sqrt{\frac{NS}{8V}} \qty(u_{\bm{k}} + v_{\bm{k}}) h^{x}_{\bm{k}},
\end{equation}
and $h^{x}_{\bm{k}}$ is the Fourier transform of the $x$-component of the microwave magnetic field $h^{x}_{\mathrm{mw}}(\bm{r})$. \\
By substituting Eq.~(\ref{eq:Hmw}) into the Heisenberg equation [Eq.~(\ref{eq:Heq})], the time evolution of the magnon annihilation operator is described by
\begin{equation}
    \frac{\mathrm{d} \hat{c}_{\bm{k}}}{\mathrm{d}t} = -i \omega_{\bm{k}} \hat{c}_{\bm{k}} - \frac{i}{\hbar} g^{*}_{\bm{k}} e^{-i\omega t}.
\end{equation}
In general, magnons in a magnetic material have lifetimes ranging from several nanoseconds to several microseconds. By introducing the magnon relaxation rate $\Gamma_{\bm{k}}$, the above equation can be rewritten as
\begin{equation}
    \frac{\mathrm{d} \hat{c}_{\bm{k}}}{\mathrm{d}t} = -i \omega_{\bm{k}} \hat{c}_{\bm{k}} - \Gamma_{\bm{k}} \hat{c}_{\bm{k}} - \frac{i}{\hbar} g^{*}_{\bm{k}} e^{-i\omega t}.
\end{equation}
Assuming $\hat{c}_{\bm{k}} \propto e^{-i\omega t}$, the expectation value of the magnon operator in the steady state can be calculated as
\begin{equation}
     \ev{\hat{c}_{\bm{k}}} = \frac{g^{*}_{\bm{k}}}{\hbar \qty(\qty(\omega - \omega_{\bm{k}}) + i \Gamma_{\bm{k}})} e^{-i\omega t},
\end{equation}
and the expectation value of the magnon number in the steady state ($\frac{\mathrm{d} n_{\bm{k}}}{\mathrm{d}t} = 0$) is
\begin{equation}
     \ev{n_{\bm{k}}} = \frac{|g_{\bm{k}}|^{2}}{\hbar^{2} \qty(\qty(\omega - \omega_{\bm{k}})^{2} +  \Gamma_{\bm{k}}^{2})}.
     \label{eq:n_steady}
\end{equation}

The derivation of magnon dynamics including nonlinear terms will be presented in the next subsection using the Heisenberg picture.
\par

Next, using the interaction picture, we demonstrate that the magnon state excited by the application of microwaves becomes a coherent state \cite{Zagury1971}. In the interaction picture, both the state and the operators evolve in time. We consider the state in the interaction basis given by
\begin{equation}
    \ket{\psi_{I}(t)} = e^{\frac{i \hat{H}_{0}t}{\hbar}} \ket{\psi(t)}.
\end{equation}
The state evolves over time according to the following interaction Hamiltonian $\hat{H}_{I}(t)$:
\begin{equation}
    i\hbar \frac{\mathrm{d}}{\mathrm{d}t} \ket{\psi_{I}(t)} = \hat{H}_{I}(t) \ket{\psi_{I}(t)},
\end{equation}
where
\begin{equation}
\begin{split}
    \hat{H}_{I}(t) &= e^{\frac{i \hat{H}_{0}t}{\hbar}} \hat{H}_{\mathrm{mw}}(t) e^{\frac{-i \hat{H}_{0}t}{\hbar}} \\
                   &= \sum_{\bm{k}} \qty(g_{\bm{k}} \hat{c}_{\bm{k}} e^{i(\omega-\omega_{\bm{k}}) t} + \mathrm{c.c}).
\end{split}
\end{equation}
The interaction state corresponding to the vacuum state $\ket{0}$ at the initial time $t = 0$ evolves into a coherent state at time $t = \tau$ as 
\begin{equation}
\begin{split}
    \ket{\psi_{I}(\tau)} &= e^{i\beta + \sum_{\bm{k}} \qty(\alpha_{\bm{k}} \hat{c}_{\bm{k}}^{\dag} - \alpha^{*}_{\bm{k}} \hat{c}_{\bm{k}})} \ket{0} \\
                      &= e^{i\beta} \prod_{\bm{k}} \ket{\alpha_{\bm{k}}},
\end{split}
\end{equation}
where $\beta$ is an appropriate phase factor and
\begin{equation}
    \alpha_{\bm{k}} = -\frac{i}{\hbar} \int^{\tau}_{0} \, \mathrm{d}t' \, g^{*}_{\bm{k}} e^{-i(\omega - \omega_{\bm{k}}) t'}.
\end{equation}
By introducing the relaxation term as $\omega_{\bm{k}} \rightarrow \omega_{\bm{k}}  - i\Gamma_{\bm{k}}$ and considering a time $\tau$ that is much longer than the relaxation time ($\tau \gg 1/\eta_{\bm{k}}$), $\alpha_{\bm{k}}$ can be calculated as
\begin{equation}
    \alpha_{\bm{k}} = -\frac{g^{*}_{\bm{k}}}{\hbar \qty(\qty(\omega - \omega_{\bm{k}}) + i \Gamma_{\bm{k}})}.
\end{equation}
\\
Therefore, the state vector at time $\tau$ is given by
\begin{equation}
    \ket{\psi(\tau)} = e^{i\beta} \prod_{\bm{k}} e^{-i \omega_{\bm{k}} \tau} \ket{\alpha_{\bm{k}}}, \\
\end{equation}
and the expectation value of the magnon number for wavevector $\bm{k}$ is 
\begin{equation}
     \ev{n_{\bm{k}}} = \frac{|g_{\bm{k}}|^{2}}{\hbar^{2} \qty(\qty(\omega - \omega_{\bm{k}})^{2} +  \Gamma_{\bm{k}}^{2})},
\end{equation}
which is consistent with the result obtained in the Heisenberg picture.

\par
\subsection{Magnon dynamics in the nonlinear region}
\label{subsec:ch6_nonlinear_theory}
This subsection discusses magnon dynamics including the nonlinear four-magnon term \cite{vov1987nonlinear,Boardman1988,Nembach2011,rezende2020fundamentals}. Specifically, we demonstrate that scattering of the excited magnons into other magnon modes becomes pronounced once the amplitude of the excited magnons exceeds a certain threshold.
\par
\begin{figure}[t]
    \centering
    \includegraphics[width=\linewidth]{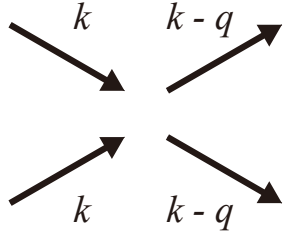}
    \caption{Schematic of the four-magnon scattering.}
    \label{fig:si_four_magnon}
\end{figure}

When the four-magnon term that creates and annihilates two magnons is included in the Heisenberg equation [Eq.~(\ref{eq:Heq})] together with the decay term, the time evolution of the magnon annihilation operator  $\hat{c}_{\bm{k}-\bm{q}}$ obeys the following differential equation:
\begin{equation}
\label{eq:4mag}
    \frac{\mathrm{d}}{\mathrm{d}t} \hat{c}_{\bm{k}-\bm{q}} = -\qty(i\omega_{\bm{k}-\bm{q}} + \Gamma_{\bm{k}-\bm{q}}) \hat{c}_{\bm{k}-\bm{q}} - i \sum_{\bm{k_{1}},\bm{k_{2}},\bm{k_{3}}} H_{\bm{k}-\bm{q},\bm{k_{1}},\bm{k_{2}},\bm{k_{3}}} \hat{c}_{\bm{k_{1}}}^{\dag} \hat{c}_{\bm{k_{2}}} \hat{c}_{\bm{k_{3}}} \delta(\bm{k} - \bm{q} + \bm{k_{1}} - \bm{k_{2}} - \bm{k_{3}}).
\end{equation}
Here, we consider the situation in which only magnons with wavevector $\bm{k}$ are excited by an external microwave field. In this case, the four-magnon scattering is dominated by the process with $\bm{k_{1}} = \bm{k} + \bm{q}$, \, $\bm{k_{2}} = \bm{k_{3}} = \bm{k}$ in Eq.~(\ref{eq:4mag}) (Fig.~\ref{fig:si_four_magnon}). Considering only this process and performing a rotating frame transformation with  $\hat{c}^{'}_{\bm{k}} = \hat{c}_{\bm{k}} e^{i\omega_{\bm{k}} t}$, Eq.~(\ref{eq:4mag}) can be rewritten as

\begin{equation}
\label{eq:4mag_trans}
    \frac{\mathrm{d}}{\mathrm{d}t} \hat{c}^{'}_{\bm{k}-\bm{q}} = - \Gamma_{\bm{k}-\bm{q}} \hat{c}^{'}_{\bm{k}-\bm{q}} - i H_{\bm{k}-\bm{q},\bm{k}+\bm{q},\bm{k},\bm{k}} \hat{c}_{\bm{k}+\bm{q}}^{' \dag} \hat{c}^{'}_{\bm{k}} \hat{c}^{'}_{\bm{k}}.
\end{equation}
Similarly, following the same discussion, the time evolution of $\hat{c}^{' \dag}_{\bm{k}+\bm{q}}$ obeys the differential equation
\begin{equation}
\label{eq:4mag_trans_dag}
    \frac{\mathrm{d}}{\mathrm{d}t} \hat{c}^{' \dag}_{\bm{k}+\bm{q}} = - \Gamma_{\bm{k}+\bm{q}} \hat{c}^{' \dag}_{\bm{k}+\bm{q}} + i H^{*}_{\bm{k}+\bm{q},\bm{k}-\bm{q},\bm{k},\bm{k}} \hat{c}_{\bm{k}-\bm{q}}^{'} \hat{c}^{' \dag}_{\bm{k}} \hat{c}^{' \dag}_{\bm{k}}.
\end{equation}

We now calculate the threshold at which the generation rate of the scattered magnon exceeds its relaxation rate. For the four-magnon scattering terms in Eqs.~(\ref{eq:4mag_trans}) and (\ref{eq:4mag_trans_dag}), we assume that $\hat{c}^{'}_{\bm{k}}$ takes the steady-state value during the excitation by resonant microwaves:

\begin{equation}
    \ev{\hat{c}^{'}_{\bm{k}}} = \frac{g^{*}_{\bm{k}}}{i\hbar \Gamma_{\bm{k}}}.
\end{equation}

By combining Eqs.~(\ref{eq:4mag_trans}) and (\ref{eq:4mag_trans_dag}), we can derive the following differential equation for $\hat{c}^{'}_{\bm{k}-\bm{q}}$:
\begin{equation}
\label{eq:4mag_trans_double}
    \qty(\frac{\mathrm{d}}{\mathrm{d}t} + \Gamma_{\bm{k}+\bm{q}}) \qty(\frac{\mathrm{d}}{\mathrm{d}t} + \Gamma_{\bm{k}-\bm{q}}) \hat{c}^{'}_{\bm{k}-\bm{q}} = H^{*}_{\bm{k}+\bm{q},\bm{k}-\bm{q},\bm{k},\bm{k}} H_{\bm{k}-\bm{q},\bm{k}+\bm{q},\bm{k},\bm{k}} \qty|\ev{\hat{c}^{'}_{\bm{k}}}|^{4} \hat{c}^{'}_{\bm{k}-\bm{q}}.
\end{equation}

We define the magnon generation rate $\gamma_{\bm{k}-\bm{q}}$ by assuming a solution of the form $\hat{c}^{'}_{\bm{k}-\bm{q}} \propto e^{\gamma_{\bm{k}-\bm{q}}t}$. By taking the limit $\bm{q} \rightarrow 0$ in the above equation, we obtain
\begin{equation}
    \qty(\gamma_{\bm{k}} + \Gamma_{\bm{k}})^{2} = \qty|H_{\bm{k},\bm{k},\bm{k},\bm{k}}|^{4} \qty|\ev{\hat{c}^{'}_{\bm{k}}}|^{4}.
\end{equation}
Thus, the magnon generation rate is given by
\begin{equation}
    \gamma_{\bm{k}} = -\Gamma_{\bm{k}} + H_{\bm{k},\bm{k},\bm{k},\bm{k}} \qty|\ev{\hat{c}^{'}_{\bm{k}}}|^{2}.
\end{equation}
This equation shows that the generation rate of scattered magnons transitions from negative to positive when the amplitude of the excited magnons exceeds a threshold, at which point the excited coherent spin waves begin to decay due to four-magnon scattering. The threshold condition is reached when

\begin{equation}
    \qty|\ev{\hat{c}^{'}_{\bm{k}}}| =  \sqrt{\frac{\Gamma_{\bm{k}}}{H_{\bm{k},\bm{k},\bm{k},\bm{k}}}}.
\end{equation}

This expression for the threshold is consistent with previous studies. For example, Ref.~\cite{Nembach2011} derives the threshold for the generation of magnons at wavevectors $\bm{k}$ and $-\bm{k}$ from two magnons with zero wavenumber (second Suhl instability) using a similar approach, yielding an equivalent expression (apart from the wavenumber dependence). Similarly, Ref.~\cite{vov1987nonlinear} calculates the threshold for instability due to scattering from two magnons at wavevector $\bm{k}$ to $\bm{k_{1}}$ and $\bm{k_{2}}$, yielding the same expression.
\par
Finally, we derive the expression for the classical magnetization amplitude at the threshold. By expressing the coefficient of the four-magnon scattering term as
\begin{equation}
\label{eq:ch6_theory_four_magnon_coeff}
    H_{\bm{k},\bm{k},\bm{k},\bm{k}} = \lambda_{\bm{k}} \frac{\hbar \gamma^{\prime 2}_{e}}{V},
\end{equation}
the magnetization amplitude in the $x$-direction at the threshold $m^{\mathrm{th}}_{x,\bm{k}}$ can be calculated from Eq.~(\ref{eq:Mclass}) as
\begin{equation}
\label{eq:ch6_Mth}
    m^{\mathrm{th}}_{x,\bm{k}} = (u_{\bm{k}} + v_{\bm{k}}) \sqrt{\frac{2\Gamma_{\bm{k}}}{\lambda_{\bm{k}} \gamma^{\prime}_{e} M_{\mathrm{s}}}} M_{\mathrm{s}}.
\end{equation}
This threshold value will be used for quantitative comparison with the experimental results.

\section{Distance calibration between the NV layer and YIG film surface}
To obtain quantitative values of the spin-wave amplitude, it is necessary to calibrate the distance between the NV layer and the YIG surface. This distance is estimated by applying a DC current to the stripline on the YIG film surface and fitting the resulting stray magnetic field distribution to an analytical model \cite{bertelli2020magnetic}. This procedure follows the approach described in Ref.~\cite{ogawa2024wideband}.
\par
\begin{figure}[H]
    \centering
    \includegraphics[width=\linewidth]{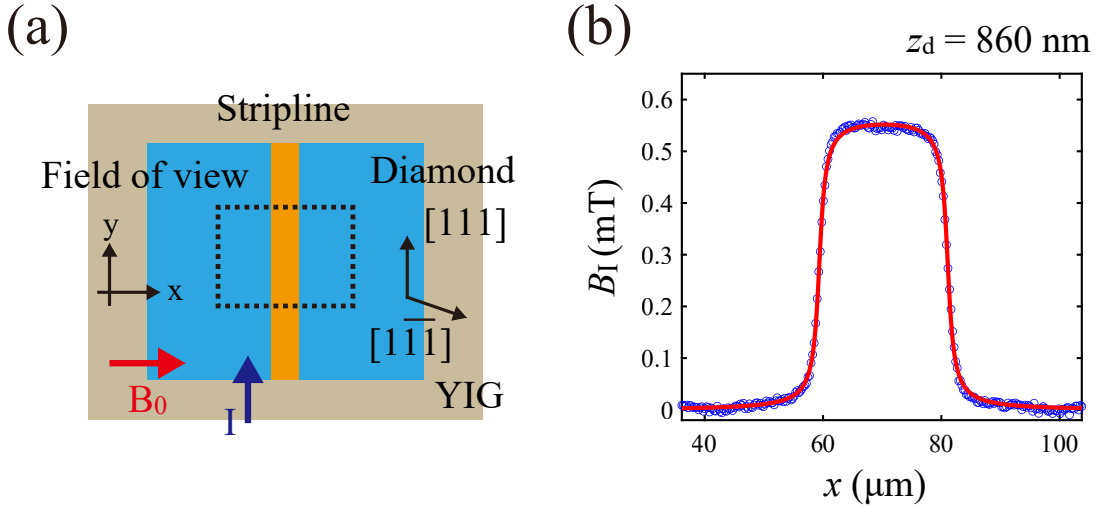}
    \caption{
     Distance calibration between the NV layer and the YIG film surface for YIG-B.
    (a) Schematic of the measurement.
    (b) Magnetic field distribution from the stripline current. (blue) Experimental results and (red) Fitting result based on the analytical model.
    }
    \label{fig:si_matesy_dist_calib}
\end{figure}

The measurement setup is shown in Fig.~\ref{fig:si_matesy_dist_calib}(a). In this configuration, the $x$-component dominates the stray magnetic field from the stripline current except near the edge of the stripline. To detect this component, we utilize NV centers oriented along the $[1\bar{1}\bar{1}]$ direction within the plane, which is $\ang{19.47}$ tilted from the $x$-direction. An external magnetic field $B_{0} = 9.62 \, \mathrm{mT}$ is applied along the $x$-direction, and a DC current $I = \pm \, 20 \, \mathrm{mA}$ is supplied to the stripline using a Yokogawa GS200. The optically detected magnetic resonance (ODMR) spectrum of the NV centers is then measured to obtain the magnetic field distribution.  We integrate the signals in the $y$-direction within the field of view, and the magnetic field distribution is calculated by estimating the two resonance frequencies of the NV spins at each position and converting them into the magnetic field \cite{Tsukamoto2021}. The magnetic field distribution due to the DC current $B_{\mathrm{I}}(x)$ is obtained by taking the difference between the distributions measured with positive and negative currents. We fit the resulting distribution using Amp\`{e}re's law:
\begin{equation}
    B_{\mathrm{I}}(x) = \int_{0}^{t} dz^{\prime} \frac{\mu_{0} I}{2\pi w t} \arctan(\frac{w(z_{d}-z^{\prime})}{(x-x_{0})^{2} + (z_{d}-z^{\prime})^{2} - (\frac{w}{2})^2}), 
\end{equation}
where $z_{d}$ represents the distance between the NV layer and the YIG film surface, $w$ is the width of the stripline, $x_{0}$ is the $x$-coordinate of the center of the stripline, and $I$ corresponds to the applied current.
The integration over $z^{\prime}$ is performed across the thickness $t$ of the stripline. For YIG-B, the stripline thickness is $116 \, \mathrm{nm}$. 
\par
In this configuration, the $z$-component of the magnetic field generated by the stripline current can be ignored because it is perpendicular to the NV axis, and its maximum expected magnitude (approximately $0.5 \, \mathrm{mT}$) is much smaller than the external bias magnetic field $B_{0} = 9.62 \, \mathrm{mT}$ in the $x$-direction. For the fitting procedure, only the parameters $w$, $x_{0}$, and $z_{d}$ are treated as free parameters; all other parameters are held fixed.
\par
For YIG-A, following this procedure, the distance between the NV layer and the YIG film surface is estimated to be $z_{d} = (878 \pm 20) \, \mathrm{nm}$ \cite{ogawa2024wideband}.
For YIG-B, Fig.~\ref{fig:si_matesy_dist_calib}(b) shows the magnetic field distribution (blue points) from the stripline along with the fitting result (red line). The fit accurately reproduces the experimental data, and the distance between the NV layer and the YIG film surface is estimated to be $z_{d} = (860 \pm 18) \, \mathrm{nm}$.
\par

\section{Details of the fitting of microwave amplitude distribution}
We describe the details of fitting the microwave amplitude distribution for both YIG-A and YIG-B. In the fitting procedure, we have to be careful that Eq.~(2) in the main text is symmetric with respect to $B_{\mathrm{sw}}(x)$ and $B_{\mathrm{strip}}(x)$; therefore, if one starts from random initial values, the two contributions cannot be distinguished. To resolve this ambiguity, we use the microwave amplitude distribution measured under the same input microwave power but with the direction of the external magnetic field reversed as a reference distribution for the stripline contribution. When the magnetic field is reversed, the magnetization of the YIG film reverses. In this situation, the spin waves propagating in the $+x$-direction become equivalent to those propagating in the $-x$-direction under the original magnetic field direction. Under the original magnetic field, the amplitude of the excited spin waves propagating in the $-x$-direction is reduced due to the polarization of the Fourier component of the microwave field from the stripline at the relevant wavenumber \cite{bertelli2020magnetic, Simon2021}. Moreover, the dipolar microwave field from spin waves propagating in the $-x$-direction under the original field does not induce transitions between the NV spins' $m_{s} = 0$ and $m_{s} = -1$ states because of its polarization \cite{bertelli2020magnetic, Rustagi2020sensing}. Therefore, we assume that the reference microwave amplitude distribution is predominantly due to the stripline. Accordingly, we fit the reference microwave distribution from $x = 0.65 \, \mathrm{\mu m}$ to $x = 26 \, \mathrm{\mu m}$ using a stretched exponential function and use the resulting fitting parameters as the initial parameters for $B_{\mathrm{strip}}(x)$ in Eq.~(2). After the fitting, we verify its validity by comparing the estimated microwave amplitude distribution from the stripline $B_{\mathrm{strip}}(x)$ with the reference microwave amplitude distribution. 

\par
\begin{figure}[H]
    \centering
    \includegraphics[width=\linewidth]{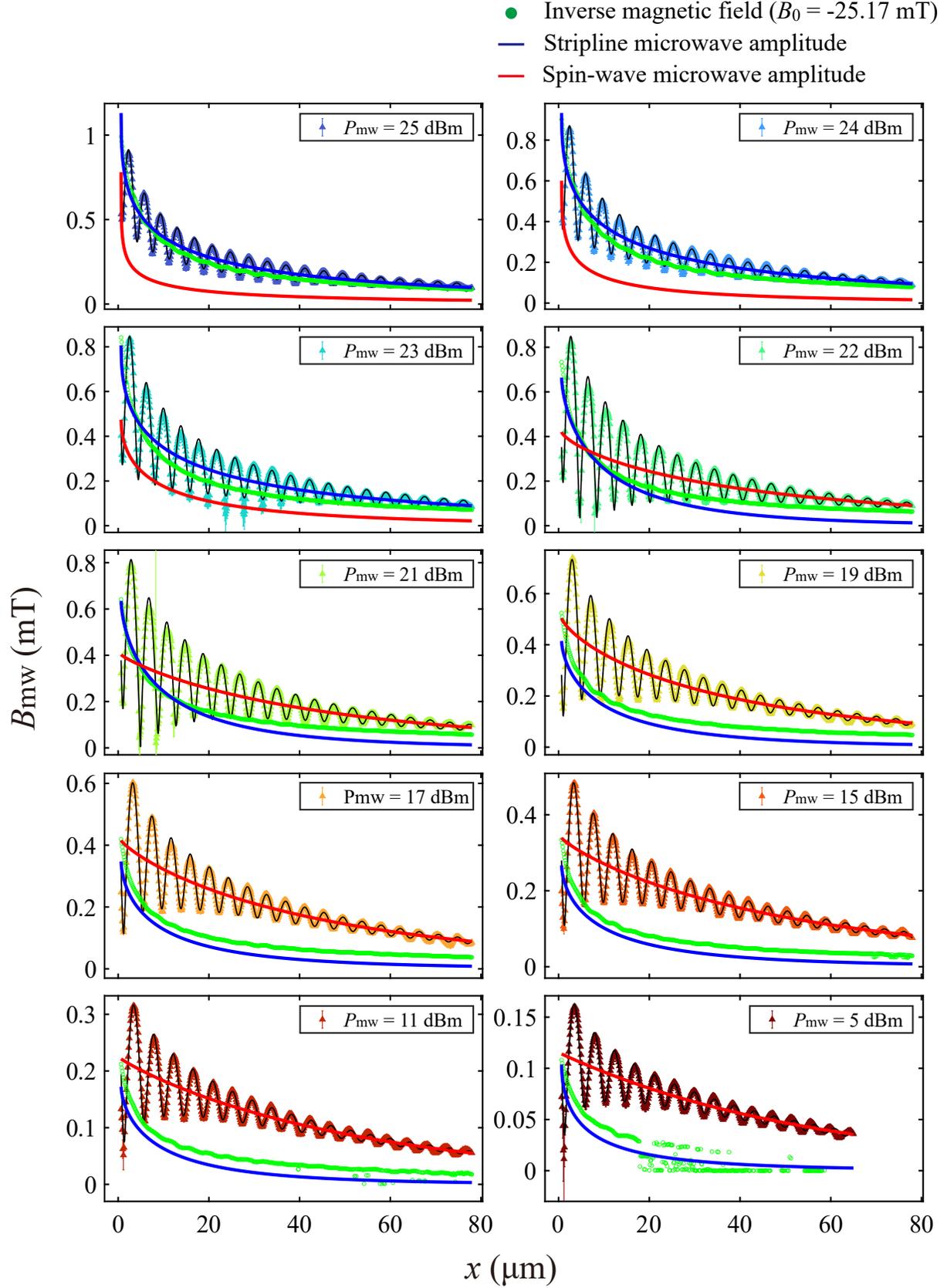}
    \caption{Fitting results of the microwave amplitude distribution for YIG-A. The triangular markers with error bars represent experimental data, the solid black lines correspond to the fitting results, the green markers represent the reference microwave amplitude distribution obtained by reversing the sign of the magnetic field, the solid blue lines correspond to the estimated microwave amplitude distribution from the stripline, and the solid red lines show the estimated microwave amplitude distribution from the spin waves.}
    \label{fig:si_matsuno_sig_fitting}
\end{figure}

Figure~\ref{fig:si_matsuno_sig_fitting} shows the fitting results for YIG-A. The triangular markers indicate the experimentally obtained microwave amplitude distribution, while the solid black lines represent the corresponding fitting results. The green markers denote the reference microwave amplitude distribution obtained by reversing the sign of the magnetic field. In addition, the solid red lines show the estimated microwave amplitude distribution arising from the spin waves, and the solid blue lines correspond to the estimated distribution from the stripline. The fitting reproduces the experimental microwave amplitude distributions well for each input power. Moreover, the estimated distribution from the stripline generally agrees with the reference distribution, thereby confirming the validity of our fitting procedure. Although some discrepancies exist between the estimated stripline microwave amplitude distribution and the reference microwave amplitude distribution, these may be attributed to factors such as imperfections in the fitting model, detuning between the NV resonance frequency and the spin-wave frequency (particularly in regions where the microwave magnetic field amplitude is very low, $B_{\mathrm{mw}} < 0.05 \, \mathrm{mT}$, which could lead to an overestimation of the reference amplitude), and the presence of spin-wave modes propagating in directions other than the $+x$-direction that might also induce transitions between the $m_{s} = 0$ and $m_{s} = -1$ states.
\par
\begin{figure}[H]
    \centering
    \includegraphics[width=\linewidth]{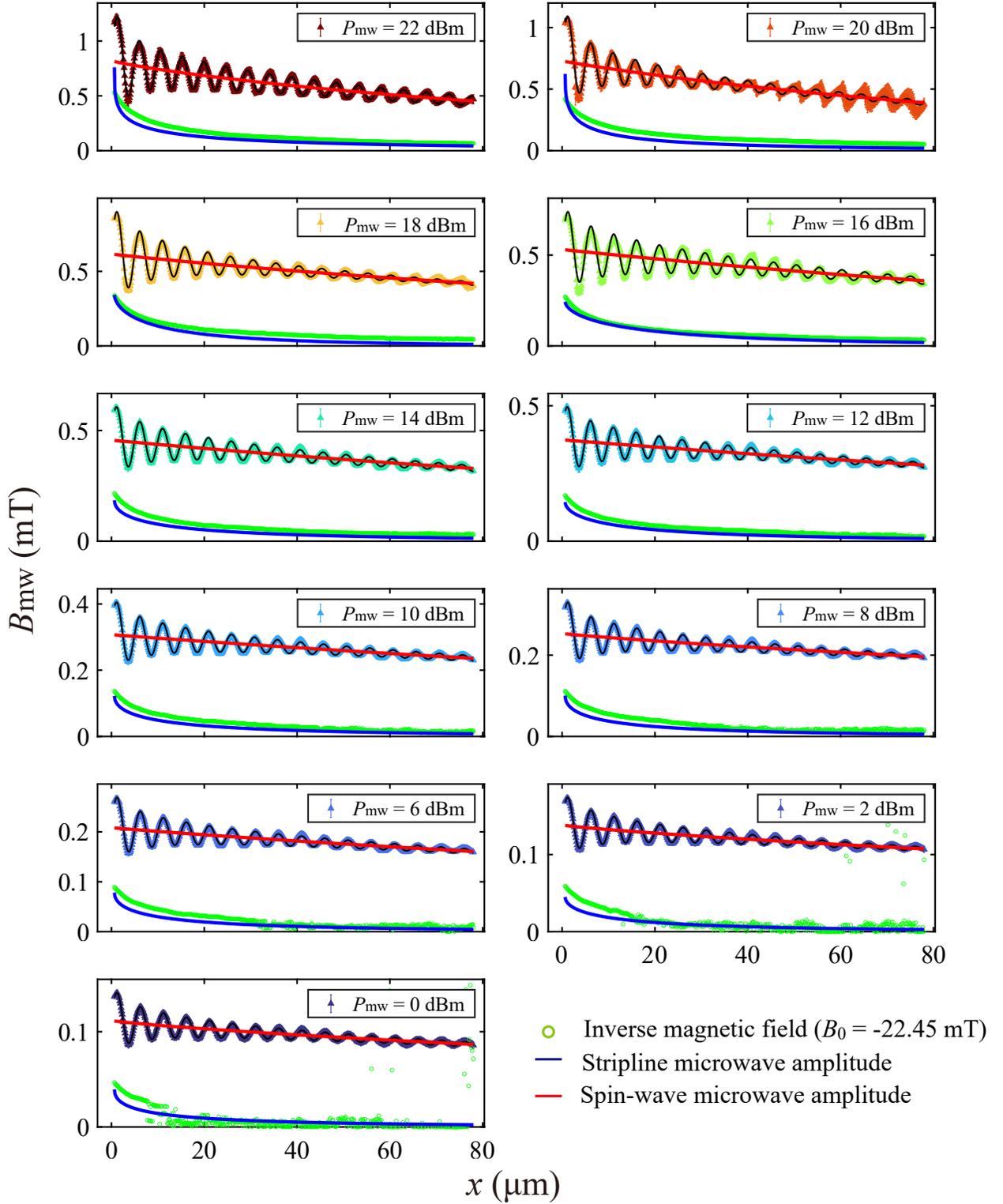}
    \caption{Fitting results of the microwave amplitude distribution for YIG-B. The triangular markers with error bars represent experimental data, the solid black lines correspond to the fitting results, the green markers represent the reference microwave amplitude distribution obtained by reversing the sign of the magnetic field, the solid blue lines correspond to the estimated microwave amplitude distribution from the stripline, and the solid red lines show the estimated microwave amplitude distribution from the spin waves.}
    \label{fig:si_matesy_sig_fitting}
\end{figure}

Figure~\ref{fig:si_matesy_sig_fitting} shows the fitting results for YIG-B, with the green markers representing the reference microwave amplitude distribution. In almost all regions and at all input microwave powers, the reference distribution is significantly smaller than the measured microwave amplitude distribution, indicating that $B_{\mathrm{sw}} \gg B_{\mathrm{strip}}$. Therefore, we use Eq.~(3) in the main text for the fitting. For each input microwave power, the fitting result (black solid line) reproduces the experimental data well. Furthermore, the reference microwave amplitude distribution generally agrees with the estimated microwave amplitude distribution from the stripline, confirming the validity of our analysis.

\clearpage
\clearpage

\clearpage

\section{Decay characteristic of the spin-wave amplitude}
In this section, we show the decay parameters of the spin-wave amplitude obtained from the fitting results for the input microwave power sweeping measurement.
\par
\begin{figure}[H]
    \centering
    \includegraphics[width=\linewidth]{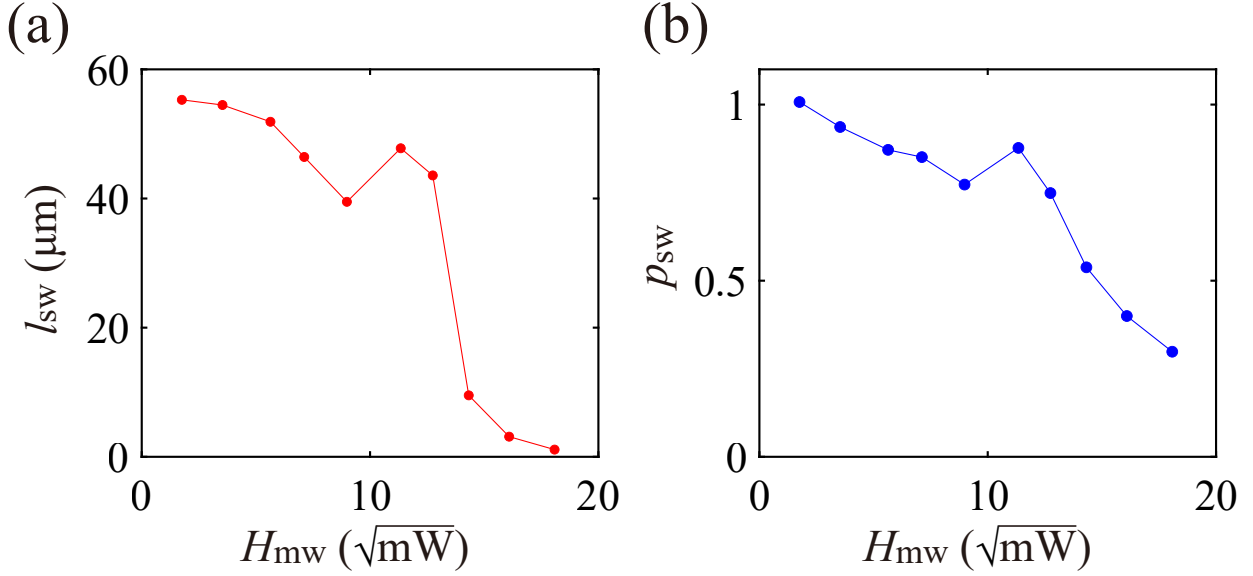}
    \caption{Decay characteristic for YIG-A. Dependence of (a) decay length $l_{\mathrm{sw}}$ and (b) stretch factor $p_{\mathrm{sw}}$ on the input microwave amplitude.}
    \label{fig:si_matsuno_decay}
\end{figure}

Figures~\ref{fig:si_matsuno_decay}(a) and (b) display the dependence of (a) the decay length $l_{\mathrm{sw}}$ and (b) the stretch factor $p_{\mathrm{sw}}$ on the input microwave amplitude $H_{\mathrm{mw}}$ for YIG-A [Fig.~2(c) in the main text]. As the input power increases, reflecting nonlinearity, the decay length decreases and the stretch factor decreases from around one to zero.

\begin{figure}[H]
    \centering
    \includegraphics[width=\linewidth]{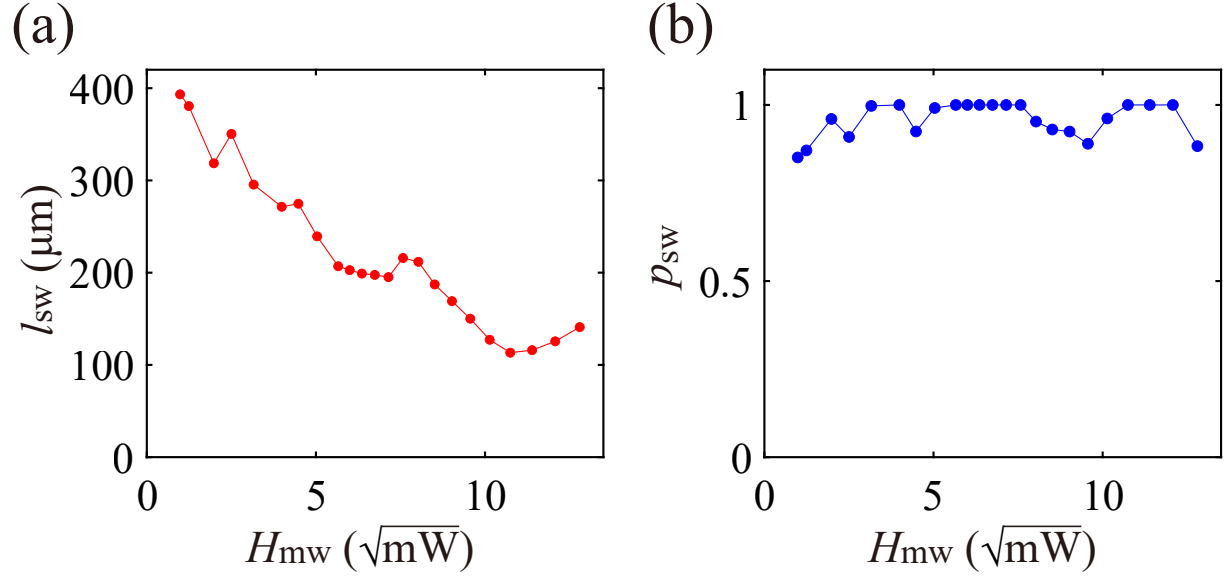}
    \caption{Decay characteristic for YIG-B. Dependence of (a) decay length $l_{\mathrm{sw}}$ and (b) stretch factor $p_{\mathrm{sw}}$ on the input microwave amplitude.}
    \label{fig:si_matesy_decay}
\end{figure}

Figures~\ref{fig:si_matesy_decay}(a) and (b) show the dependence of (a) $l_{\mathrm{sw}}$ and (b) $p_{\mathrm{sw}}$ on $H_{\mathrm{mw}}$ for YIG-B [Fig.~5(c) in the main text]. As the input power increases, the stretch factor remains approximately one for all data, whereas the decay length shortens with increasing input power.

\clearpage

\section{Considerations of multi-magnon scattering}
The nonlinear decay of the spin-wave amplitude is attributed to multi-magnon scattering, in which excited magnons with (frequency, wavevector) = ($f_{\mathrm{sw}}$, $\bm{k}$) are scattered into magnons with different frequencies and wavevectors while conserving total energy and momentum \cite{Schultheiss2012, Tobias2020, Mohseni2021, vov1987nonlinear, Boardman1988, Mark2004, Lake2022}. Experimentally, Brillouin light scattering (BLS) measurements have revealed that, above a certain threshold in magnetic waveguides, additional signals appear at frequencies different from that of the excited spin waves \cite{Schultheiss2012, Tobias2020, Mohseni2021}. 
\par
\begin{figure}[H]
    \centering
    \includegraphics[width=\linewidth]{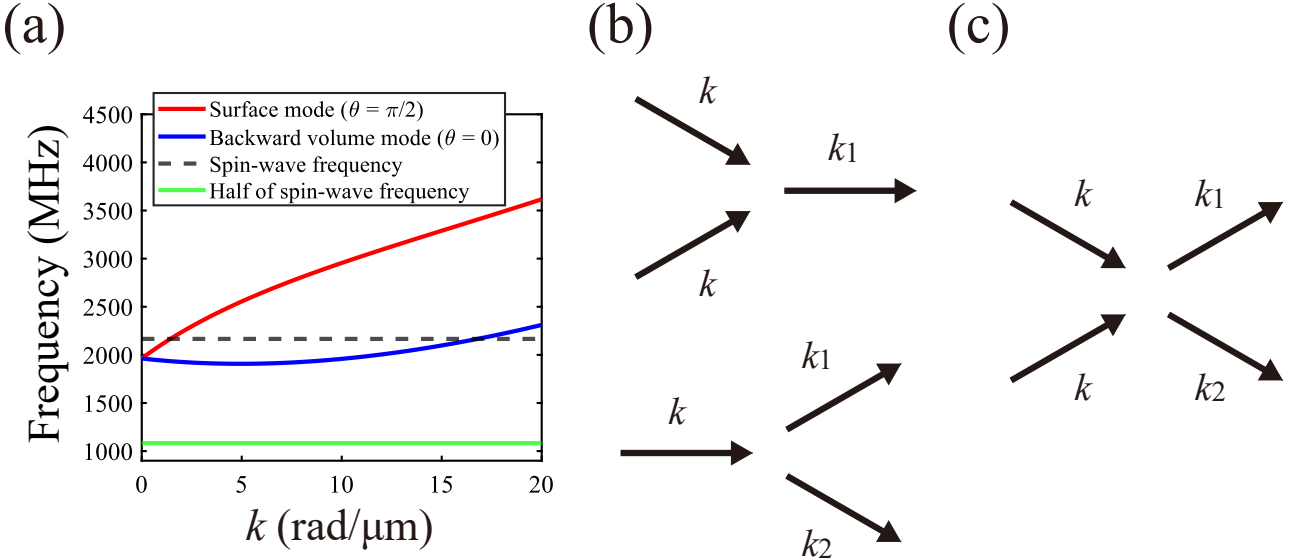}
    \caption{(a) Spin-wave dispersion relation of the surface mode and backward volume mode under the present experimental condition. (b) Schematics of the three-magnon scattering. (c) Schematic of the four-magnon scattering.}
    \label{fig:si_matsuno_three}
\end{figure}

From the magnon Hamiltonian given in Eq.~(\ref{eq:H_all_bogo}), the lowest-order nonlinear process is three-magnon scattering. This process can occur in two ways: (i) two excited magnons are annihilated to create a new magnon with ($f_{1}, \bm{k}_{1}$) satisfying $2\bm{k} = \bm{k}_{1}$ and $2f_{\mathrm{sw}} = f_{1}$, or (ii) an excited magnon is annihilated to create two new magnons with ($f_{1}, \bm{k}_{1}$) and ($f_{2}, \bm{k}_{2}$), satisfying $\bm{k} = \bm{k}_{1} + \bm{k}_{2}$ and $f_{\mathrm{sw}} = f_{1} + f_{2}$ [Fig.~\ref{fig:si_matsuno_three}(b)]. Regarding the dispersion relation of spin waves, for a fixed $|\bm{k}|$, the lowest frequency mode is the backward volume mode at $\theta_{k} = 0$, while the highest frequency mode is the surface mode at $\theta_{k} = \frac{\pi}{2}$ [see Eq.~(\ref{eq:ch6_dispersion})]. Figure~\ref{fig:si_matsuno_three}(a)  plots the dispersion relations for these modes. In calculating the dispersion, the effects of exchange interactions become significant at larger wavenumbers; therefore, we use the literature value of the exchange stiffness coefficient $\alpha_{\mathrm{ex}} = 3.0 \times 10^{-16} \, \mathrm{m^{-2}}$ \cite{Serga2010}.
\par
From the dispersion relation, however, it is evident that neither process (i) nor (ii) can simultaneously satisfy the simultaneously conservation of wavenumber and frequency. In particular, the minimum frequency at the bottom of the dispersion is considerably larger than the half of the excited spin-wave frequency $f_{\mathrm{sw}}/2 = 1082.9 \, \mathrm{MHz}$, and the wavenumber of the spin waves corresponding to twice the excited frequency, $2f_{\mathrm{sw}} = 4331.6 \, \mathrm{MHz}$, exceeds $20 \, \mathrm{rad/\mu m}$.
\par
Therefore, under the present experimental conditions, the dominant nonlinear effect is four-magnon scattering. Specifically, this process involves the annihilation of two excited magnons and the creation of two new magnons with ($f_{1}, \bm{k}_{1}$) and ($f_{2}, \bm{k}_{2}$), satisfying the conservation relations $2\bm{k} = \bm{k}_{1} + \bm{k}_{2}$ and $2f = f_{1} + f_{2}$ [Fig.~\ref{fig:si_matsuno_three}(c)].

\section{Considerations of the Joule heating effect}
In the main text, we observe a significant modulation of the spin-wave wavenumber within the field of view for YIG-A at large input microwave power. Here, we demonstrate that the Joule heating effect does not contribute to this spatial variation in the wavenumber.
\par
The wavenumber modulation is attributed to a reduction in the static magnetization along the direction of the saturation magnetization. Two possible causes for this reduction are: (i) a reduction in the static magnetization due to nonlinear spin-wave excitation, and (ii) a temperature gradient induced by strong microwave application, leading to an inhomogeneous reduction in static magnetization.
\par
To determine the dominant factor, we examine the wavenumber distribution at an external magnetic field of $B_{0} = 22.45 \, \mathrm{mT}$, lower than that used in the main text ($B_{0} = 25.17 \, \mathrm{mT}$), under strong input microwave power. At this magnetic field, the wavenumber of the surface spin waves is approximately $3.3 \, \mathrm{\mathrm{rad/ \mu m}}$, which is larger than the value of $1.4 \, \mathrm{\mathrm{rad/ \mu m}}$ at $B_{0} = 25.17 \, \mathrm{mT}$. Due to the increased wavenumber, the excitation efficiency of the spin waves, which is determined by the geometric size of the stripline, is reduced, leading to a smaller spin-wave amplitude.  Because the Joule heating effect is expected to be similar at both magnetic fields for the same input microwave power, this experiment allows us to discern the contributions from these factors.
\par
\begin{figure}[H]
    \centering
    \includegraphics[width=0.8\linewidth]{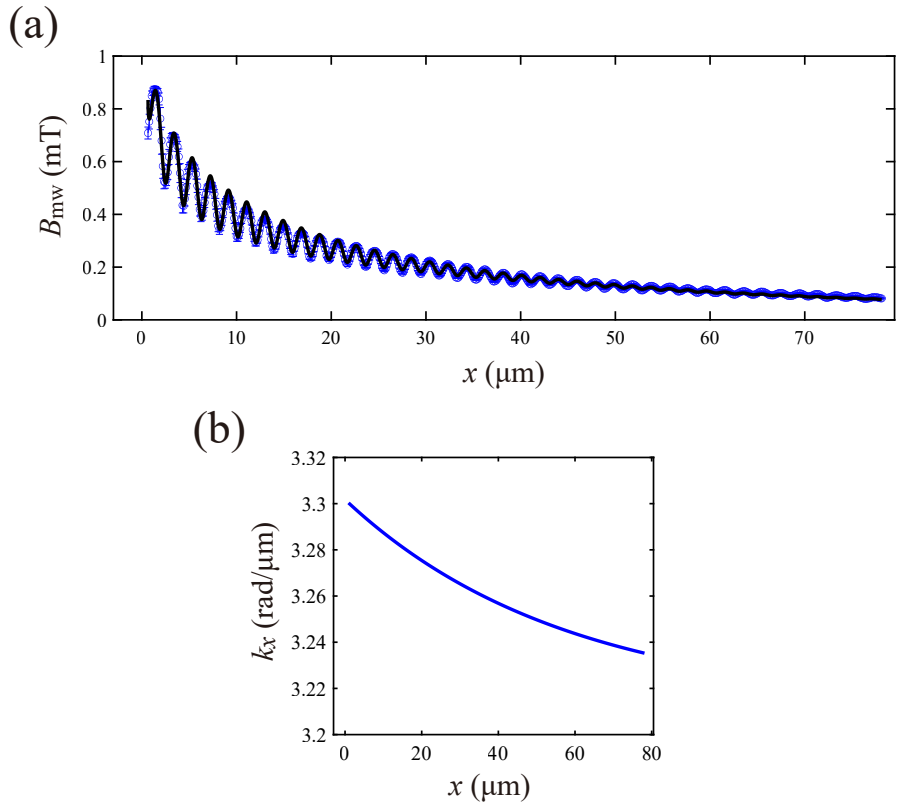}
    \caption{(a) Microwave amplitude distribution obtained at the external magnetic field of $B_{0} = 22.45 \, \mathrm{mT}$, the spin-wave frequency of $f_{\mathrm{sw}} = 2242.7 \, \mathrm{MHz}$, and the input microwave power of $P_{\mathrm{mw}} = 23 \, \mathrm{dBm}$ for YIG-A. Blue markers represent the experimental result, and the black line is the fitting line. (b) Fitting result of the wavenumber distribution.}
    \label{fig:si_matsuno_temp}
\end{figure}

Figure~\ref{fig:si_matsuno_temp}(a) shows the microwave amplitude distribution for YIG-A at $B_{0} = 22.45 \, \mathrm{mT}$ and $P_{\mathrm{mw}} = 23 \, \mathrm{dBm}$. Blue markers represent the experimental result, and the black line
is the fit. Figure~\ref{fig:si_matsuno_temp}(b) displays the corresponding wavenumber distribution extracted from the fit. Although there is some spatial variation in the wavenumber, the change is only about 3\% from the left to the right end of the field of view. In contrast, at $B_{0} = 25.17 \, \mathrm{mT}$, the wavenumber varies by approximately 20\% across the field of view [see Fig.~2(d) in the main text].
\par
These results indicate that the Joule-heating effect does not produce significant spatial variations in the static magnetization (or, if heating occurs, its impact is uniform across the field of view). Therefore, the wavenumber variation observed at $B_{0} = 25.17 \, \mathrm{mT}$ in the main text is attributed to the intrinsic nonlinear dynamics of the spin waves rather than to extrinsic effects such as Joule heating.

\section{Details of YIG-B}
\subsection{Magnetic field sweep measurement}
\begin{figure}[H]
    \centering
    \includegraphics[width=\linewidth]{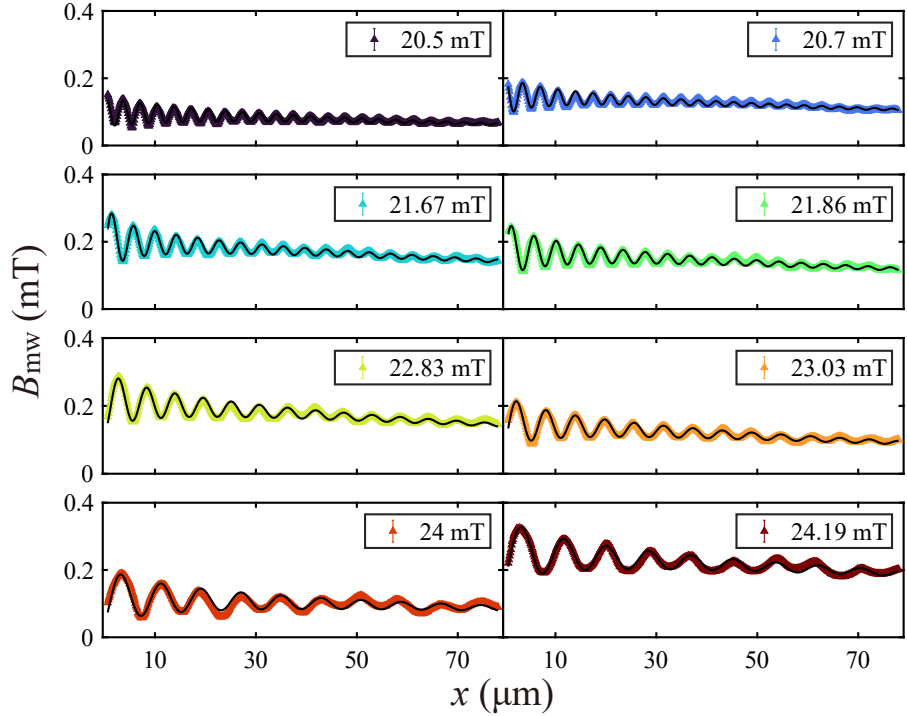}
    \caption{Microwave amplitude distribution $B_{\mathrm{mw}}(x)$ obtained by sweeping the external magnetic field with the input microwave power of $P_{\mathrm{mw}} = 8 \, \mathrm{dBm}$.}
    \label{fig:si_matesy_Bsweep_Bmw}
\end{figure}

This section presents the results of measurements in which the external magnetic field is swept for YIG-B to calibrate the effective magnetization. The external magnetic field is varied at 23 values between $B_{0} = 20.50 \, \mathrm{mT}$ to $B_{0} = 24.58 \, \mathrm{mT}$ while the input microwave power is fixed at $P_{\mathrm{mw}} = 8 \, \mathrm{dBm}$. At each magnetic field, we obtain the microwave amplitude distribution by measuring Rabi oscillations. The triangular markers in Fig.~\ref{fig:si_matesy_Bsweep_Bmw}(a) represent the measured microwave amplitude distributions at eight representative external magnetic fields. As the magnetic field increases, the wavelength of the spin waves becomes longer. The black solid lines are the corresponding fitting curves. For the fitting, Eq.~(3) in the main text is used under the assumption that $B_{\mathrm{sw}} \gg B_{\mathrm{strip}}$. Since the input microwave power is within the linear regime, we assume that the wavenumber is spatially uniform and that the spin-wave amplitude decays exponentially with $p_{\mathrm{sw}} = 1$ [see Eq.~(4) in the main text]. The fitting results accurately reproduce the experimental data for all magnetic fields.
\par
\begin{figure}[H]
    \centering
    \includegraphics[width=0.8\linewidth]{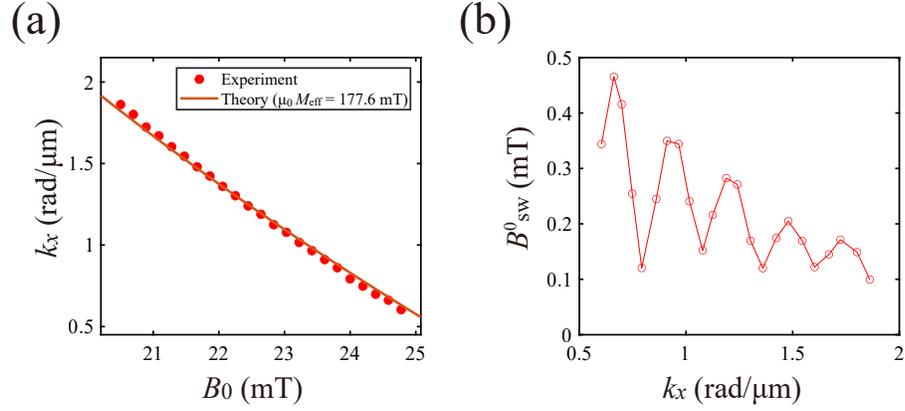}
    \caption{(a) Relationship between the external magnetic fields and the experimentally obtained spin-wave wavenumbers. The solid line represents the theoretical line at the effective magnetization $\mu_{0} M_{\mathrm{eff}} = 177.6 \, \mathrm{mT}$ determined by the fitting. (b) Wavenumber dependence of the microwave amplitude from the spin waves at the left edge of the field of view $B^{0}_{\mathrm{sw}}$.}
    \label{fig:si_matesy_Bsweep_para}
\end{figure}

Figure~\ref{fig:si_matesy_Bsweep_para}(a) shows the relationship between the external magnetic field $B_{0}$ and the wavenumber $k_{x}$ of the spin waves obtained from the fitting results.
The solid line represents the fitting line using the dispersion relation described in the main text and Ref.~\cite{ogawa2024wideband}.
We set the external magnetic field $B_{0}$, the obtained wavenumber of the spin waves $k_{x}$, and the spin-wave frequency $f_{\mathrm{sw}}$ as input variables, and the only free parameter is the effective magnetization $M_{\mathrm{eff}}$. The fitting curve agrees well with the experimental data, and the obtained effective magnetization is $\mu_{0} M_{\mathrm{eff}} = (177.6 \pm 0.3) \, \mathrm{mT}$.
\par
Figure~\ref{fig:si_matesy_Bsweep_para}(b) illustrates the dependence of the microwave amplitude from the spin waves at the left end of the field of view $B^{0}_{\mathrm{sw}}$ on the wavenumber. As the wavenumber increases, the microwave amplitude decreases while exhibiting periodic oscillations. This behavior originates from the periodic oscillation of the Fourier transform of the microwave field produced by the stripline, which excites the spin waves.

\subsection{Detailed condition of the input microwave power sweep measurement}
\begin{figure}[H]
    \centering
    \includegraphics[width=\linewidth]{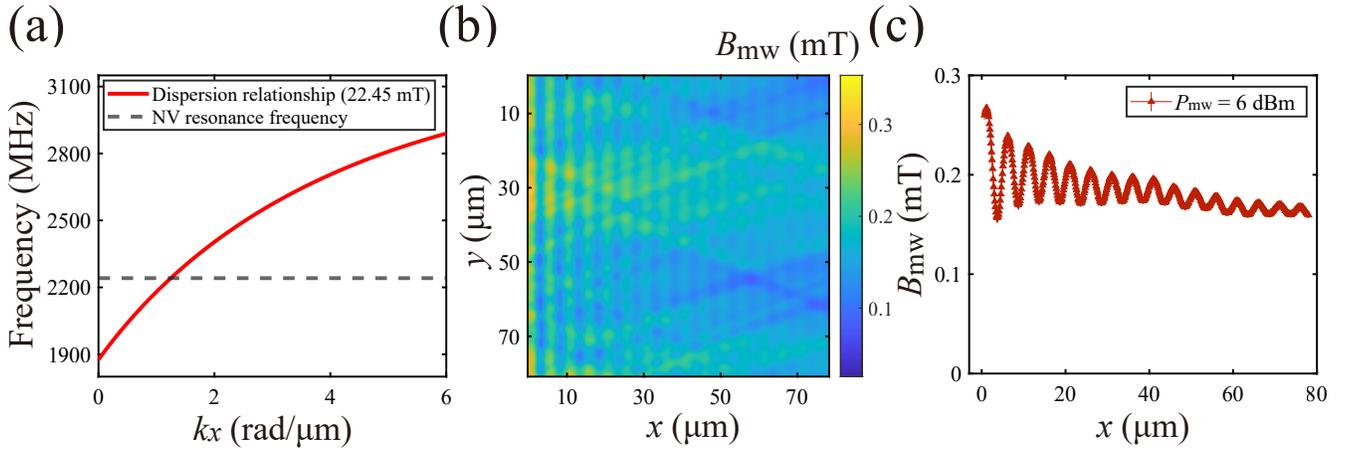}
    \caption{
    (a) The dispersion relation of the surface spin waves at external magnetic field $B_{0} = 22.45 \, \mathrm{mT}$ for YIG-B. The horizontal dashed line corresponds to the resonance frequency of the NV spin.
    (b) Two-dimensional image of microwave amplitude at input microwave power $P_{\mathrm{mw}} = 6 \, \mathrm{dBm}$ for YIG-B.
    (c) One-dimensional microwave amplitude distribution $B_{\mathrm{mw}}(x)$ for the data in (b).
    }
    \label{fig:matesy_pre}
\end{figure}

In the input microwave power sweep measurement for YIG-B, the external magnetic field is set at $B_{0} = 22.45 \, \mathrm{mT}$. Figure~\ref{fig:matesy_pre}(a) shows the dispersion relation of the surface spin waves under this magnetic field along with the resonance frequency of the NV spins. The resonance frequency of the NV spins $f_{\mathrm{NV}}$ is $2243.1 \, \mathrm{MHz}$, and the wavenumber of the surface spin waves at this frequency is approximately $1.3 \, \mathrm{rad/\mu m}$. Figure~\ref{fig:matesy_pre}(b) presents the two-dimensional microwave amplitude distribution at an input microwave power of $6 \, \mathrm{dBm}$. In the analysis, signals from each $520 \, \mathrm{nm}$ in the $x$- and $y$-directions are integrated and treated as a single pixel. Similar to YIG-A, oscillations in the microwave amplitude along the $x$-direction are observed. In addition, characteristic patterns, ``caustics,'' are observed along diagonal directions for this sample as well. By integrating signals along the $y$-direction, we focus on the spin waves propagating in the $x$-direction. Figure~\ref{fig:matesy_pre}(c) shows the one-dimensional microwave amplitude distribution $B_{\mathrm{mw}}(x)$ in one dimension along the $x$-axis, obtained by analyzing the integrated signals over a width of $13 \, \mathrm{\mu m}$ from $y = 31.87 \, \mathrm{\mu m}$ to $y = 44.87 \, \mathrm{\mu m}$ in the $y$-direction. The minimum $x$-coordinate in the experimental data is $x = 0.65 \, \mathrm{\mu m}$. Compared to YIG-A, decay of the microwave amplitude from the left to the right end of the field of view is small, reflecting a lower relaxation rate of the spin waves. Compared to YIG-A, the decay of the microwave amplitude from the left to the right end is small, reflecting a lower relaxation rate of the spin waves. Specifically, the envelope of the microwave amplitude distribution attenuates by approximately 20\% across the field of view, which corresponds to a decay length of approximately
$300 \, \mathrm{\mu m}$.

\subsection{Analysis of the wavenumber distribution}
\begin{figure}[H]
    \centering
    \includegraphics[width=\linewidth]{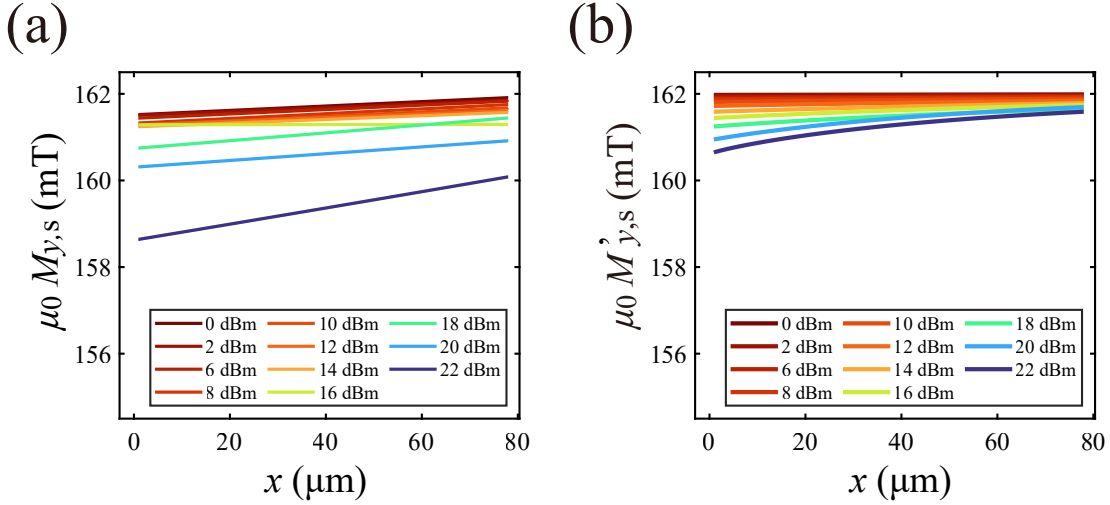}
    \caption{(a) Static magnetization distribution obtained by converting the wavenumber distribution for YIG-B. (b) Estimation of the static
    magnetization distribution considering only contributions from the
    excited coherent spin waves for YIG-B.}
    \label{fig:matesy_k}
\end{figure}

We present a detailed analysis of the spatial distribution of the wavenumber for YIG-B. Figure~\ref{fig:matesy_k}(a) shows the distribution of the static magnetization in the $y$-direction $M_{y,\mathrm{s}}(x)$, estimated from the wavenumber distribution, for each input microwave power. Although the static magnetization decreases with increasing microwave power, even at maximum power the variation within the field of view remains limited to about $1 \, \mathrm{mT}$, which is smaller than that in YIG-A [see Fig.~4(c) in the main text]. This difference is likely due to the lower amplitude of the excited spin waves in YIG-B. Figure~\ref{fig:matesy_k}(b) shows the estimated reduction in the static magnetization considering only the contribution from the surface spin waves. As in YIG-A, the contribution from surface spin waves is minor for all input microwave powers [see Fig.~4(d) in the main text].

\section{Modulation of the Rabi oscillation signal in the nonlinear region}
In the main text, we have discussed the nonlinear dynamics of spin waves based on the microwave amplitudes obtained from fitting the Rabi oscillation signals. In this section, we report that the Rabi oscillation signals themselves also exhibit modulations in the nonlinear regime.
\par
Figures~\ref{fig:si_matesy_rabi_mod}(a), (b), and (c) display the raw Rabi oscillation signals for YIG-B at five positions ranging from $x = 13 \, \mathrm{\mu m}$ to $x = 65 \, \mathrm{\mu m}$ for three different input microwave powers: $P_{\mathrm{mw}} = 0 \, \mathrm{dBm}$ (minimum power), $P_{\mathrm{mw}} = 16 \, \mathrm{dBm}$ (intermediate power), and $P_{\mathrm{mw}} = 22 \, \mathrm{dBm}$ (maximum power). In these figures, the markers represent the experimental results, and the solid lines correspond to the fitting results obtained using a function that is the product of a cosine function and an exponential decay with an added constant term.
\par
At the minimum input microwave power ($P_{\mathrm{mw}} = 0 \, \mathrm{dBm}$), the fitted curve closely reproduces the raw signal up to the maximum sweep time of $\tau = 0.6 \, \mathrm{\mu s}$. However, at the intermediate power ($P_{\mathrm{mw}} = 16 \, \mathrm{dBm}$), the raw signal shows a noticeable slowdown in the oscillation period around $\tau = 0.15 \, \mathrm{\mu s}$ [indicated by the black arrow in Fig.~\ref{fig:si_matesy_rabi_mod}(b)]. Moreover, at the maximum power ($P_{\mathrm{mw}} = 22 \, \mathrm{dBm}$), a noticeable dip like a bump is observed around $\tau = 0.01 \, \mathrm{\mu s}$ [shown by the black arrow in Fig.~\ref{fig:si_matesy_rabi_mod}(c)]. In this nonlinear regime, one possible explanation for the modulation of the Rabi oscillation signals is the generation of magnons at frequencies near the NV spin resonance frequency as a result of four-magnon scattering \cite{Tobias2020, Merbouche2022}. If there is a broadening of the spin-wave frequency spectrum around the NV resonance, the coherent microwave amplitude will decay over time as the phases of the contributions from magnons with different frequencies increasingly dephase. The observed slowing of the oscillations at $P_{\mathrm{mw}} = 16 \, \mathrm{dBm}$ may reflect such a broadening, while at $P_{\mathrm{mw}} = 22 \, \mathrm{dBm}$, the nonlinearity may be even more pronounced, potentially leading to a loss of coherence and more complex modulations in the Rabi oscillation signals.
\par
Since the nonlinear dynamics are directly reflected in the Rabi oscillation signals, more detailed analyses of the oscillation shapes and decay times in future studies could provide further insights into these nonlinear effects.

\begin{figure}[H]
    \centering
    \includegraphics[width=\linewidth]{fig_matesy_rabi_mod_rev1.pdf}
    \caption{Raw signals of the Rabi oscillation at the input microwave power (a) $P_{\mathrm{mw}} = 0 \, \mathrm{dBm}$ (b) $P_{\mathrm{mw}} = 16 \, \mathrm{dBm}$ (c) $P_{\mathrm{mw}} = 22 \, \mathrm{dBm}$.}
    \label{fig:si_matesy_rabi_mod}
\end{figure}

\section{Estimation of energy conversion efficiency of the spin waves}
In this section, we provide a detailed calculation of the estimation of the energy conversion efficiency of the spin waves for each sample. \par
The spin-wave energy per unit volume $E_{\mathrm{sw}}$ can be obtained by considering the Zeeman energy. Let the transverse components of the spin-wave amplitude be denoted as $(m_{x},m_{y})$ and the saturation magnetization as $M_{\mathrm{s}}$. The deviation of the magnetization $\Delta M_{z}$ along the external magnetic field $B_{0}$ can be calculated as the reduction of the longitudinal component of the magnetization due to spin-wave excitation, i.e,
\begin{equation}
\Delta M_{z} = M_{s} - \sqrt{M^{2}_{s} - m^{2}_{x} - m^{2}_{y}},
\end{equation}
and for the case $m_{x}, \, m_{y} \ll M_{\mathrm{s}}$, it can be approximated as
\begin{equation}
\Delta M_{z} \simeq \frac{m^{2}_{x} + m^{2}_{y}}{2 M_{\mathrm{s}}}.
\end{equation}
By using this expression, the spin-wave energy per unit volume is given by
\begin{equation}
E_{\mathrm{sw}} = B_{0} \Delta M_{z} \simeq \frac{B_{0}}{2 M_{\mathrm{s}}} \qty(m^{2}_{x} + m^{2}_{y}).
\end{equation}
Assuming the spin-wave time evolution $m_{x} = m_{0} \cos \omega t, \, m_{y} = m_{0} \eta \sin \omega t$, where $m_{0}$ is the spin-wave amplitude, $\omega$ is the angular frequency, $\eta$ is the ellipticity of the spin waves, we obtain
\begin{equation}
    E_{\mathrm{sw}} \simeq \frac{B_{0}}{4 M_{\mathrm{s}}} \qty(1 + \eta^{2}) m^{2}_{0}.
\end{equation}
The spin-wave power converted per unit time, $P_\mathrm{sw}$ can then be expressed as
\begin{equation}
    P_{\mathrm{sw}} = E_{\mathrm{sw}} v_{\mathrm{g}} l_{\mathrm{st}} d,
\end{equation}
where $v_{\mathrm{g}}$ is the group velocity, $l_{\mathrm{st}}$ is the stripline length, and $d$ is the film thickness. 
\par
Using this expression, we compute the power converted into spin waves for YIG-A at $P_\mathrm{mw} = 5 \, \mathrm{dBm}$ and for YIG-B at $P_\mathrm{mw} = 6 \, \mathrm{dBm}$, together with the parameters listed in the Table~\ref{tb:exp_params}. As a result, we obtain $P_\mathrm{sw} = 179 \, \mathrm{pW}$ for YIG-A and $P_\mathrm{sw} = 474 \, \mathrm{pW}$ for YIG-B. The conversion efficiencies with respect to the input power are thus estimated to be $5.8 \times 10^{-8}$ for YIG-A and $1.2 \times 10^{-7}$ for YIG-B. We attribute the difference between the two samples to variations in wiring (delivery) losses and from differences in relaxation rates, which lead to different excitation amplitudes (see Eq.~(\ref{eq:n_steady}) in the Supplemental Material).

\begin{table}[t]
    \begin{center}
    \begin{tabular}{|c|c|c|} \hline
       & YIG-A & YIG-B  \\ \hline
       Input microwave power $P_{\mathrm{mw}}$ & $3.1 \, \mathrm{mW}$ & $3.9 \, \mathrm{mW}$ \\ \hline
       External magnetic field $B_{0}$ & $25.17 \, \mathrm{mT}$ &  $22.45 \, \mathrm{mT}$ \\ \hline
       Saturation magnetization $M_{\mathrm{s}}$ &  $141 \, \mathrm{mT}$ & $162 \, \mathrm{mT}$ \\ \hline
       Ellipticity $\eta$ & 0.40 & 0.42 \\ \hline
       Spin-wave amplitude $\mu_{0} m_{0}$ & $7.2 \mathrm{mT}$ & $6.6 \mathrm{mT}$ \\ \hline
       Group velocity $v_{\mathrm{g}}$ & $0.78 \, \mathrm{km/s}$ & $1.5 \, \mathrm{km/s}$ \\ \hline
       Film thickness $d$ & $54 \, \mathrm{nm}$ & $109 \, \mathrm{nm}$ \\ \hline
       Stripline length $l_{\mathrm{st}}$ & $2 \, \mathrm{mm}$ & $2 \, \mathrm{mm}$ \\ \hline
    \end{tabular}
    \end{center}
    \caption{Experimental parameters of YIG-A and YIG-B for calculation of the energy conversion.}
    \label{tb:exp_params}
\end{table}

\section{Numerical simulation of the nonlinear spin-wave dynamics}
This section presents numerical simulation results aimed at validating the nonlinear dynamics observed experimentally when the amplitude of the excitation microwaves becomes large. In particular, the simulation demonstrates both a nonlinear decay of the spin-wave amplitude above a certain microwave amplitude and the generation of spin waves at frequencies different from the excitation frequency.
\par

\subsection{Details of the numerical simulation}
\begin{figure}[H]
    \centering
    \includegraphics[width=0.7\linewidth]{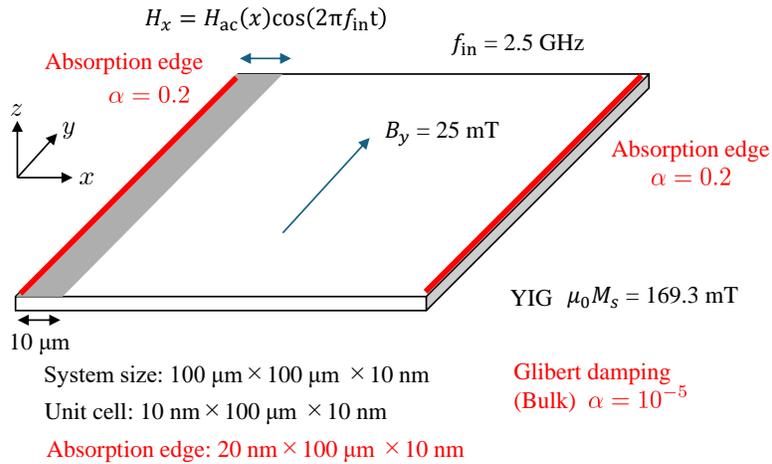}
    \caption{Schematic of the numerical simulation.}
    \label{fig:ch6_numerical_setup}
\end{figure}

Figure~\ref{fig:ch6_numerical_setup} shows the schematic of the numerical simulation. The system is a two-dimensional thin film measuring $100 \, \mathrm{\mu m} \times 100 \, \mathrm{\mu m} \times 10 \, \mathrm{nm}$, which is discretized into unit cells measuring $10 \, \mathrm{nm} \times 100 \, \mathrm{\mu m} \times 10 \, \mathrm{nm}$.
The simulation considers one-dimensional propagation of magnetostatic spin waves along the $x$-direction whose wavelength is typically on the micrometer scale. In addition, a stripline with a width of $10 \, \mathrm{\mu m}$ is positioned along the left edge of the film, and the spin waves are excited by applying a microwave magnetic field in the region of the stripline.
\par

The temporal evolution of the magnetization is simulated by numerically solving the Landau-Lifshitz-Gilbert (LLG) equation \cite{rezende2020fundamentals}:
\begin{equation}
\label{eq:ch6_LLG}
    \frac{d \bm{M}}{dt} = - \gamma^{\prime}_{e} \mu_{0} \bm{M} \times \bm{H}_{\mathrm{eff}} + \frac{\alpha}{M_{\mathrm{s}}} \bm{M} \times \frac{d \bm{M}}{dt}.
\end{equation}
Here, $\bm{H}_{\mathrm{eff}}$ represents the effective magnetic field, which includes both the external magnetic field $\bm{H}_{\mathrm{ext}}$ and the dipolar magnetic field $\bm{H}_{\mathrm{dip}}$.
\par
For the external magnetic field, we consider a static field in the $y$-direction combined with a microwave magnetic field in the $x$-direction at frequency $f_{\mathrm{in}}$:
\begin{equation}
    \bm{H}_{\mathrm{ext}} = \qty(H_{\mathrm{ac}} \cos \qty(2\pi f_{\mathrm{in}}t), \, H_{y}, \, 0),
\end{equation}
where the static magnetic field is set to $\mu_{0} H_{y} = 25 \, \mathrm{mT}$, and the microwave amplitude $H_{\mathrm{ac}}$ is a parameter that is swept in the simulation. The saturation magnetization is set to $\mu_{0}  M_{\mathrm{s}} = 169.3 \, \mathrm{mT}$.
The Gilbert damping coefficient $\alpha$ is set to $\alpha = 10^{-5}$ in the bulk, and the regions at both ends of the $x$-direction (measuring $20 \, \mathrm{nm} \times 100 \, \mathrm{\mu m} \times 10 \, \mathrm{nm}$) are designated as absorbing boundaries with $\alpha = 0.2$ to prevent spin-wave reflection at the edges.
\par
The dipole field at the site $i$ is calculated as
\begin{equation}
    \bm{H}_{\mathrm{dip}, i} = \frac{1}{4\pi} \sum_{j} \qty(\frac{\bm{M}_{j} - 3\bm{r}_{ij} \qty(\bm{M}_{j} \cdot \bm{r}_{ij})}{r^{3}_{ij}}),
\end{equation}
where $\bm{r}_{ij}$ is the vector from site $j$ to site $i$ and $r_{ij} = |\bm{r}_{ij}|$.
\par
The temporal evolution of the magnetization is computed using the fourth-order Runge-Kutta method with a time step of $\Delta t = 0.1 \, \mathrm{ps}$.
\par

\subsection{Derivation of the dispersion relation}
\begin{figure}[H]
    \centering
    \includegraphics[width=\linewidth]{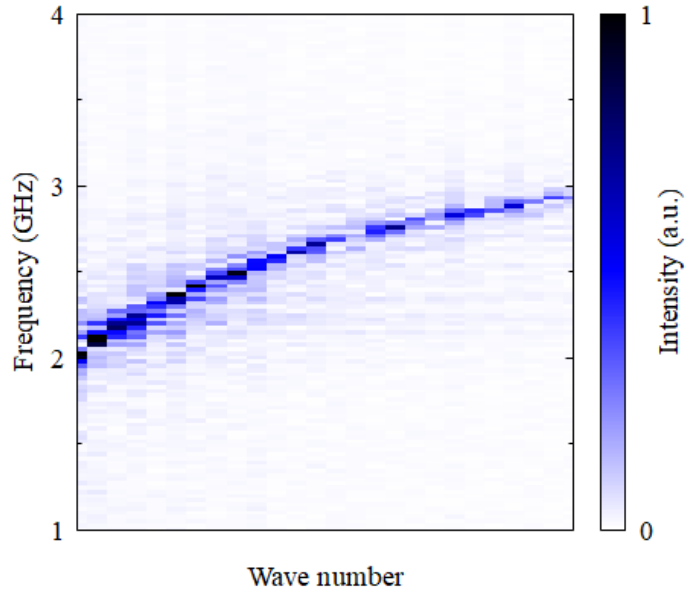}
    \caption{Spin-wave amplitude map in wavenumber and frequency space under excitation by white noise.}
    \label{fig:sim_disp}
\end{figure}

First, we calculate the dispersion relation of the spin waves by applying a random white noise, which contains multiple frequency components, over the entire film. In Fig.~\ref{fig:sim_disp}, the spin-wave amplitude is shown in wavenumber and frequency space. This amplitude is obtained by performing a Fourier transform on the complex signal $M_{x}(x, t) + i M_{z}(x,t)$. The figure shows that the spin-wave frequencies range from 2 GHz (uniform mode) to 3 GHz.

\subsection{Main results of the numerical simulations}
Next, we numerically investigate how spin-wave propagation dynamics depend on the amplitude of the excitation microwave field. We excite spin waves at a frequency of $f_{\mathrm{in}} = 2.5 \, \mathrm{GHz}$ by applying microwaves in the stripline region, and sweep the microwave amplitude $B_{\mathrm{ac}} \, (= \mu_{0}H_{\mathrm{ac}})$ from $1 \, \mathrm{mT}$ to $50 \, \mathrm{mT}$. For this frequency and geometry, the spin-wave wavenumber is approximately $20 \, \mathrm{rad/\mu m}$. To more closely resemble the experimental conditions, we use an excitation microwave field with smoothed edges rather than a perfect rectangular profile, as depicted in Fig.~\ref{fig:sim_strip_dist}. The spatial distribution of the microwave field is assumed to follow the function
\begin{equation}
    H_{\mathrm{ac}}(x) = \frac{H_{\mathrm{ac}}}{2} \qty(1 + \tanh \frac{(x-x_{\mathrm{L}})}{\xi} \tanh \frac{-(x-x_{\mathrm{R}})}{\xi}), 
\end{equation}
where $x_{\mathrm{L}} = 1 \, \mathrm{\mu m}$, $x_{\mathrm{R}} = 9 \, \mathrm{\mu m}$, and $\xi = 1 \, \mathrm{\mu m}$.

\begin{figure}[H]
    \centering
    \includegraphics[width=\linewidth]{fig_sim_strip_dist.pdf}
    \caption{Excitation microwave amplitude distribution.}
    \label{fig:sim_strip_dist}
\end{figure}

We present the results for the spin-wave amplitude at the excitation microwave frequency. Figure~\ref{fig:sim_sw_amp}(a) shows the distribution of the spin-wave amplitude at the input microwave frequency ($f_{\mathrm{in}} = 2.5 \, \mathrm{GHz}$) for various excitation microwave amplitudes. Here, the amplitude distribution is obtained by Fourier transforming the magnetization time trace at each spatial position and extracting the component at the excitation frequency. The figure indicates that at the minimum excitation amplitude ($B_{\mathrm{ac}} = 1 \, \mathrm{mT}$), the amplitude decays gradually in space, whereas for excitation amplitudes of approximately $B_{\mathrm{ac}} = 5 \, \mathrm{mT}$ and above, the decay length becomes shorter, signaling the onset of nonlinear decay. 
\par
Figure~\ref{fig:sim_sw_amp}(b) illustrates the dependence of the spin-wave amplitude at the excitation frequency on the excitation microwave amplitude $B_{\mathrm{ac}}$ at $x = 20 \, \mathrm{\mu m}$. In this plot, the markers represent the simulation results, and the dashed line corresponds to a linear response calibrated at the minimum excitation amplitude ($B_{\mathrm{ac}} = 1 \, \mathrm{mT}$). It can be seen that for $B_{\mathrm{ac}} \gtrsim 5 \, \mathrm{mT}$, the spin-wave amplitude deviates from linear behavior, continues to increase, and eventually saturates. These behaviors are consistent with our experimental observations (see Figs.~3 and 6 in the main text).

\begin{figure}[H]
    \centering
    \includegraphics[width=\linewidth]{fig_sim_sw_amp_rev1.pdf}
    \caption{(a) Dependence of the spin-wave amplitude distribution ($f_{\mathrm{sw}} = f_{\mathrm{in}}$) on the excitation microwave amplitude. (b) Dependence of the spin-wave amplitude ($f_{\mathrm{sw}} = f_{\mathrm{in}}$) at $x = 20 \, \mathrm{\mu m}$ on the excitation microwave amplitude.}
    \label{fig:sim_sw_amp}
\end{figure}

\begin{figure}[H]
    \centering
    \includegraphics[width=\linewidth]{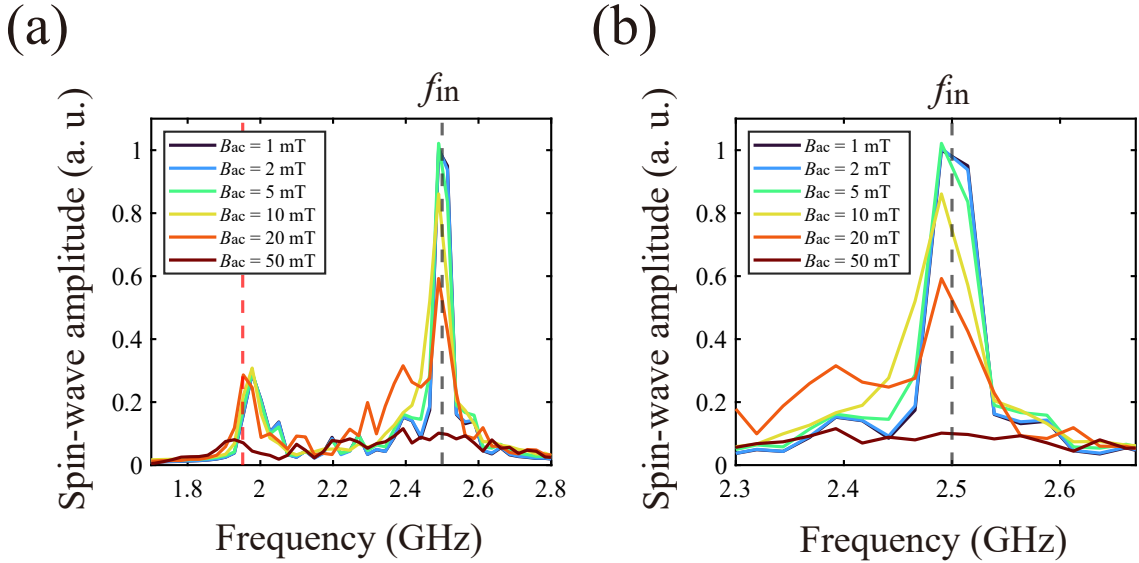}
    \caption{Frequency dependence of spin-wave amplitude in the range in the range of $x = 10 \, \mathrm{\mu m}$ to $x = 20 \, \mathrm{\mu m}$ for each excitation microwave amplitude.
    (a) Wide frequency range. (b) Small frequency range around the excitation frequency.
    The block dotted lines represent the input microwave frequency and the red dotted line in (a) represents the ferromagnetic resonance frequency.}
    \label{fig:sim_sw_amp_freq}
\end{figure}

Next, we present the results for the amplitude of spin waves at frequencies other than the excitation frequency. Figure~\ref{fig:sim_sw_amp_freq}(a) shows the frequency dependence of the spin-wave amplitude in the spatial range from $x = 10 \, \mathrm{\mu m}$ to $x = 20 \, \mathrm{\mu m}$ for each excitation microwave amplitude. Figure~\ref{fig:sim_sw_amp_freq}(b) shows a magnified view of the area around the excitation frequency. Note that the amplitude for each excitation power is normalized by dividing by the spin-wave amplitude by the microwave amplitude and subsequently by the spin-wave amplitude at $B_{\mathrm{ac}} = 1 \, \mathrm{mT}$. From Fig.~\ref{fig:sim_sw_amp_freq}(b), we observe that in the nonlinear regime (when $B_{\mathrm{ac}}$ exceeds 10 mT), the spin-wave amplitude at the excitation frequency (2.5 GHz) decreases, while the frequency distribution of the spin-wave amplitude broadens. This result is consistent with theoretical predictions that spin waves at other frequencies are generated via multi-magnon scattering in the nonlinear region.

\subsection{Summary of the numerical simulation}
We have conducted numerical simulations of the spin-wave dynamics and demonstrated that the spin-wave amplitude at the excitation frequency deviates from linearity above a certain excitation microwave amplitude. Furthermore, the simulations confirmed that, in the nonlinear regime, spin waves at frequencies other than the excitation frequency are generated. These results are consistent with our experimental observations, supporting the validity of our interpretation.

For simplicity, our simulations were performed in one dimension along the $x$-direction for a thin film with a thickness of 10 nm, which only partially replicates the geometry of the experimental sample. By adjusting the simulation conditions to better match the actual sample and performing further validation, a more quantitative comparison should be achievable. This represents a promising direction for future work.

\bibliography{citation_SM}